\documentclass{aa}
\usepackage[colorlinks=true,citecolor=blue, urlcolor=blue, linkcolor=red]{hyperref}
\usepackage[varg]{txfonts}

\makeatletter
\renewcommand*\aa@pageof{, page \thepage{} of \pageref*{LastPage}}
\makeatother

\usepackage{amsmath,amssymb,listings,upgreek,nicefrac}
\usepackage{mathptmx}   %
\usepackage{grffile}
\usepackage{xcolor,xspace}
\usepackage[normalem]{ulem}  %

\usepackage[T1]{fontenc}
\usepackage{textcomp}     %

\def\mH{m_\mathrm{H}}                              %
\def\MJ{M_{\mathrm{J}}}                            %
\def\RJ{R_{\mathrm{J}}}                            %
\def\MSun{{M_\odot}}                               %
\def\LSun{{L_\odot}}                               %

\def\Lya{Ly\,$\alpha$\xspace}                             %
\def\Ha{H\,$\alpha$\xspace}                               %
\def\Hb{H\,$\beta$\xspace}                                %
\def\Paa{Pa\,$\alpha$\xspace}                             %
\def\Pab{Pa\,$\beta$\xspace}                              %
\def\Bra{Br\,$\alpha$\xspace}                             %
\def\Brg{Br\,$\gamma$\xspace}                             %
\def\Ks{$K_{\mathrm{s}}$\xspace}                          %
\def\LkCab{LkCa\,15\,b\xspace}                            %
\def\PDS{PDS\,70\xspace}                                  %
\def\PDSb{PDS\,70\,b\xspace}                              %
\def\PDSc{PDS\,70\,c\xspace}                              %
\def\Dlrmb{Delorme\,1\,(AB)b\xspace}                      %

\def\kms{\mathrm{km}\,\mathrm{s^{-1}}}             %
\def\MdotUJ{\MJ\,\mathrm{yr}^{-1}}                 %
\def\MdotUS{\MSun\,\mathrm{yr}^{-1}}               %

\def\Mdot{\ensuremath{\dot{M}}\xspace}             %
\def\MP{\ensuremath{M_{\mathrm{p}}}}\xspace        %
\def\RP{\ensuremath{R_{\mathrm{p}}}}\xspace        %
\def\ffill{f_{\mathrm{fill}}}                      %
\def\ff{\ffill}                                    %
\newcommand{\Lacc}{\ensuremath{{L_{\textnormal{acc}}}}\xspace}  %
\newcommand{\Facc}{\ensuremath{{F_{\textnormal{acc}}}}\xspace}  %
\def\LHa{\ensuremath{{L_{\textnormal{H}\,\alpha}}}\xspace}    %
\def\LHamax{{L_{\textnormal{H}\,\alpha,\,\textnormal{max}}}}  %
\def\FHa{{F_{\textnormal{H}\,\alpha}}}             %
\def\FHamod{{F_{\textnormal{H}\,\alpha}^{\textnormal{mod}}}}             %
\def\FHb{{F_{\textnormal{H}\,\beta}}}              %
\def\FBra{{F_{\textnormal{Br}\,\alpha}}}           %
\def\fluxratobs{\varphi_{\textnormal{obs}}}        %
\def\fluxratth{\varphi_{\textnormal{theo,\,surf}}}        %
\def\fluxratthred{\varphi_{\textnormal{theo,\,rednd}}}        %
\def\pUV{p_{\textnormal{UV}}}                      %
\def\pNIR{p_{\textnormal{NIR}}}                    %
\newcommand{\RAkk}{{R_{\textnormal{acc}}}}         %
\newcommand{\RHill}{{R_{\textnormal{Hill}}}}       %
\newcommand{\RBondi}{{R_{\textnormal{Bondi}}}}     %
\newcommand{\Tint}{{T_{\textnormal{int}}}}         %
\newcommand{\Teff}{{T_{\textnormal{eff}}}}         %
\def\Mach{\mathcal{M}}                             %
\def\Pram{P_{\textnormal{ram}}}                    %
\def\AR{A_R}                              %
\def\AU{A_U}                                       %
\def\AHa{\ensuremath{A_{\mathrm{H}\,\alpha}}\xspace}  %
\def\AV{A_V}                              %
\def\RV{R_V}                            %
\def\AK{A_K}                              %
\def\DF{\Delta F}                                  %
\def\lmbdHa{\lambda_{\mathrm{H}\,\alpha}}          %
\def\lmbdHb{\lambda_{\mathrm{H}\,\beta}}           %
\def\lmbdBra{\lambda_{\mathrm{Br}\,\alpha}}        %
\def\taudHa{\tau_{\mathrm{dust,\,H}\,\alpha}}
\def\kappaStbint{\kappa_{\bullet,\,\mathrm{H}\,\alpha}}  %
\def\kapStbfpg{\kappa_{\mathrm{dust,\,H}\,\alpha}} %
\def\Sigmad{\Sigma_{\mathrm{dust}}}
\def\Sigwhered{\tilde{\Sigma}_{\textrm{gas}}}      %
\def\vK{v_{\mathrm{K}}}                            %
\def\taumax{{\tau_{\mathrm{max}}}}                 %
\def\fC{f_\mathrm{C}}                              %
\def\thmin{\theta_\mathrm{min}}                    %
\def\thmax{\theta_\mathrm{max}}                    %

\usepackage{etoolbox}

\makeatletter

\patchcmd{\NAT@citex}
  {\@citea\NAT@hyper@{%
     \NAT@nmfmt{\NAT@nm}%
     \hyper@natlinkbreak{\NAT@aysep\NAT@spacechar}{\@citeb\@extra@b@citeb}%
     \NAT@date}}
  {\@citea\NAT@nmfmt{\NAT@nm}%
   \NAT@aysep\NAT@spacechar\NAT@hyper@{\NAT@date}}{}{}

\patchcmd{\NAT@citex}
  {\@citea\NAT@hyper@{%
     \NAT@nmfmt{\NAT@nm}%
     \hyper@natlinkbreak{\NAT@spacechar\NAT@@open\if*#1*\else#1\NAT@spacechar\fi}%
       {\@citeb\@extra@b@citeb}%
     \NAT@date}}
  {\@citea\NAT@nmfmt{\NAT@nm}%
   \NAT@spacechar\NAT@@open\if*#1*\else#1\NAT@spacechar\fi\NAT@hyper@{\NAT@date}}
  {}{}

\makeatother
\def\degr{{\mbox{\textdegree}}}

\makeatletter
\let\jnl@style=\rm
\def\ref@jnl#1{{\jnl@style#1}}

\def\aj{\ref@jnl{AJ}}                   %
\def\actaa{\ref@jnl{Acta Astron.}}      %
\def\araa{\ref@jnl{ARA\&A}}             %
\def\apj{\ref@jnl{ApJ}}                 %
\def\apjl{\ref@jnl{ApJ}}                %
\def\apjs{\ref@jnl{ApJS}}               %
\def\ao{\ref@jnl{Appl.~Opt.}}           %
\def\apss{\ref@jnl{Ap\&SS}}             %
\def\aap{\ref@jnl{A\&A}}                %
\def\aapr{\ref@jnl{A\&A~Rev.}}          %
\def\aaps{\ref@jnl{A\&AS}}              %
\def\azh{\ref@jnl{AZh}}                 %
\def\baas{\ref@jnl{BAAS}}               %
\def\bac{\ref@jnl{Bull. astr. Inst. Czechosl.}}
\def\caa{\ref@jnl{Chinese Astron. Astrophys.}}
\def\cjaa{\ref@jnl{Chinese J. Astron. Astrophys.}}
\def\icarus{\ref@jnl{Icarus}}           %
\def\jcap{\ref@jnl{J. Cosmology Astropart. Phys.}}
\def\jrasc{\ref@jnl{JRASC}}             %
\def\memras{\ref@jnl{MmRAS}}            %
\def\mnras{\ref@jnl{MNRAS}}             %
\def\na{\ref@jnl{New A}}                %
\def\nar{\ref@jnl{New A Rev.}}          %
\def\pra{\ref@jnl{Phys.~Rev.~A}}        %
\def\prb{\ref@jnl{Phys.~Rev.~B}}        %
\def\prc{\ref@jnl{Phys.~Rev.~C}}        %
\def\prd{\ref@jnl{Phys.~Rev.~D}}        %
\def\pre{\ref@jnl{Phys.~Rev.~E}}        %
\def\prl{\ref@jnl{Phys.~Rev.~Lett.}}    %
\def\pasa{\ref@jnl{PASA}}               %
\def\pasp{\ref@jnl{PASP}}               %
\def\pasj{\ref@jnl{PASJ}}               %
\def\rmxaa{\ref@jnl{Rev. Mexicana Astron. Astrofis.}}%
\def\qjras{\ref@jnl{QJRAS}}             %
\def\skytel{\ref@jnl{S\&T}}             %
\def\solphys{\ref@jnl{Sol.~Phys.}}      %
\def\sovast{\ref@jnl{Soviet~Ast.}}      %
\def\ssr{\ref@jnl{Space~Sci.~Rev.}}     %
\def\zap{\ref@jnl{ZAp}}                 %
\def\nat{\ref@jnl{Nature}}              %
\def\iaucirc{\ref@jnl{IAU~Circ.}}       %
\def\aplett{\ref@jnl{Astrophys.~Lett.}} %
\def\apspr{\ref@jnl{Astrophys.~Space~Phys.~Res.}}
\def\bain{\ref@jnl{Bull.~Astron.~Inst.~Netherlands}} 
\def\fcp{\ref@jnl{Fund.~Cosmic~Phys.}}  %
\def\gca{\ref@jnl{Geochim.~Cosmochim.~Acta}}   %
\def\grl{\ref@jnl{Geophys.~Res.~Lett.}} %
\def\jcp{\ref@jnl{J.~Chem.~Phys.}}      %
\def\jgr{\ref@jnl{J.~Geophys.~Res.}}    %
\def\jqsrt{\ref@jnl{J.~Quant.~Spec.~Radiat.~Transf.}}
\def\memsai{\ref@jnl{Mem.~Soc.~Astron.~Italiana}}
\def\nphysa{\ref@jnl{Nucl.~Phys.~A}}   %
\def\physrep{\ref@jnl{Phys.~Rep.}}   %
\def\physscr{\ref@jnl{Phys.~Scr}}   %
\def\planss{\ref@jnl{Planet.~Space~Sci.}}   %
\def\procspie{\ref@jnl{Proc.~SPIE}}   %

\def\ptp{\ref@jnl{Prog.~Th.~Phys.}}   %

\makeatother

\defcitealias{goli04}{G04}
\defcitealias{marl07}{M07}
\defcitealias{scvh}{SCvH}
\defcitealias{bl94}{BL94}
\defcitealias{mc14}{MC14}
\defcitealias{ensman94}{E94}
\defcitealias{lever81}{LP81}

\newcommand{\K}[1]{}
\def\fpg{f_{\textrm{d/g}}}
\def\amax{a_{\rm max}}
\def\amin{a_{\rm min}}
\newcommand{\sigSB}{{\sigma}}

\def\Lint{\ensuremath{L_{\mathrm{int}}}\xspace}    %
\def\Lphot{\ensuremath{L_{\mathrm{phot}}}\xspace}  %
\def\Lshock{\ensuremath{L_{\mathrm{shock}}}\xspace}  %
\def\Hi{\textnormal{H\,\textsc{i}}}

\newcommand{\MStern}{{M_{\star}}}
\newcommand{\MSternen}{{M_{\star,\,1}}}
\newcommand{\TZerst}{{T_{\textnormal{dest}}}}
\newcommand{\RZerst}{{R_{\textnormal{dest}}}}
\newcommand{\TPop}{{T_{\textnormal{pop}}}}
\newcommand{\TGas}{{T_{\textnormal{gas}}}}
\newcommand{\rmax}{{r_{\rm max}}}

\def\Tmax{T_{\rm max}}

\begin{document}

\title{Accreting protoplanets: Spectral signatures and magnitude\\of gas and dust extinction at {\boldmath H\,$\alpha$}}
\titlerunning{Accreting protoplanets: Extinction by gas and dust at {\boldmath H\,$\alpha$}}

\author{%
G.-D.~Marleau\inst{\ref{Tue},\ref{Bern},\ref{MPIA}}\and
Y.~Aoyama\inst{\ref{Bei1},\ref{Bei2}}\and
R.~Kuiper\inst{\ref{HD},\ref{Tue},\ref{MPIA}}\and
K.~Follette\inst{\ref{Amherst}}\and
N.~J.~Turner\inst{\ref{JPL}}\and
G.~Cugno\inst{\ref{ETH}}\and
C.~F.~Manara\inst{\ref{ESO}}\and
S.~Y.~Haffert\inst{\ref{AZ}}\thanks{NASA Hubble Fellowship Program Sagan Fellow}\and
D.~Kitzmann\inst{\ref{CSH}}\and
S.~C.~Ringqvist\inst{\ref{SU}}\thanks{N\'e S.\ C.\ Eriksson}\and
K.~R.~Wagner\inst{\ref{AZZ}}\footnotemark[1]$^,$\thanks{NASA NExSS Alien Earths Team}\and   %
R.~van Boekel\inst{\ref{MPIA}}\and
S.~Sallum\inst{\ref{Irvine}}\and
M.~Janson\inst{\ref{SU}}\and
T.~O.~B.~Schmidt\inst{\ref{HS}}\and
L.~Venuti\inst{\ref{NASA},\ref{SETI}}\and
Ch.~Lovis\inst{\ref{UGe}}\and
C.~Mordasini\inst{\ref{Bern}}
}

\authorrunning{Marleau et al.}

\institute{%
Institut f\"ur Astronomie und Astrophysik,
Universit\"at T\"ubingen,
Auf der Morgenstelle 10,
D-72076 T\"ubingen, Germany\\
\email{gabriel.marleau@uni-tuebingen.de}
\label{Tue}\and
Physikalisches Institut,
Universit\"{a}t Bern,
Gesellschaftsstr.~6,
CH-3012 Bern, Switzerland
\label{Bern}\and
Max-Planck-Institut f\"ur Astronomie,
K\"onigstuhl 17,
D-69117 Heidelberg, Germany
\label{MPIA}\and
Institute for Advanced Study,
Tsinghua University,
Beijing 100084,
People's Republic of China
\label{Bei1}\and
Department of Astronomy,
Tsinghua University,
Beijing 100084,
People's Republic of China%
\label{Bei2}\and
Institut f\"ur Theoretische Astrophysik (ITA),
Universit\"at Heidelberg,
Albert-Ueberle-Str.~2,
D-69120 Heidelberg,
Germany 
\label{HD}\and
Physics and Astronomy Department,
Amherst College,
25 East Drive,
Amherst,
MA 01002,
USA
\label{Amherst}\and
Jet Propulsion Laboratory,
California Institute of Technology,
4800 Oak Grove Drive,
Pasadena,
CA 91109,
USA
\label{JPL}\and
European Southern Observatory,
Karl-Schwarzschild-Stra\ss{}e 2,
D-85748 Garching bei M\"unchen,
Germany
\label{ESO}\and
Institute for Particle Physics and Astrophysics,
ETH Z\"urich,
Wolfgang-Pauli-Strasse 27,
CH-8093 Z\"urich,
Switzerland
\label{ETH}\and
University of Arizona,
1200~E University Blvd.,
Tucson, AZ 85721,
USA
\label{AZ}\and
Centre for Space and Habitability,
Universit\"{a}t Bern,
Gesellschaftsstr.~6,
CH-3012 Bern,
Switzerland
\label{CSH}\and
Institutionen f\"or astronomi,
AlbaNova universitetscentrum,
Stockholms universitet,
SE-106 91 Stockholm,
Sweden
\label{SU}\and
Department of Astronomy and Steward Observatory,
University of Arizona,
933~N Cherry Ave,
Tucson,
AZ 85719,
USA
\label{AZZ}\and
Department of Physics and Astronomy,
University of California,
4129 Frederick Reines Hall,
Irvine, CA 92697-4575,
USA
\label{Irvine}\and
Hamburger Sternwarte,
Gojenbergsweg 112,
D-21029 Hamburg,
Germany
\label{HS}\and
NASA Ames Research Center,
Moffett Blvd,
Mountain View,
CA 94035,
USA
\label{NASA}\and
SETI Institute,
189 Bernardo Avenue, Suite 200,
Mountain View,
CA 94043,
USA
\label{SETI}
\and
Observatoire astronomique de l'Universit\'e de Gen\`eve,
51 ch.\ des Maillettes,
CH-1290 Versoix,
Switzerland
\label{UGe}
}

\date{ -- %
/ 
17 November 2021%
}

\abstract%
{Accreting planetary-mass objects have been detected at H\,$\alpha$, but targeted searches have mainly resulted in non-detections. Accretion tracers in the planetary-mass regime should originate from the shock itself,
making them particularly susceptible to extinction by the accreting material.
High-resolution ($R>50,000$) spectrographs operating at H\,$\alpha$\ should soon enable one to study how the incoming material shapes the line profile.
}%
{We calculate how much the gas and dust accreting onto a planet reduce the H\,$\alpha$ flux from the shock at the planetary surface and how they affect the line shape. We also study the absorption-modified relationship between the H\,$\alpha$ luminosity and accretion rate.
}%
{We computed the high-resolution radiative transfer of the H\,$\alpha$ line using a one-dimensional velocity--density--temperature structure for the inflowing matter in three representative accretion geometries: spherical symmetry, polar inflow, and magnetospheric accretion. For each, we explored the wide relevant ranges of the accretion rate and planet mass. We used detailed gas opacities and carefully estimated possible dust opacities.
}%
{At accretion rates of $\dot{M}\lesssim3\times10^{-6}~M_{\mathrm{J}}\,\mathrm{yr}^{-1}$, gas extinction is negligible for spherical or polar inflow and at most $A_{\mathrm{H}\,\alpha}\lesssim0.5$~mag for magnetospheric accretion. Up to $\dot{M}\approx3\times10^{-4}~M_{\mathrm{J}}\,\mathrm{yr}^{-1}$, the gas contributes $A_{\mathrm{H}\,\alpha}\lesssim4$~mag. This contribution decreases with mass. We estimate realistic dust opacities at H\,$\alpha$ to be $\kappa\sim0.01$--10~cm$^2$\,g$^{-1}$, which is 10--$10^4$ times lower than in the interstellar medium.
Extinction flattens the $\LHa$--$\dot{M}$ relationship, which becomes non-monotonic with a maximum luminosity $\LHa\sim10^{-4}~\LSun$ towards $\dot{M}\approx10^{-4}~M_{\mathrm{J}}\,\mathrm{yr}^{-1}$ for a planet mass $\sim10~M_{\mathrm{J}}$.
In magnetospheric accretion, the gas can introduce features in the line profile, while the velocity gradient smears them out in other geometries.
}%
{For a wide part of parameter space, extinction by the accreting matter should be negligible, simplifying the interpretation of observations, especially for planets in gaps. At high $\dot{M}$, strong absorption reduces the H\,$\alpha$ flux, and some measurements can be interpreted as two $\dot{M}$ values. Highly resolved line profiles ($R\sim10^5$) can provide (complex) constraints on the thermal and dynamical structure of the accretion flow.}


\maketitle

\section{Introduction}

In order to test and constrain planet formation models, it is crucial to detect planets not only shortly after their formation, but also during.
The timeline of the mass assembly of gas giants---their accretion rate history---is
an important and barely constrained aspect of planet formation with consequences for the migration history
and thus chemical composition as well as dynamics of forming planetary systems,
and thus of the assembly of possibly life-bearing planets.
The detection of accretion tracers such as \Ha\ provides a unique window into this key phase.

So far, a few low-mass accreting objects have been detected through their \Ha emission. Overall, dedicated surveys with different strategies have not yet been successful in revealing accreting planets. However, upcoming instrumental upgrades offer the hope of soon uncovering a large population of forming planets. We review this in Section~\ref{sec:instr}.

The interpretation of both detections and non-detections
requires models of the emission and radiation transfer of the accretion tracers.
Despite decades of work by a few groups (e.g.\ \citealp{muzerolle01,romanova04,kurosawa12}),
the equivalent problem in low-mass star formation is not entirely solved.
The planetary case is even less well understood but there have been recent theoretical developments.
Indeed, \citet{Aoyama+2018} and \citet{Aoyama+2020} presented detailed models of the emission of tracers from an accretion shock on the circumplanetary disc (CPD) or the planet surface, respectively.
\citet{Thanathibodee+2019} presented the first predictions of a model of magnetospheric accretion for low-mass stars (Classical T~Tauri Stars; CTTSs) applied to planets.

What these studies have not modelled in detail is absorption by the material,
both gas and dust, accreting onto the planet.
This is the subject of this study for the case of the planet-surface accretion shock.
Recently, in \citet{Sanchis+2020} and \citet{szul20}, this was undertaken using global disc simulations 
with a very different density and temperature structure near the planet.
We discuss this in Section~\ref{sec:cfSanchisSzul}.

We study \Ha since it is the most commonly used accretion tracer and indeed usually exhibits the strongest signal.
Estimating the strength of the absorption will inform studies of accretion tracers
both in a static and in a (more realistic) time-varying picture.

The paper is laid out as follows:
Section~\ref{sec:instr} highlights some aspects of the current observational and instrumental landscape.
In Section~\ref{sec:model} we 
present three possible accretion geometries and the details of the macroscopic and microscopic quantities for the accretion flow.
The effect of the absorption by the gas is studied in Section~\ref{sec:abs gas},
which presents integrated fluxes and line profiles at extremely high spectral resolution across the parameter space.
Then, we deal with absorption by dust in Section~\ref{sec:cont abs dust}.
In Section~\ref{sec:obscons}, we discuss the resulting \Ha-luminosity--accretion rate relationship and apply our results to known accreting low-mass objects.
Section~\ref{sec:disc}
presents a discussion of our model
and in Section~\ref{sec:summ} we summarise and conclude.
Appendix~\ref{sec:cffull} discusses %
the contribution of
the emission from the accreting material in the radiative transfer, and Appendix~\ref{sec:line profiles COLD} presents additional line profiles.
\section{Current and future instrumentation}
 \label{sec:instr}

Thanks to instrumentation advances (VLT/SPHERE, VLT/MUSE, LBT/LBTI, Magellan/MagAO; \citealp{Bacon+2010,Close+2014,Close+2014b,Schmid+2017,Schmid+2018}),
a handful of low-mass companions to young stars have been
detected and explored that (possibly) show signs
of ongoing accretion: \LkCab (\citealp{Sallum+2015}; but see also \citealt{mendigut18,Currie+2019}), \PDSb\ and~c \citep{keppler18,mueller18,wagner18,Haffert+2019,Zhou+2021}, and \Dlrmb \citep{Eriksson+2020}.
On the other hand,
recent surveys searching for further accreting planets through their \Ha signal have returned null results,
from five \citep{Cugno+2019} and eleven \citep{Zurlo+2020} different sources\footnote{%
   \citet{Zurlo+2020}
   do not include \PDS in order to analyse the known planets in that system separately.%
} (see also \citealt{xie20}).

However, more companions might be revealed by ongoing and future searches at \Ha\ with various instruments.
One is Subaru/SCExAO+VAMPIRES (\citealp{lozi18,Uyama+2020}), with $R\gtrsim1000$. Another is VLT/MUSE, which provides at \Ha in narrow-field mode (NFM) the currently highest available spectral resolution of 
$R=2516$, corresponding to an instrumental spectral full width at half maximum (FWHM) of $\approx0.26$~nm (see for example Figure~B.3 of \citealt{Eriksson+2020}).
A similar resolution will be afforded by
HARMONI, the first-light spectrograph on the Extremely Large Telescope (ELT)\footnote{See \url{http://harmoni-elt.physics.ox.ac.uk}.}
\citep{thatte16,rodrigues18},  %
with $R\approx3300$ blueward of 0.8~$\upmu$m.

Also, there are hopes from upgrades to existing instruments or upcoming or proposed instruments. Indeed, several efforts focussed on pushing visible direct imaging (photometry) or spectroscopy to the limit are being developed with \Ha as their main science case. Imaging instruments include:
Magellan/MagAO-X \citep{Males+2018,Close+2018} and 
GMagAO-X on the Giant Magellan Telescope (GMT) \citep{males19},
the Keck Planet Imager and Characterizer (KPIC) at the Keck~II telescope \citep{Jovanovic+2019},
and also possibly the near-infrared (NIR) spectrograph NIRSpec\footnote{\label{Fn:NIRSpec}Only the $R\sim100$ prism can access \Ha (see \url{https://jwst-docs.stsci.edu}), making it challenging but still possible.} on the \textit{James Webb Space Telescope} (JWST)
or the Coronograph Instrument\footnote{\url{https://wfirst.ipac.caltech.edu/sims/Param_db.html}} (CGI) aboard the \textit{Nancy Grace Roman Space Telescope} (formerly known as WFIRST). %
A few spectrographs covering \Ha are also being planned and developed:
an integral-field spectrograph of $R=15,000$  %
named Visible Integral-field Spectrograph eXtreme (VIS-X) %
at MagAO-X
\citep{haffert21visx};
a proposed
Extreme-Adaptive-Optics- (XAO)-assisted high-resolution spectrograph for the VLT,
dubbed RISTRETTO\footnote{See \url{https://zenodo.org/record/3356296}.} \citep[PI: Ch.~Lovis;][]{chazelas20},
with $R\geqslant130,000$--$150,000$;  %
as well as the Replicable High-resolution Exoplanet and Asteroseismology (RHEA) spectrograph for
Subaru/SCExAO
\citep[][]{rains16,rains18,anagnos20}, which should provide $R\approx60,000$.
For both RISTRETTO and RHEA, a low throughput could, however, make observations challenging.
We note that VLT/UVES \citep{dekker00} has a resolution of approximately $R=100,000$ at \Ha (and $R=80,000$ at \Hb), with some variations depending on the exact setting. One of the main limitations for using UVES for exoplanet purposes is that it is seeing limited, so that unlike AO-assisted instruments such as MUSE, UVES generally cannot spatially resolve the planetary flux. In that case, the only way to study the \Ha line would be if it were strong enough to be comparable to the flux of the primary star.
However, in cases where a companion is a few arcseconds (i.e.\  a few hundreds of astronomical units at 150~pc) away and its \Ha line is reasonably strong, it could still be possible to get a spatially resolved signature with UVES, \Dlrmb \citep{Eriksson+2020} being one potential example.%

Finally, one should also mention %
the Extremely Large Telescope (ELT), expected to come online in the next decade, with its second-generation High Resolution Spectrograph (HIRES).
While it does not cover \Ha, its Integral Field Unit (IFU) is sensitive to $1.0$--1.8~$\upmu$m at $R\approx100,000$--$150,000$ (\citealp{Marconi+2016,Marconi+2018,tozzi18}), which includes shock emission lines such as \Pab \citep{Aoyama+2018} and He\,\textsc{i}~$\lambda$\,10830. Thanks to this high resolution, it could also be a powerful help in characterising accreting planets.
\section{Physical model: Macro- and microphysics}
 \label{sec:model}

Here we detail
the geometries we consider for the accretion (Section~\ref{sec:geo}),
the parameter space of accretion rate, planet mass, and planet radius  (Section~\ref{sec:par space}),
the structure of the flow (Section~\ref{sec:preshock}),
the calculation of the input line profiles (Section~\ref{sec:Aoyamasumm}),
the radiative transfer in the accretion flow (Section~\ref{sec:radtrans}),
and the gas opacities used (Section~\ref{sec:gas}).

\subsection{Accretion geometry}
 \label{sec:geo}

The geometry of accretion onto forming gas giants is an open question.
In the classical, %
simplified picture of accreting gas giants,
matter begins at a finite starting radius $\RAkk$ and falls freely onto the planet
in a spherically symmetric fashion (e.g.\ \citealp{pollack96,boden00,morda12_I,helled14ppvi,m16Schock,m18Schock}).
Based on the arguments in, for exmaple, \citet{ginzburg19a}, may be a reasonable model, at least on the scale of the Bondi radius $\RBondi$ for sub-thermal planets,
that is, for planets whose Hill radius $\RHill=a(\MP/3\MStern)^{1/3}$
is smaller than the pressure scale height of the circumstellar disc (CSD),
with $a$ the semi-major axis and $\MStern$ the stellar mass.

More realistically, or at least for higher-mass planets,
radiation-hydrodynamical simulations have suggested the presence of
meridional circulation around an accreting planet:
motion from the upper layers of the protoplanetary disc towards the planet at its location,
where it has opened at least a partial gap
(e.g.\ \citealp{Kley+2001,Machida+2008,tanigawa12,gressel13,morbidelli14,fung15,fung16,szul16,szul19II,dong19,schulik19,schulik20,bailey21,rabago21};
see also \citealp{b19}). %
Thanks to the exquisite sensitivity of ALMA,
\citet{teague19} provided the first observational evidence
for such a pattern over a scale of a few Hill radii in a young disc thought to contain accreting planets, HD~163296, and recently obtained a similar result for HD~169142 \citep{yu21}.
For such a geometry to hold, however, the gap opened by the planet needs to be not too wide,
and once higher masses have been reached, the large width of the gap
($\propto{\MP}^{0.5}$; \citealp{kanagawa17}, but see also \citealp{bergez20}, who explore how time-dependent disc models lead to a different opening criterion than in equilibrium discs) should lead to accretion
from the CSD onto a circumplanetary disc (CPD),
and from this onto the planet.
This will hold especially in multiple-planet systems in which forming gas giants have opened a common gap \citep{Close2020}.  %
As \citet{ginzburg19b} point out, this might be a long phase of the accretion process,
over which a significant fraction of the planet mass could be assembled.

How the gas then ultimately reaches the forming planet, with a size of order $\RP\sim(0.01\textrm{--}0.001)\RHill$, is an open question.
From angular momentum conservation, and by analogy with objects across a large range of mass scales, a part of the matter likely goes through a CPD (which in its outer regions may be a decretion disc, i.e.\  exhibit an outflow; \citealp{tanigawa12}; the general theory of decretion discs is presented in \citealp{lyndenbell74,pringle91,nixon21}) while the rest could fall directly onto the proto-gas-giant with a polar shock (see \citealp{b19b,b19}). However, the fraction of the gas processed through a CPD is unknown, as is
how the gas leaves the CPD to be incorporated into the planet. Obvious possibilities are processes invoked for CTTSs (see review in \citealp{hartmann16}), including
boundary-layer accretion (e.g.\ \citealp{lyndenbell74,kley96,piro04}; see brief review in \citealp{geroux16}),
or
magnetospheric accretion
(\citealp{lovelace11,batygin18,Cridland2018}; see also \citealp{fendt03}).
As in the stellar context (see reviews in \citealp{romanova15,mendigut20}), which of these two mechanisms dominates will depend on the magnetic field strength of the forming planet and the coupling of the gas to the magnetic field.
In the boundary-layer accretion (BLA) scenario there is no accretion shock on the planet surface. Therefore, we do not treat BLA here; in this case the observed \Ha\ would need to come from a shock on the CPD \citep{Aoyama+2018}. %

There are suggestions that both the magnetic field and the coupling of the gas to the magnetic field could be strong enough for magnetospheric accretion to occur \citep{lovelace11,batygin18}. Applying the \citet{christensen09} scaling to the luminosities of young planets, \citet{katarzy16} found that they should have a surface dipole field $B\sim0.5$--1~kG. Using typical accretion rates \citep{morda12_I}, the resulting estimate of the magnetospheric (or Alfv\'en) radius is usually larger than the planet radius, which holds even more for forming planets, which might have an even higher luminosity \citep{mordasini17}. Taken together with the estimate of a relatively low magnetic diffusivity, this suggests that planets could indeed accrete magnetospherically \citep{batygin18,hasegawa21}.  %
Nevertheless, one should keep in mind
that the validity of the \citet{christensen09} scaling for forming planets has yet to be shown, especially since they are not necessarily fully convective \citep{berardo17,berardocumming17}.
Estimating whether planets can accrete magnetospherically might also depend on the field topology,
which for CTTSs has significant non-dipole components \citep{hartmann16}.%

If magnetospheric accretion onto planets does take place, 
it could be in analogy to the stellar case,
in which the material is lifted from the disc, following the magnetic field lines connecting it to the protostar. Alternatively, or simultaneously,
mass newly supplied could be coming from above, in the downward part of a meridional flow (of size $\gtrsim\RHill$), and thus falling onto the apex of the magnetic fields lines \citep{batygin18}.
However,
for the shock this detail should not matter much
since in both cases the velocity of the infalling matter will be essentially
the free-fall velocity, even though possibly starting effectively from different accretion radii.

\begin{figure*}[t] %
 \centering
 \includegraphics[width=0.95\textwidth]{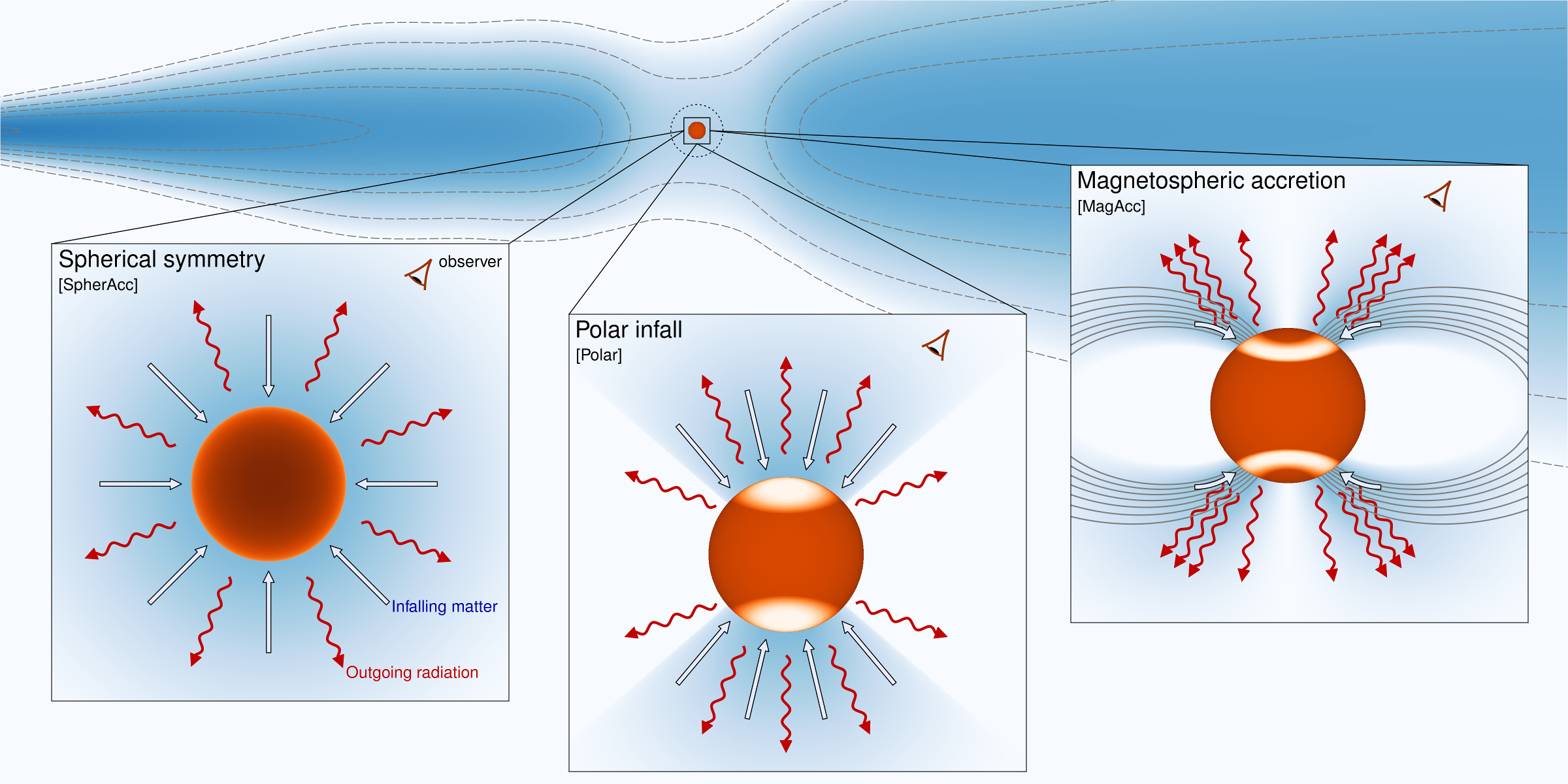}
\caption{%
Schematic view of the accretion scenarios considered in this work.
In spherical symmetry (\texttt{SpherAcc}), the gas (straight thick lines)
is assumed to be distributed uniformly on the surface of the planet ($\ffill=100~\%$), %
whereas in the case of polar infall (\texttt{Polar}) the accreting gas is concentrated near the pole regions
(assumed to be uniform within $\ffill\approx30~\%$, and zero outside of this).
In the magnetospheric accretion case (\texttt{MagAcc}), the magnetic field of the protoplanet
leads the gas onto a small fraction of the surface (here shown as polar-ring hot spots; $\ffill\sim10~\%$). See Table~\ref{tab:scenarios}.
At slightly larger scales (the insets show out to $\RAkk$) the gas could be coming from above, falling onto the apex of the magnetic field lines \citep{batygin18}, or from a circumplanetary disc (purposefully not shown).
In all cases, the radiation (\Ha-coloured squiggly lines) reaching the observer
passes through the matter (gas and dust) accreting onto the planet and 
bears the spectral imprint of this material,
which is the subject of this paper. \textit{Illustration by Th.\ M\"uller (MPIA/Haus der Astronomie).}
}
\label{fig:scenarios}
\end{figure*}

Since in all, the accretion geometry is very uncertain, we consider 
three basic geometries for the accretion shock on the planet surface,
as illustrated in Figure~\ref{fig:scenarios}:
\begin{enumerate}
 \item \texttt{SpherAcc}: spherically symmetric accretion,
 \item \texttt{Polar}: accretion concentrated towards the planet's magnetic poles, with no accretion outside,
 \item \texttt{MagAcc}: magnetospheric accretion, with accretion only along one or several column(s), with overall a small filling factor $\ffill$, which is the fraction of the planet's surface covered by the (global) accretion rate.
\end{enumerate}
The \texttt{MagAcc} case differs from the \texttt{Polar} case by the filling factor (only quantitatively) but also by the length of the accretion flow over which the gas can affect the radiation (see the end of this subsection).

We treat the hydrodynamics and the radiative transfer approximately with one-dimensional models
by following in each geometry the radiative transfer along the flow.
For \texttt{SpherAcc} this is natural,
for \texttt{Polar} this is effectively averaged over the cone angle with a non-zero accretion rate,
and for \texttt{MagAcc} it is as for \texttt{Polar}, but for an even smaller region (a thin accretion stream or several).
Table~\ref{tab:scenarios} describes these different scenarios
in terms of the filling factor $\ffill$, the size of the ``accretion radius'' $\RAkk$
from which the gas is starting at rest \citep{boden00,morda12_I},
and the integration outer limit $\rmax$ for the optical depth calculation
(see Equation~(\ref{eq:L pure abs})).

\begin{table}
\caption{Accretion geometries considered in this work.}
\label{tab:scenarios}
\centering
\begin{tabular}{r|ccccc} %
\hline 
\hline 
Model name  &  $\ffill$\tablefootmark{a}   & $\thmin$\tablefootmark{b}  & $\thmax$\tablefootmark{b}       &  $\RAkk$\tablefootmark{c}  &  $\rmax$\tablefootmark{d} \\
\hline 
\multicolumn{6}{c}{\textit{Spherical accretion}} \\  %
\hline 
  \texttt{SpherAcc} &   1.0                   & $0\degr$  & $90\degr$      &  $\infty$  &  $100~\RP$ \\  %
\hline 
\multicolumn{6}{c}{\textit{Polar inflow}} \\
\hline 
  \texttt{Polar}    &   0.3  & $0\degr$  & $46\degr$  &  $\infty$  &  $100~\RP$ \\
\hline 
\multicolumn{6}{c}{\textit{Magnetospheric accretion}} \\
\hline 
  \texttt{MagAcc}   &   0.1                   & $20\degr$  & $50\degr$   &  $5~\RP$  &  $\approx\frac{1}{2}\RAkk$ \\
\hline
\end{tabular}
\tablefoot{
The different geometries are illustrated in Figure~\ref{tab:scenarios}.
Model names in the paper are suffixed with \texttt{Warm} or \texttt{Cold} depending on whether the radius fit $\RP(\Mdot,\MP)$ 
to cold-accreting or warm-accreting planets (see Section~\ref{sec:par space}) is used.
\tablefoottext{a}{Fraction of the planet's surface covered by the accretion flow.}
\tablefoottext{b}{Opening angles of the accretion cone covering each pole (Equation~(\ref{eq:ffilltheta})); for illustration.}
\tablefoottext{c}{Starting position of the infalling gas with a radial velocity $v_r=0$ there.}
\tablefoottext{d}{Outer limit of the optical radiative transfer calculation, limited to $\rmax\approx100~\RP$ due to uncertainty in the structure of the far layers. The exact value is however inconsequential for \texttt{SpherAcc} and \texttt{Polar}.}
}
\end{table}

In the polar case, the accretion is assumed to be uniform within an axisymmetric cone or ring and zero outside, with the observer looking down into this region.
The accreting region is defined by
\begin{equation}
 \label{eq:ffilltheta}
 \ffill = \cos\thmin-\cos\thmax,
\end{equation}
where $\thmin=0$ %
for
the pole ($\thmin\neq0$ for a ring; see e.g.\ \citealt{kulkarni13}) and $\thmax\leqslant90\degr$ is the equivalent opening angle of the 
accreting region 
at each pole. For $\ffill=0.3$ (e.g.\ 15~\%\ of the total area at the north pole and 15~\%\ at the south pole), the equivalent opening angle from $\thmin=0\degr$ is $\thmax=46\degr$.
For a ring with $\ffill=0.1$ starting and at $\thmin=20\degr$ (for example), $\thmax=50\degr$, while $\ffill=0.01$ would require $\thmax=21.6\degr$, or $\thmax=12.9\degr$ if instead $\thmin=10\degr$.
In reality, there is evidence for multiple accretion components in observations, in agreement with simulations (e.g.\ \citealp{Ingleby+2013,robinson19,robinson21}), but our assumption of a constant local accretion rate and an infinitely sharp transition between the accreting and the non-accreting region is a minor one.
Also, given that very edge-on systems are less likely to be detected or observed, there is a relatively large probability of viewing the planet indeed within
$45\degr$ of pole-on. %
Therefore, we assume that the radiation is travelling only radially (i.e.\ away from the planet) through the accretion region towards the observer,
and neglect the possibility of scattering out of the accretion cone into the line of sight towards an observer not looking down into the accretion cone.

For the magnetospheric case, in analogy to CTTSs, we consider $\RAkk=5~\RP$ \citep{calvetgull98},
but also $\RAkk=2~\RP$ for comparison, as the magnetic field may be weaker, whether in total or in its dipole component.
We assume that the flux coming from the postshock region (the accretion ``hot spots'') passes tangentially through the accretion arc to the observer and thus travels a distance $\rmax\approx 1/3 \times\pi \RAkk/2 \approx \RAkk/2$, as illustrated in Figure~\ref{fig:scenarios}. Therefore, curvature can be ignored at the level of our approximation, and the radiative transfer can be performed also here purely radially.

This assumption of observing along the accretion column leads to a strong estimate of the amount of absorption by the accreting gas. If the optical depth is high along the accretion column, it might be more realistic for the photons to escape preferentially at an angle from the base of the accretion footpoints (see the pole-side photons in Figure~\ref{fig:scenarios}). In this case, they would travel relatively unimpeded towards the observer. Scattering out of the accretion arc close to the planet could also contribute to the flux, depending on the system's inclination. Thus our approach of integrating along a segment of the flow is a simple one that should maximise potential signatures.
\subsection{Parameter space}
 \label{sec:par space}

For a geometry given by $(\ffill,\RAkk,\rmax)$, the main quantities defining the parameter space are
the mass accretion rate $\Mdot$, the planet mass $\MP$, and the planet radius $\RP$.
We consider a range of accretion rates $\Mdot\approx3\times10^{-8}$--$3\times10^{-4}~\MdotUJ$. %
The high end of the $\Mdot$ range %
is higher than usually found in planet formation models \citep{morda12_I,Tanigawa+Tanaka2016} but could be relevant to an accretion outburst akin to the situation of FU Orionis stars \citep{audard14}.
With these high accretion rates, we can address whether we would indeed observe a signal in the (perhaps unlikely) event a planet were caught outbursting.
This range also covers the accretion rate inferred for the \PDS\ planets
\citep{Haffert+2019,manara19}.

Concerning the mass, we focus on the gas-giant and low-mass-brown-dwarf regime with $\MP\approx1$--$20~\MJ$. At much higher masses, the velocity of the gas at the shock is too high for \Ha\ to be generated in the postshock region, and the hydrogen lines are thought to be emitted by the accreting gas \citep{hartmann16,AMIM21L}.

Both to reduce the dimensionality and to provide guidance
as to a possible trend, we do not keep the planet radius as a free parameter
but instead adopt for definiteness the fits\footnote{For convenience, they can be found in a few popular languages in the ``Suite of Tools to Model Observations of accRetIng planeTZ'' at \url{https://github.com/gabrielastro/St-Moritz}.
Radii of up to $\RP\approx9$--12~$\RJ$ are reached for $\Mdot=3\times10^{-4}~\MdotUJ$.
This might seem high but is in line with the high-entropy models of \citet{Spiegel+Burrows2012} and \citet{mc14}.} by \citet{Aoyama+2020} of the radius as a function of accretion rate and planet mass in
the population synthesis calculations\footnote{Since the planet structure model is the same for gas giants, the newest-generation population synthesis of \citet{Emsenhuber+2020a,Emsenhuber+2020b} yields the same results, only with fewer synthetic gas giants and therefore less statistically robust fits.} of \citet{morda12_I,morda12_II},
for their scenarios of ``cold accretion'' and of ``hot-accretion'' (see also \citealp{mordasini17}). We call these fits respectively ``Cold-population fit'' and ``Warm-population fit''.
Towards high accretion rates and masses, the radius increases with either quantity, and the warm-population fit is larger than the cold-population fit.
Typical values are $\RP=1.5$--3~$\RJ$ and 2--4~$\RJ$, with a stronger dependence of $\RP$ on $\Mdot$ in the ``hot'' case.
The ``cold-'' and ``hot accretion'' scenarios represent extreme outcomes of the accretion process
in which the entire accretion energy is respectively radiated away at the shock or on the contrary brought into the planet \citep{marl07,morda12_I,m16Schock,m18Schock}.
We assume that the radius depends only on the total accretion rate and the mass but not on the filling factor.

A complication is that in reality, the accretion history, not only the instantaneous values, likely sets the radius. This is suggested by the results of \citet{berardo17}, who find that during formation, planets can have a radiative (and not convective) structure for a large fraction of their outer mass layers. Thus the results of classical convective-planet structures that let us write $\RP=\RP(\Mdot,\MP)$, with no time dependence, might be a simplification. However, we effectively mitigate this by considering the two different fits (``cold'' and ``hot''), and will find that the radius does not have a major influence in any case.

Finally, there is another, minor parameter: the interior flux from the planet, characterised by an effective temperature $\Tint$.
We assumed somewhat arbitrarily that $\Tint\approx1100$~K for all models.
However, the interior flux is relevant only at the lowest accretion rates,
for which there will be essentially no absorption, so that in practice it is not important.

\subsection{Structure of the accretion flow}
 \label{sec:preshock}

\begin{figure}[t] %
 \centering
 \includegraphics[width=0.5\textwidth]{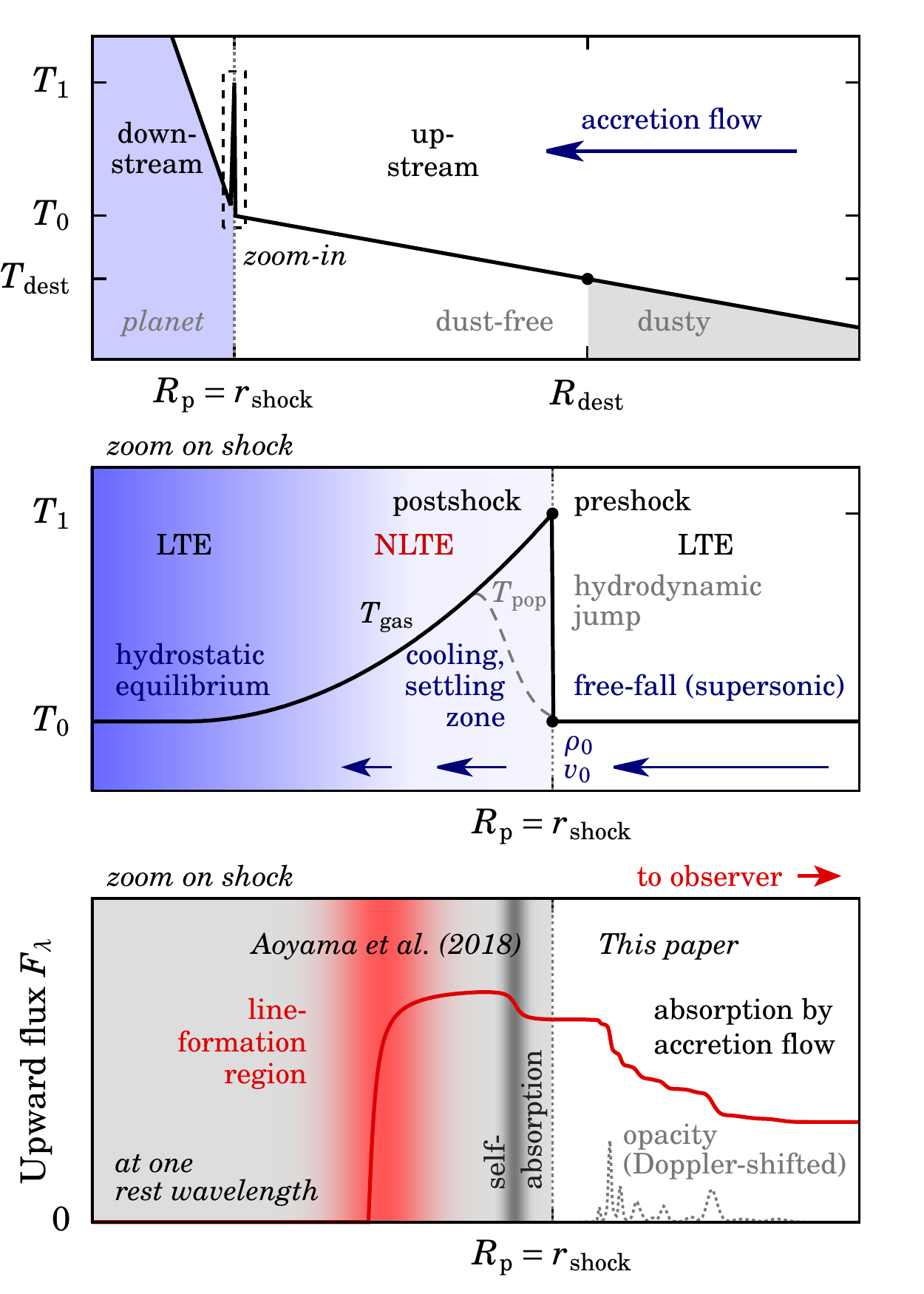}
\caption{%
Overview of the structure of the accretion flow %
and of the shock region (see labels). %
\textit{Top panel:} temperature (double logarithmic scale), %
showing the shock temperature $T_0$, the Zel'dovich spike up to $T\approx T_1$, %
and the dust destruction front at radius $\RZerst$. %
The shock defines the planet's radius.
The temperature flattening at $T\approx\TZerst$ (e.g.\ \citealp{m18Schock}) is not shown.
The next two panels focus on the dashed box.
\textit{Middle panel:} temperature near the shock on a linear scale. Below the hydrodynamic jump, in the NLTE cooling and settling zone, the ``population temperature'' of the electrons $\TPop$ and the kinetic temperature $\TGas$ differ at first.
The temperature structure is simplified;
Figure~8 of \citet{vaytetgonz13} gives a more accurate schematic.
\textit{Bottom panel:} upward-moving 
flux at one restframe wavelength within the \Ha line (red line),  %
from the microphysical models of \citet{Aoyama+2018}.
We illustrate the line-formation region (red) and the self-absorption (dark grey) in the settling zone, and the
absorption by the accretion flow,
calculated in this work.
The features in the opacity (dotted grey curve; arbitrary linear scale)
come from the Doppler-shift gradient (Equation~(\ref{eq:nu vel}); Figure~\ref{fig:kappa2D}).
}
\label{fig:Teresa}
\end{figure}

For all accretion geometries, the supersonic infalling matter makes up the accretion flow\footnote{In
    \citet{m16Schock,m18Schock}, this whole region (out to $\RAkk$) is often called the ``preshock region'', meant as a synonym. However, for clarity, we keep here the term ``preshock'' for the layers immediately before (i.e.\ upstream of) the shock, as is more common in the literature.}
before hitting the planet's surface, defined as the location of the radiative hydrodynamical shock \citep{m16Schock,m18Schock}.
This is illustrated in Figure~\ref{fig:Teresa}.
Immediately below the hydrodynamical shock is the postshock region proper, a spatially thin ``settling zone'' and cooling region. There, the gas is subsonic and gradually slows down, reaching hydrostatic equilibrium at depths, where
the structure is that of a non-accreting planet. The %
postshock region %
consists of
the layers close to but below the shock, especially the thin cooling region. Note that an accreting planet does not %
have an atmosphere in the classical sense (for an isolated object), or does so at most only on the parts of its surface that are not accreting.

We now describe the mechanical and thermal structure of the accreting matter.
We assume that the supersonic accretion flow is essentially spherically symmetric with purely radial motion. In general, the accretion is non-zero within a region whose boundaries in angle are set by $\ffill$.
For $\ffill=1$ we have true spherical symmetry.
Therefore, the velocity and density in the accretion flow are given by
\begin{subequations}
\label{eq:v}
\begin{align}
  v(r) =&~ \sqrt{2G\MP\left(\frac{1}{r}-\frac{1}{\RAkk}\right)}\\
     =&~ 73~\kms \left(\frac{\MP}{3~\MJ}\right)^{1/2}  \left(\frac{r}{2~\RJ}\right)^{-1/2}\zeta^{1/2}, \label{eq:vspelledout} %
\end{align}
\end{subequations}
\begin{subequations}
\label{eq:rho}
\begin{align}
  \rho(r) =&~ \frac{\Mdot}{4\pi r^2 \ffill v(r)} \label{eq:rhoa}\\
          = &~3\times10^{-12}~\mathrm{g\,cm}^{-3}\frac{1}{\ffill} \left(\frac{\Mdot}{\upmu\MdotUJ}\right)  \notag\\ %
          &\left(\frac{\MP}{3~\MJ}\right)^{1/2} 
          \left(\frac{r}{2~\RJ}\right)^{-3/2}\zeta(r)^{-1/2},  \label{eq:rhob}
\end{align}
\end{subequations}
where
\begin{equation}
 \zeta(r)\equiv\left(1-\frac{r}{\RAkk}\right),
\end{equation}
with $\zeta=0.8$ for $\RAkk=5\RP$ and $\zeta\rightarrow1$ when $\RAkk\gg r$. For a given accretion rate, the density depends on the filling factor but the velocity does not. These are the classical formulae for an accretion flow (e.g.\ \citealp{calvetgull98,zhu15}) and have been verified to apply to planets by \citet{m16Schock,m18Schock} through one-dimensional radiation-hydrodynamical simulations. For reference, the shock Mach number is in the range $\Mach\approx10$--40 \citep{m18Schock}, and the ram pressure of the incoming gas is given by (e.g.\ \citealp{berardo17})
\begin{subequations}
\begin{align}
\label{eq:Pram}
 \Pram =&~ \rho(\RP) v(\RP)^2 = \frac{\Mdot v(\RP)}{4\pi\RP^2\ffill}\\
  \approx &~171~\mathrm{erg\,cm}^{-3}\frac{1}{\ffill} \left(\frac{\Mdot}{\upmu\MdotUJ}\right)  \notag\\ %
          &\left(\frac{\MP}{3~\MJ}\right)^{1/2} 
          \left(\frac{\RP}{2~\RJ}\right)^{-5/2}\zeta^{1/2}.
\end{align}
\end{subequations}

For the temperature structure of the flow, we use the results of \citet{m16Schock,m18Schock}. They have found that the radiative precursor \citep{drake06,vaytetgonz13} of the shock extends to the Hill sphere (formally to infinity). This means that the infalling gas is preheated by the radiation escaping from the shock, which the work of \citet{Aoyama+2018} suggests occurs mainly through the absorption of \Lya\ photons since they carry most of the shock emission flux. This contrasts to the stellar case, in which the precursor region is thin and located close to the star (see e.g.\ \citealt{calvetgull98,colombo19,deS19}) and is due to the absorption of photoionising radiation.
This infinite precursor implies that the temperature profile is given by
\begin{equation}
     T(r) = T_0 \left(\frac{r}{\RP}\right)^{-1/2}.\label{eq:T} %
\end{equation}
In turn, the equilibrium gas temperature at the shock\footnote{A quick tool
  to compute these temperature and density profiles is provided at \url{https://github.com/gabrielastro/St-Moritz},
  which also computes the time since the beginning of free-fall and provides a more accurate temperature profile than Equation~(\ref{eq:T}) when the opacity is high (see Section~\ref{sec:Tstructr}).} $T_0$
has the value needed to radiate the mechanical energy converted to radiation as well as the interior flux:
\begin{subequations}
\label{eq:T0 not historical...}
\begin{align}
 ac T_0^4 &= \Facc + \sigSB \Tint^4\\
          &= \frac{G\MP\Mdot}{4\pi\RP^3\ffill}
          \zeta
          + \sigSB \Tint^4,\label{eq:T0b}
\end{align}
\end{subequations}
where $v_0=v(\RP)$, $\rho_0=\rho(\RP)$, and
$a$ and $\sigSB$ are respectively the radiation and Stefan--Boltzmann constants.
\citet{m16Schock} found that the entire mechanical energy goes into radiation. Thus the radiation flux from accretion is $\Facc = 0.5 \rho_0 v_0^3$, leading to %
Equation~(\ref{eq:T0b}). %
When $\Mdot\approx0$, the choice of $\Tint\approx1100$~K sets a minimum of $T_0\approx800$~K (since $4\sigSB=ac$; $\Tint$ is an effective temperature but $T_0$ a material temperature).%

Equation~(\ref{eq:T0 not historical...}) uses the result from \citet{m16Schock,m18Schock} that the planetary accretion shock is supercritical and thus isothermal, in the sense that upstream and downstream of the Zel'dovich spike the gas temperature is set by the balance of the incoming flux of material energy and the outgoing radiative flux.
This single temperature $T_0$ is shown in Figure~\ref{fig:Teresa}.
The shock is a ``thick--thin'' shock in the usual classification (e.g.\ \citealp{drake06}): the radiation is diffusive below and free-streaming above the shock.
(For recent discussion of sub- and supercritical shocks, see \citet{commer11} and \citet{vaytet13,vaytetgonz13}, and importantly \citet{drake07} for a brief critical review of inadequate uses of these terms in the literature.) In the limit that the kinetic energy dominates over the internal flux, the shock temperature is \citep{m18Schock}  %
\begin{align}
 \label{eq:T0approx} %
T_0 \approx
    &~1280~{\rm K}~\frac{1}{\ffill^{1/4}}\left(\frac{\RP}{2~\RJ}\right)^{-3/4} \notag \\
    &\times\left(\frac{\Mdot}{\upmu\MdotUJ}\right)^{1/4}  %
    \left(\frac{\MP}{3~\MJ}\right)^{1/4}\zeta^{1/4}.
\end{align}
Since we focus on a line, \Ha, that carries a negligible fraction of the total flux \citep{Aoyama+2018},
we assume that the temperature in the accretion flow is fixed and independent of the
absorption or non-absorption of the \Ha\ photons.

Note that Equation~(\ref{eq:T}) assumes a radially constant luminosity in the accretion flow given by
\begin{subequations}
\begin{align}
    L =&~\Lacc+\Lint=\frac{G\MP\Mdot}{\RP}\zeta + \Lint,\\
      =&~4\times10^{-4}~\LSun~\left(\frac{\Mdot}{\upmu\MdotUJ}\right)  \left(\frac{\MP}{3~\MJ}\right)\left(\frac{\RP}{2~\RJ}\right)^{-1}\zeta + \Lint,
\end{align}
\end{subequations}
where $\Lint=4\pi r^2 \sigSB \Tint^4$ is the interior luminosity from the planet.
\citet{m18Schock} showed that a constant $L(r)$ is a good approximation. However, Equation~(\ref{eq:T}) does not reflect the region of a flatter, almost constant, $T(r)$ profile where the dust is destroyed, near a destruction temperature $\TZerst=1220\times{\rho_{-11}}^{0.0195}$~K, where $\rho_{-11}\equiv\rho\times10^{11}~\mathrm{cm}^3\,\mathrm{g}^{-1}$ \citep{isella05}. In particular, Equation~(\ref{eq:T}) assumes that the frequency-averaged radiation is free-streaming throughout. Thus, if the accretion flow is devoid of dust, Equation~(\ref{eq:T}) will hold, while in the presence of dust this will slightly underestimate the temperature at a given radial position (and thus density). However, at these low temperatures $T\lesssim1200$~K, the gas opacity is small anyway and we find (see Section~\ref{sec:abs gas}) that, in the relevant part of the parameter space, the \Ha\ is not extincted. This justifies approximately the simplification.

\begin{figure}[t] %
 \centering
 \includegraphics[width=0.47\textwidth]{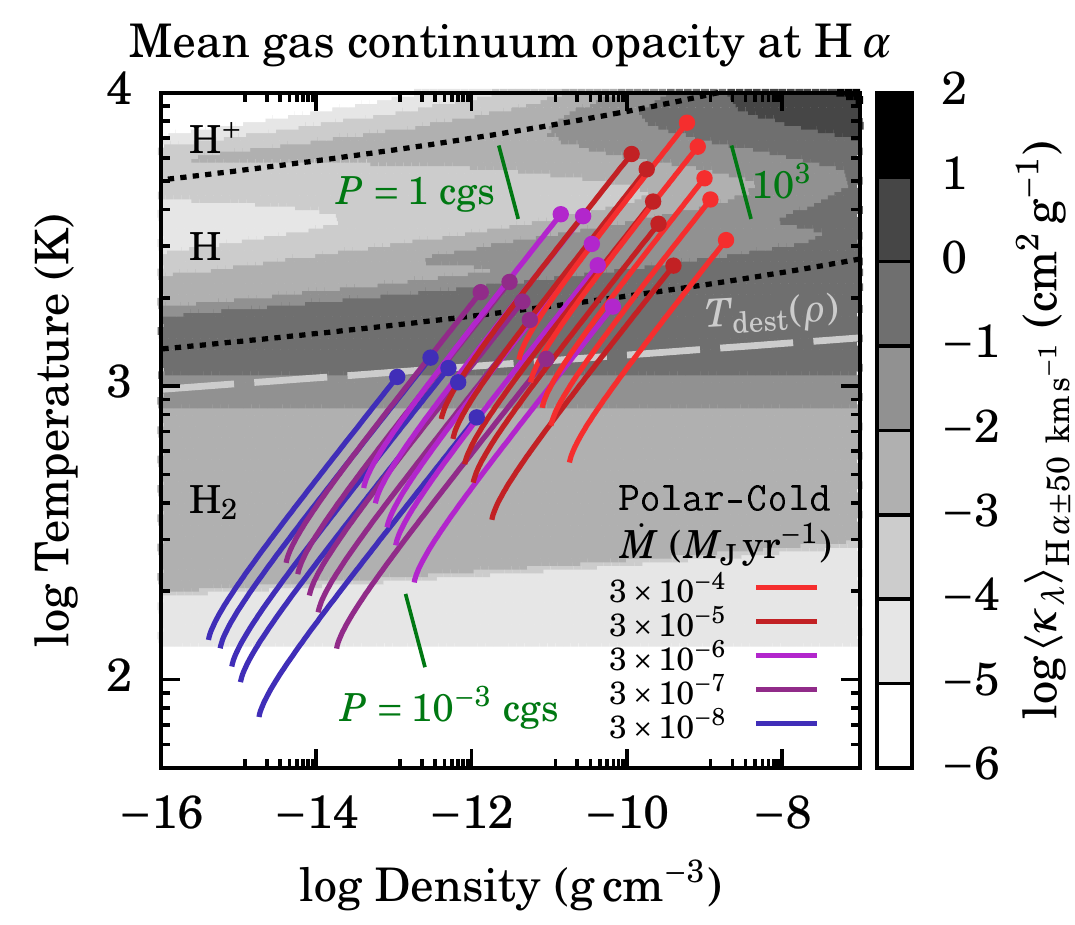} %
\caption{
Structure of the accretion flow (solid lines) and gas opacity (background greyscale).
The profiles are for a grid of accretion rates (isochromatic curve groups; from $\Mdot=3\times10^{-4}$ (top right)
to $3\times10^{-8}~\MdotUJ$ (bottom left)) and masses ($\MP=1$, 3, 5, 10, 20~$\MJ$ from bottom to top
within each group) in the \texttt{Polar-Cold} case.
Each profile begins on the left at $\rmax\approx100\RP$ and ends at the shock at $\RP$ (dot), at $(\rho_0,T_0)$.
Only the continuum opacity at \Ha\ (no resonance; see text) is shown averaged over $\Delta v=\pm50~\kms$.
The dust destruction line $\TZerst(\rho)$ is indicated (\citealp{isella05}; long-dashed line), and line segments show pressures $P=10^{-3}$, 1, $10^3$~erg\,cm$^{-3}$.
Black dotted lines show where half of the hydrogen is molecular, atomic, or ionised.
The opacity depends mostly on temperature, and most models $(\Mdot,\MP)$ include a region of high opacity, often near the shock.%
}
\label{fig:kappa gas Ha}
\end{figure}

We have scaled Equations~(\ref{eq:v}),~(\ref{eq:rho}), and~(\ref{eq:T0approx}) using reasonable values (with $\Mdot$ close to the minimum $\Mdot\gtrsim5\times10^{-7}~\MdotUJ$ derived by \citealp{Hashimoto+2020} for \PDSb) but one should remember that the parameter space is large, with especially $\Mdot$, $\MP$, and $\ffill$ varying by orders of magnitude. Profiles in $\rho$--$T$ space are shown for a range of parameters in Figure~\ref{fig:kappa gas Ha}. Densities are $\rho\sim10
^{-16}$--$10^{-9}$~g\,cm$^{-3}$ and temperatures $T\sim100$--$10^4$~K, with approximately $T\propto\rho^{1/3}$, as can be seen from Equations~(\ref{eq:rho}) and~(\ref{eq:T0approx}) for $\zeta\approx1$. An extensive discussion of the profiles, in particular against radial distance from the planet, is given in \citet{m18Schock}.

\subsection{Spectral profiles of shock line emission}
 \label{sec:Aoyamasumm}

The hydrogen line emission at the shock is taken from the models of \citet{Aoyama+2020}. These apply the non-LTE radiation-hydrodynamical simulations of \citet{Aoyama+2018} to the shock at the surface of an accreting planet. Through detailed calculations of chemical reactions and electron transitions in hydrogen atoms, the \citet{Aoyama+2018} models calculate the cooling in the disequilibrium region immediately below the hydrodynamical shock, corresponding roughly to the downstream part of the Zel'dovich spike \citep{vaytetgonz13}.
See Figure~\ref{fig:Teresa}c.
This provides high-resolution profiles and line-integrated fluxes for 55~hydrogen lines in the series of Lyman, Balmer, Paschen, Brackett, etc.,
which are emitted from the shock towards the observer. These line profiles thus serve as the input for our calculation of 
the radiative transfer
(Equation~(\ref{eq:L pure abs}) below), 
and will be seen as the black dashed lines in Figures~\ref{fig:profilesWARM100}--\ref{fig:profilesMagAkk}. The line-integrated luminosity $\LHa$ as a function of $\Mdot$ and $\MP$ is shown in Figure~7 of \citet{Aoyama+2020} and constitutes one of their main results.%

The \citet{Aoyama+2018} microphysical shock models depend only on
the number density of hydrogen protons $n_0$ and the velocity $v_0$, both evaluated immediately before the shock. Thus $v_0=v(\RP)$ from Equation~(\ref{eq:v}) and $n_0=X\rho(\RP)/\mH$ from Equation~(\ref{eq:rho}), where $X$ is the hydrogen mass fraction and $\mH$ is the hydrogen atomic mass.
The number ratios are given by
$\textrm{H}:\textrm{He}:\textrm{C}:\textrm{O}=1:10^{-1.07}:10^{-3.48}:10^{-3.18}$ \citep{Allen2000}.
The present iteration of the models \citep{Aoyama+2018} covers a range $n_0=10^9$--$10^{14}$~cm$^{-3}$ and $v_0=20$--200~$\kms$.

With the fits of the planet radius $\RP=\RP(\Mdot,\MP)$ that were mentioned in Section~\ref{sec:par space}, Equations~(\ref{eq:v}) and~(\ref{eq:rho}) evaluated at $r=\RP$ relate the macrophysical parameters $(\Mdot,\MP,\ffill)$ to the microphysical ones, $(n_0,v_0)$.
For $\ffill=1$, this leads to ranges
of $n_0\sim10^{10}$--$10^{14}$~cm$^{-3}$
and $v_0\sim50$--200~$\kms$
for typical $(\Mdot,\MP,\RP)$ values.
For small filling factors ($\ffill\lesssim10~\%$) or the largest accretion rates ($\Mdot/\ffill\gtrsim 1\times10^{-4}~\MdotUJ$ for the cold-population radius fit
and $\Mdot/\ffill\gtrsim3\times10^{-3}~\MdotUJ$ for $\MP\gtrsim10~\MJ$ in the warm-population fit), the resulting $n_0$ was higher than available in the previously mentioned grid and we extrapolated the spectra in preshock density. Nevertheless, this should not introduce much inaccuracy on the spectral shape or total flux.

The \Ha line is surrounded by a continuum, set by the flux from the planet's interior. However, it is usually insignificant compared to the strong line emission. Therefore, we will neglect the continuum in the main part of this work, but do discuss this approximation in Section~\ref{sec:impcont} and Appendix~\ref{sec:cffull}.%

\subsection{Radiative transfer}
 \label{sec:radtrans}

We calculate the radiative transfer only within the accreting region, assuming spherical symmetry. For the \texttt{Polar} or \texttt{MagAcc} case, this ignores edge effects for the radiation travelling close to the walls of the accretion cone or the accretion columns, respectively; the matter and radiation properties are assumed to be independent of the angle within that region. This simplification should lead only to a modest overestimate of the amount of absorption, and is in line with other approximations in our approach.

In spherical symmetry, 
the radial radiative flux $F_\lambda$ is given in general by
\begin{equation}
 \label{eq:F}
 F_\lambda(r) = 2\pi\int_{-1}^1 I_\lambda(r,\mu)\mu\,d\mu,
\end{equation}
where $I_\lambda$ is the specific intensity in a given direction and $\mu=\cos\xi$,
with $\xi$ the angle between the given direction and the radial direction.
The intensity $I_\lambda$ is set by the radiation transfer equation, which reads \citep{davis12}
\begin{equation}
 \label{eq:dIds}
 \hat{n}\cdot \nabla I_\lambda(\hat{n}) = \frac{dI_\lambda}{ds} = \alpha_\lambda\left(S_\lambda - I_\lambda\right),
\end{equation}
where 
$\hat{n}$ is a unit vector defining a direction,
$S_\lambda=j_\lambda/\alpha_\lambda$ is the source function with $j_\lambda$ the emissivity,
$\alpha_\lambda$ is the coefficient of extinction including scattering and true absorption,
and $s$ is the position along the ray defined by $\hat{n}$.
The middle term in Equation~(\ref{eq:dIds}) is written for the specific intensity in the direction of the ray.  %

We assume that the accretion flow is in local thermodynamic equilibrium (LTE), i.e.\ that collisions determine the electron populations in the accretion flow. Therefore,
$S_\lambda=B_\lambda$ by Kirchhoff's law, with $B_\lambda$ the Planck function.

In general, Equation~(\ref{eq:dIds}) must be integrated numerically.
However, the peak intensity of the \Ha\ line %
corresponds to that of a blackbody usually at a much higher temperature
than the gas anywhere in the accretion flow, so that $I_\lambda \gg S_\lambda = B_\lambda$.
Therefore, in Equation~(\ref{eq:dIds}) the absorption term dominates over the emission, leading to
\begin{equation}
  \frac{dI_\lambda}{ds} \approx -\alpha I_\lambda. \label{eq:dIdsabs}
\end{equation}

Since we are concerned with the flux at the observer (at infinity), $r\gg\RP$ and $dI_\lambda/ds$ needs to be integrated only along a radial ray.
This effectively neglects limb darkening.  %
As mentioned before, we assume that the radiative quantities are independent of angle within the accretion flow. Therefore, 
we only need to integrate Equation~(\ref{eq:dIdsabs})
radially outwards from the shock at $\RP$.
The solution is $I_\lambda(r) = I_\lambda(\RP)\exp\left(-\Delta\tau_\lambda(r)\right)$,
where %
\begin{equation}
\label{eq:dtau}
 \Delta\tau_\lambda(r)\equiv\int_{\RP}^{r} \alpha_\lambda(r') \, dr'. 
\end{equation}
With this,
Equation~(\ref{eq:F}) implies that the flux at large $r$ is
\begin{subequations}
\begin{align}
 F_\lambda & = 2\pi\int_{-1}^1 I_\lambda(\RP) e^{-\Delta\tau_\lambda} \mu\, d\mu\\
   &\approx 2\pi I_\lambda(\RP) e^{-\Delta\tau_\lambda} \int_{1-0.5(\RP/r)^2}^{1} \mu\,d\mu  \label{eq:L middle step}\\
   &= F_\lambda(\RP) \frac{\RP^2}{r^2} e^{-\Delta\tau_\lambda},  \label{eq:L pure abs}  %
\end{align}
\end{subequations}
taking $I_\lambda(\RP)$ for Equation~(\ref{eq:L middle step}) to be constant (neglecting limb darkening) and non-zero only
over the small angle $\Delta\xi\approx\RP/r$ (around $\hat{n}$)
subtended by the planet, %
so that also $F_\lambda(\RP)=\pi I_\lambda(\RP)$.
We use Equation~(\ref{eq:L pure abs}) to calculate the flux at the observer throughout this work and do in Appendix~\ref{sec:cffull} a comparison with the full solution to Equation~(\ref{eq:dIds}). For most cases the approximation is excellent.

\subsection{Gas opacity}
 \label{sec:gas}

The coefficient of absorption $\alpha_\lambda=\kappa_\lambda\rho$ (with dimensions of inverse length),
where $\kappa_\lambda$ is the opacity,
is divided into two contributions for the gas:
the \Ha resonant (i.e.\ particularly strong) opacity due to the electrons in the $n=2$ quantum energy level,
as well as
a \mbox{(pseudo-)}continuum, made up of a true continuum and
the superposition of many line wings. The resonant and continuum components are described in the following subsections.

The Doppler shift due to the bulk motion of the infalling gas
is taken into account by evaluating for a given observer-frame (restframe) frequency $f$
the opacity at frequency
\begin{equation}
\label{eq:nu vel}
 f'(v) = f\left(1+\frac{v}{c}\right),
\end{equation}
where $v$ is the velocity at a given position.
This explains the strong variations of the monochromatic opacity shown as a grey dashed line in Figure~\ref{fig:Teresa}c.
Given the massive uncertainties on the dust absorption,
we treat it separately in Section~\ref{sec:cont abs dust}.
Note that for simplicity we do not include scattering as this would introduce a disproportionate level of complexity (especially for the realistic case of anisotropic scattering) compared to the rest of our approach.

\subsubsection{Resonant opacity}
 \label{sec:reskap}

We use standard formulae to calculate the resonant opacity (e.g.\ \citealp{carson88,hilborn02,sharpburrows07,Wiese+Fuhr2009,Hubeny+Mihalas2014}). Since we deal with temperatures much lower than what corresponds to \Ha, we do not include stimulated emission, and approximate the ground state to be dominantly populated, which implies that the partition function is $Q(T)\approx2$. With this approximation, we do not need to handle the well-known divergence of $Q$, which, however, can be corrected easily by the occupation probability formalism \citep{Hubeny+1994}. We assume a Doppler line profile, appropriate for our temperature regime.

The strength of the resonant opacity is proportional to the number of absorbers, calculated from the Saha equation (e.g.\ \citealp{D'Angelo+Bodenheimer2013}). Only for the highest values of $\Mdot/\ffill$ and $\MP$ is the gas at least partially ionised when reaching the shock; for most $(\Mdot,\MP)$ combinations, the hydrogen is atomic (see black dotted lines in Figure~\ref{fig:kappa gas Ha}). At low $\Mdot/\ffill$ the hydrogen is molecular.

\subsubsection{Continuum opacity}
 \label{sec:contkap}

For the continuum gas opacity,
we use very-high-resolution LTE absorption coefficients with a constant step size in wavenumber of 0.01~cm$^{-1}$, corresponding to 
a spectral resolution $R=1.5\times10^6$. This is  sufficient to resolve line cores over the relevant $(P,T)$ domain \citep{moll15,moll19}. We assume a solar-metallicity \citep{Asplund+2009} mixture. The opacities of the individual species were calculated with \texttt{HELIOS-K} \citep{Grimm+Heng2015,grimm21}.  %
All atoms and ions up to a proton number $Z=40$ are included,
and so are the continua of H$^-$ and collision-induced absorption (CIA) of
H$_2$--H$_2$, He--H, and H$_2$—He.
Molecular absorption is included for H$_2$O , CO, TiO, SH, and VO.
Abundances were determined by chemical equilibrium calculations for the gas phase as provided by \texttt{FastChem} \citep{Stock+2018}.
Where needed, we extrapolate the fit of the equilibrium constants beyond their tabulated range of $T=100$--6000~K to calculate opacities from $T=100$ to~$10^4$~K.
A detailed description of chemical and opacity data for the atoms and ions can be found in \citet{Hoeijmakers+2019}.
Molecular absorption coefficients are based on the Exomol line lists where available \citep{polyansky18,mckemmisch16,mckemmisch19,gorman19} and on HITEMP \citep{li15} otherwise.
The CIA data are taken from the HITRAN database \citep{karman19}.

Assuming solar metallicity might not be accurate because (i)~important opacity sources could be locked up in large dust grains that remain in the midplane despite meridional circulation and thus do not accrete onto the planet; (ii)~molecular abundances at a given position might not correspond to the chemical-equilibrium values at the local density and temperature, if the dynamical timescale is shorter than the chemical timescale \citep{boothilee19,cridland20}; and (iii)~the accretion luminosity of the planet, in particular in the UV via photochemical reactions,
can also affect the chemical abundances \citep{rab19}.
Whether any of these effects will increase or decrease the opacity 
cannot be said in general and is likely not robust against the details of the modelling. %

The opacity tables used in this work assume that all molecules and atoms are in the gas phase even at low temperatures. Whether for equilibrium or non-equilibrium abundances, this should not be of consequence for the absorption by the gas because below $T\approx1000$~K, the gas opacity is very low (see Figure~\ref{fig:kappa gas Ha}). Thus we keep this simplification, keeping in mind that the exact molecular abundances are uncertain, as discussed previously.

A formal limitation is that the absorption coefficients of molecules are tabulated only up to $T\approx3000$~K,
with the value at the highest temperature used for higher temperatures.
However, in practice this is not an issue because molecules are usually not important anymore at such high temperatures due to dissociation.
For atoms and ions, the tables go up to $T=6100$~K.
We only consider thermal broadening and the natural line widths, except for the case of the Na and K resonance line wings,
where we use the pressure-broadened line profiles
provided by \citet{Allard+2016} and \citet{Allard+2019}.
We note, however, that pressure broadening is not important for this study given the range of relevant pressures 
($P\sim10^{-12}$--10$^{-2}$~bar; see Figure~\ref{fig:kappa gas Ha}, discussed below).

We will compute the monochromatic radiative transfer (Equation~(\ref{eq:L pure abs})) and then integrate over the line width to obtain the total line extinction. However,
the (flux-)averaged opacity provides an estimate of the strength of the absorption.
The wavelength range is small (the width of the emerging line is of the order of 1/2000th of the wavelength, which is $\lmbdHa=656.464$~nm in vacuum; \citealp{Wiese+Fuhr2009}),
so that it is even sufficient to compute the direct mean opacity because
the Planck function does not vary much.
Indeed, at $T\approx10^4$~K, the highest temperatures, the Planck function changes near \Ha 
over a scale
of $H_B\equiv B_\lambda/(dB_\lambda/d\lambda)\approx260~\mathrm{nm}$, which is much larger than the line width $\Delta v/c\times\lmbdHa\approx0.1$~nm for $\Delta v=50~\kms$. Even at $T\approx10^5$~K, the scale would still be $H_B\approx160$~nm.

Figure~\ref{fig:kappa gas Ha} displays the continuum gas opacity near $\lmbdHa$
averaged directly\footnote{%
I.e., $\langle\kappa_\lambda\rangle=\int\kappa_\lambda\,d\lambda/\int d\lambda$. Our resolution is high enough for this to yield the correct result \citep{malygin14}.}
over $\Delta v=\pm50~\kms$.
Shown is also
the $\rho$--$T$ structure of the accreting gas for \texttt{SpherAcc-Cold} as an example. This wavelength range covers most of the flux emerging from the shock for all models, as Figures~\ref{fig:profilesWARM100}--\ref{fig:profilesMagAkk} will show.
The opacity is at most $\kappa\sim0.1$~cm$^2$\,g$^{-1}$ at \Ha\ (greyscale),
but note that for $T\approx800$--1200~K the wavelength dependence (not shown)
is large.
In Figure~\ref{fig:kappa gas Ha}, the resonant opacity is not included because of its strong wavelength and temperature dependence;
because of the Doppler shift (Equation~(\ref{eq:nu vel})), even an average would not provide a meaningful estimate of the typical opacity at a $(\rho,T)$ position in the structure.

We can now combine these elements to calculate the line shape of accreting planets, which we do in the following sections.

\section{Absorption by the gas}
 \label{sec:abs gas}

We have calculated the absorption of \Ha (Equation~(\ref{eq:L pure abs})) for a grid of accretion rates $\Mdot$ and masses $\MP$ 
for the different accretion geometries
discussed in Section~\ref{sec:geo}: a spherically symmetric inflow (\texttt{SpherAcc}), accretion onto the polar regions (\texttt{Polar}), and magnetospheric accretion (\texttt{MagAcc}). They are illustrated in Figure~\ref{fig:scenarios} and summarised in Table~\ref{tab:scenarios}.
In each case we take the emerging flux from the \citet{Aoyama+2018} models and integrate the extinction along a radial radiative path starting at the shock and going out to $\rmax$. The density and temperature are as shown in Figure~\ref{fig:kappa gas Ha} for one geometry. %

Section~\ref{sec:oneex} presents and discusses in detail one combination of \Mdot, $\MP$, and $\RP$ in the \texttt{Polar-Cold} geometry. Then, Sections~\ref{sec:line fluxes} and following present the results for the grid of models.

\subsection{One example of gas absorption}
 \label{sec:oneex}

\begin{figure*} [th] %
 \centering
 \includegraphics[width=0.9\textwidth]{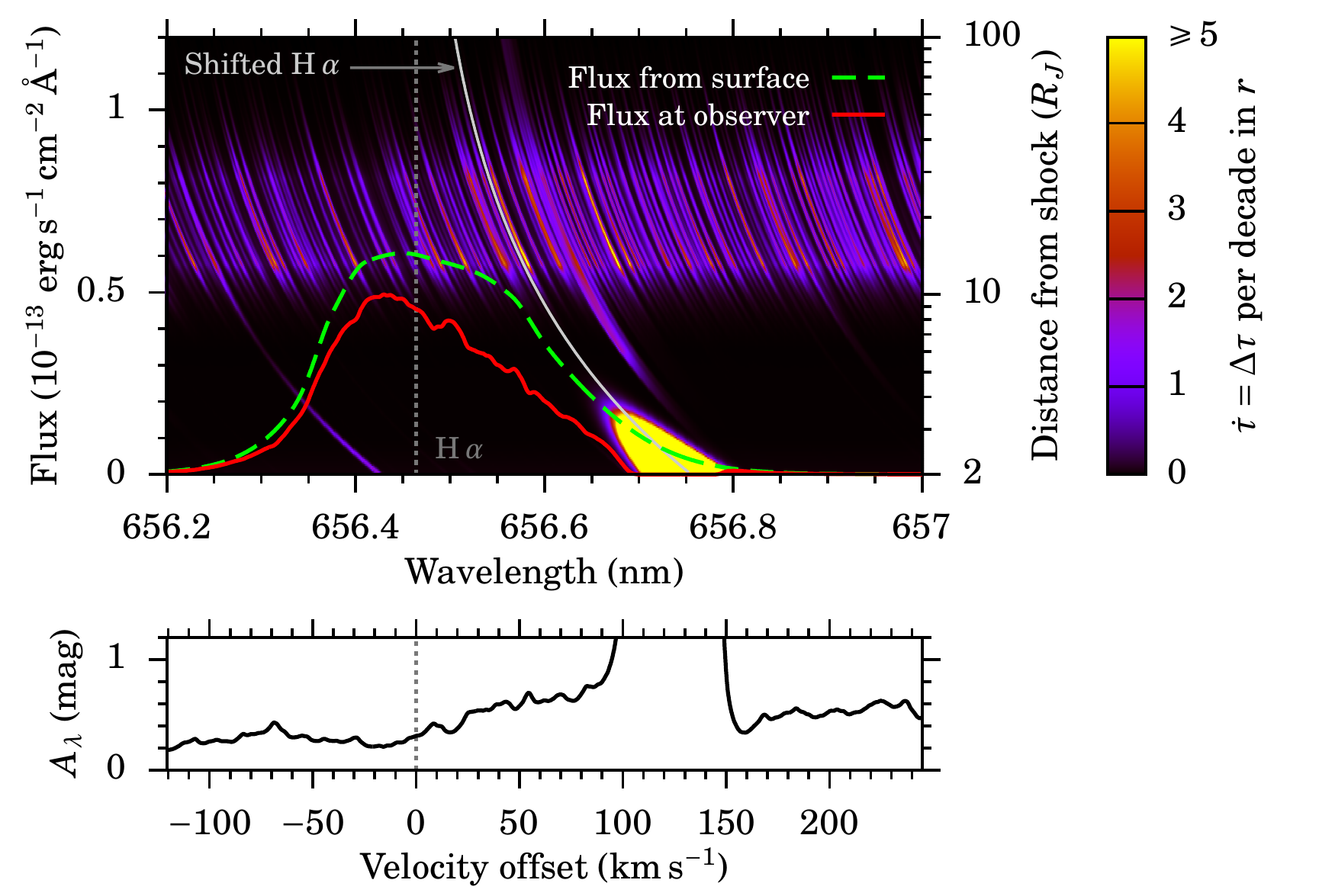}
\caption{
Top panel:
\Ha line profile at spectral resolution $R=1.5\times10^6$ for one case in the \texttt{Polar-Cold} scenario %
with $\Mdot=3\times10^{-5}~\MdotUJ$, $\MP=10~\MJ$, and $\RP=2.04~\RJ$ 
(solid red line) when observing into the accreting area at 150~pc. The flux in the absence of absorption by the incoming matter is also shown (dashed green line).
The background colour shows as a function of distance from the planet (right vertical axis) and observer-frame wavelength %
the strength of the absorption by the gas $\dot{\tau}(r,\lambda)$ (Equation~(\ref{eq:dottau})), capped at $\dot{\tau}=5$.
Few regions have a higher $\dot{\tau}$,
the main one being the \Hi\ resonance near the shock (reaching $\dot{\tau}\approx5\times10^3$). %
Indicated are also the central peak of the \Ha\ line in the rest frame $\lmbdHa$ (vertical dotted grey line) and, as a function of radial distance from the planet, the wavelength that becomes blueshifted to $\lmbdHa$ (see Equation~(\ref{eq:nu vel}); curved solid grey line).
No dust nor ISM extinction is included.
Over the brighter region at $r=15$--30~$\RJ$, the temperature is $T\approx2000$--1500~K and the pressure $P\approx1$--0.2~erg\,cm$^{-3}$ (cf.\ Figure~\ref{fig:kappa gas Ha}).
Bottom panel: Extinction $A_\lambda$ from the accretion flow against velocity offset relative to \Ha.
}
\label{fig:kappa2D}
\end{figure*}

We show the spectrally resolved line flux in Figure~\ref{fig:kappa2D}
for one case in the \texttt{Polar} accretion geometry,
with $\Mdot=3\times10^{-5}~\MdotUJ$, $\MP=10~\MJ$, and $\RP=2.04~\RJ$.
The flux is the one seen by an observer looking %
into the accretion cone (within $\thmax=46\degr$ of the pole; see Figure~\ref{fig:scenarios})  %
and at 150~pc,
similar to the distance of well-known young star-forming regions such as Taurus, Lupus, or $\rho$~Ophiucus--Upper Scorpius.
As throughout this work, no absorption by the interstellar medium (ISM) is included, to separate the two effects\footnote{For
   the PDS~70 planets ($d=113$~pc), \citet{mueller18} derived a $V$-band extinction $\AV\approx0.05$~mag, %
or $\AR=0.04$~mag (at the $R$~band, which is around \Ha) using \citet{cardelli89} with $\RV=5$.}.
Due to the absorption by the accreting gas, the line-integrated flux drops by about 32~\%,
corresponding to an extinction\footnote{Note that $A\equiv2.5\log_{10}(F_0/F)$
	and $\Delta\tau=\ln(F_0/F)=\ln(10)\times\log_{10}(F_0/F)$ are nearly equal ($\Delta\tau=0.92A$) since $\ln(10)=2.30\approx2.5$.%
}
$\AR/\left(\mathrm{mag}\right)\approx\Delta\tau = 0.4$. We now discuss the wavelength-dependent extinction.

In contrast with the line shape leaving the shock, the observable line shows small-scale structures.
What dominates the absorption
can be easily seen by looking at a measure of the strength of the local absorption in the flow $\dot{\tau}$,
the optical depth per decade in $r$. It is given by
\begin{equation}
 \label{eq:dottau}
 \dot{\tau}(r,\lambda) \equiv \frac{d\tau}{d\log_{10}r} = \ln(10)\,\kappa(r,\lambda)\rho(r) r.
\end{equation}
This is indicated as a colourscale in Figure~\ref{fig:kappa2D}. Thus, on a logarithmic radial scale every unit $\dot{\tau}$ contributes equally to the total extinction at a given wavelength.
A radial cut at one restframe wavelength is shown in Figure~\ref{fig:Teresa}c. The obvious curvature of the opacity features is due to the Doppler shift gradient, described by Equation~(\ref{eq:nu vel}).

The quantity $\dot{\tau}$ %
reveals that most of the absorption occurs at $r=15$--30~$\RJ$.
There, the temperatures are $T\approx2000$--1500~K respectively (not shown),
a factor $\approx2$ lower than the shock temperature.
The extinction
comes mainly from the continuum opacity with a small contribution from the resonant hydrogen absorption.
In general, because of the high temperatures in the postshock region \citep{Aoyama+2018},
the emerging line is usually much broader than the thermal broadening
of the incoming gas.
Equation~(\ref{eq:nu vel})) shows that
there are only redshifted absorbers;
layers moving towards the planets absorb photons blueshifted to \Ha in the frame of the accreting gas (see the curved grey line in Figure~\ref{fig:kappa2D}).
Therefore,
resonant absorption (by the incoming $n=2$ electrons) can occur only redward of the rest central wavelength,
approximately equal to the centre of the line emerging from the postshock region. This is the yellow region in Figure~\ref{fig:kappa2D}.

Because the velocity decreases away from the planet,
there is the possibility that the complete red wing be absorbed by the incoming $n=2$ electrons.
However, the temperature and thus the number fraction of absorbers decrease outwards much faster than the Doppler shift.
Therefore,
only a small wavelength range can be absorbed by the incoming $n=2$ electrons.%
Also, as we calculate in Appendix~\ref{sec:cffull}, the emission by the accretion gas in that wavelength range would be important, so that in reality the increase in the extinction would be smaller than what Figure~\ref{fig:kappa2D}b suggests (see Figure~\ref{fig:fullI}a).

Figure~\ref{fig:kappa2D} shows that the Doppler shift gradient smears the strong wavelength dependence of $\dot{\tau}(r,\lambda)$
for the integrated extinction
and that the resulting absorption (bottom panel) has less structure.
However, in this example, there happens to be a clear downward slope in the extinction from 656.45 to 656.55~nm ($\Delta v\approx0$--$100~\kms$),
which exacerbates the asymmetry in the input line profile\footnote{%
  This is partly due to a clear resonance near $\lmbdHa+0.05$~nm, present from large $r$ down to $r\approx4~\RJ$ ($T\approx3000$--4000~K). Comparing with \citet{sharpburrows07}, it likely comes from TiO or maybe VO.}.
The result is a crudely gaussian-looking blue wing but a linearly decreasing red wing. This leads to an apparent offset in the line peak but only by 25~$\kms$, much less than the line width.

Our temperature structure is based on frequency-integrated equations and supported by the grey flux-limited diffusion (FLD) simulations of
\citet{m16Schock,m18Schock}.
FLD is an approximate method (e.g.\ \citealp{ensman94,turnerstone01}), but the temperature structure
should be robust because it reflects global energy conservation\footnote{As a corroboration,
in the stellar context \citet{vaytet13} found that non-grey collapse simulations, which also feature a shock, yielded the same structures and evolution as a frequency-averaged approach.}.
In particular, in our case
the temperature must drop from a value of order $\approx T_0$ (Equation~(\ref{eq:T0 not historical...})) to a much lower value in the CSD
and therefore cross at some point this range $T\approx2000$--1500~K where the opacity is particularly strong. It depends on the density only weakly (see Figure~\ref{fig:kappa gas Ha}),
and for a given temperature profile $T(r)$ and thus\footnote{This ignores the density dependence of dust opacity transitions.} opacity profile $\kappa(r)$, $\dot{\tau}$ goes as $\rho r \propto r^{-1/2}$ (in the limit $\zeta=1$; Equations~(\ref{eq:rho}) and~(\ref{eq:dottau})),
which is not a strong dependence.
Thus if the temperature of maximum absorption were at another distance from the planet, the integrated absorption should be similar. The frequency dependence would be different (because of the radial dependence of the velocity) but only quantitatively.
Within the other assumptions, these results are somewhat robust since the maximum absorption does not occur in the outermost parts of the flow, where the accretion geometry is much more uncertain.

We have not considered absorption by the dust here.
Taking the approximate expression for $\rho$ valid in the limit $\RAkk=\infty$, Equation~(\ref{eq:rho}) with $\zeta=1$,
the gas column density is
\begin{equation}
\label{eq:Siginf}
 \Sigma = \int_{\RP}^\rmax \rho \,dr \approx \frac{\Mdot}{2\pi\ff\sqrt{2G\MP\RP}}  %
\end{equation}
since $\rmax\gg\RP$,
which yields 
$\Sigma=5$~g\,cm$^{-2}$
for this example.
Thus,
the dust will not be able to absorb much radiation if the dust opacity (cross-section per gram of gas, not of dust) $\kapStbfpg\equiv\fpg\kappaStbint\ll1/\Sigma\approx0.2$~cm$^2$\,g$^{-1}$, where $\kappaStbint$ is the dust material opacity (cross-section per gram of dust) at \Ha.
Comparing to the opacity values of \citet{woitke16}, who study the effect of the variation of several relevant opacity parameters, this situation seems possible. However, the parameter space is large and the opacity itself is very uncertain. We explore the absorption by the dust in more detail and systematically in Section~\ref{sec:cont abs dust}.

In the following subsections, we discuss the results from the grid of models covering the relevant planet-formation parameter space (see Section~\ref{sec:par space}), as discussed at the beginning of this section.

\subsection{Line-integrated fluxes}
 \label{sec:line fluxes}

\begin{figure*} %
 \centering
 \includegraphics[width=0.47\textwidth]{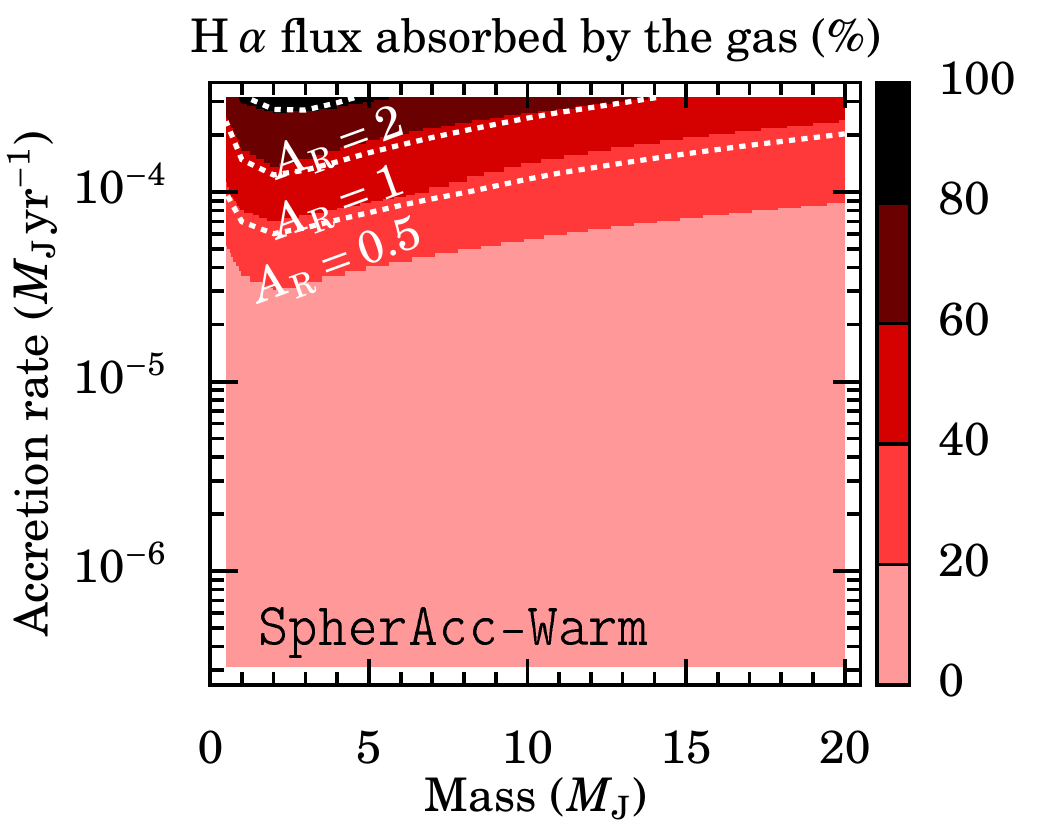}
 \includegraphics[width=0.47\textwidth]{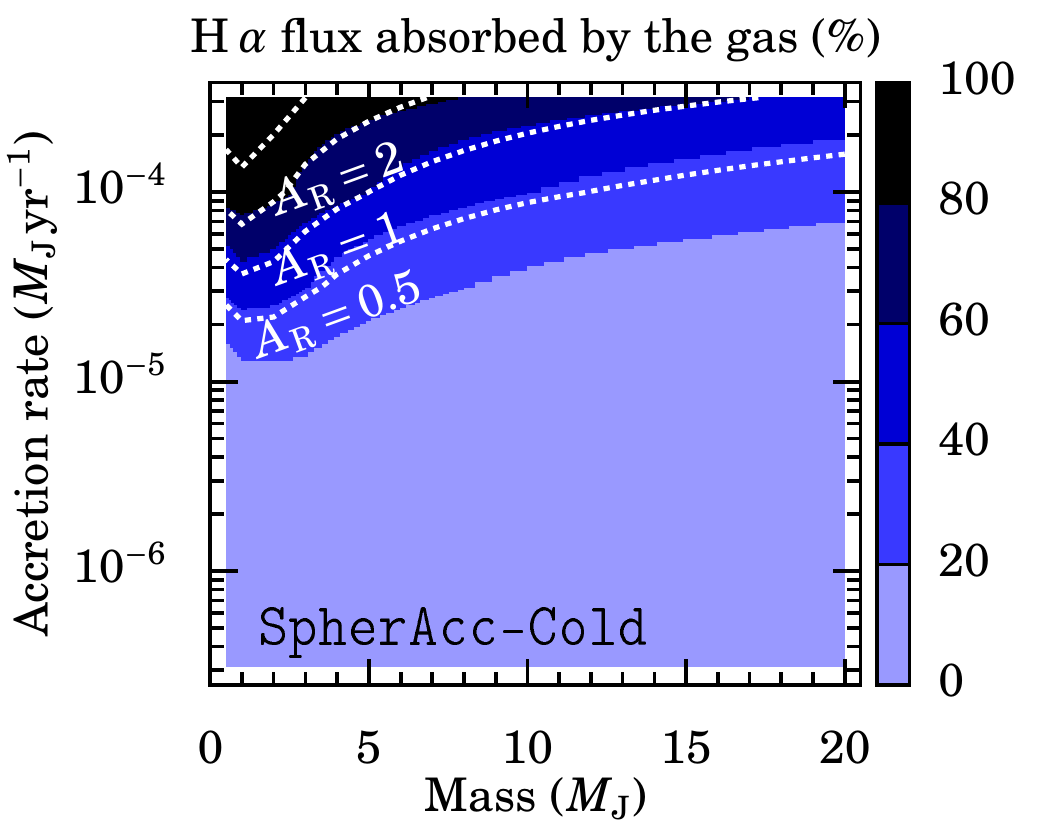}\\
 \includegraphics[width=0.47\textwidth]{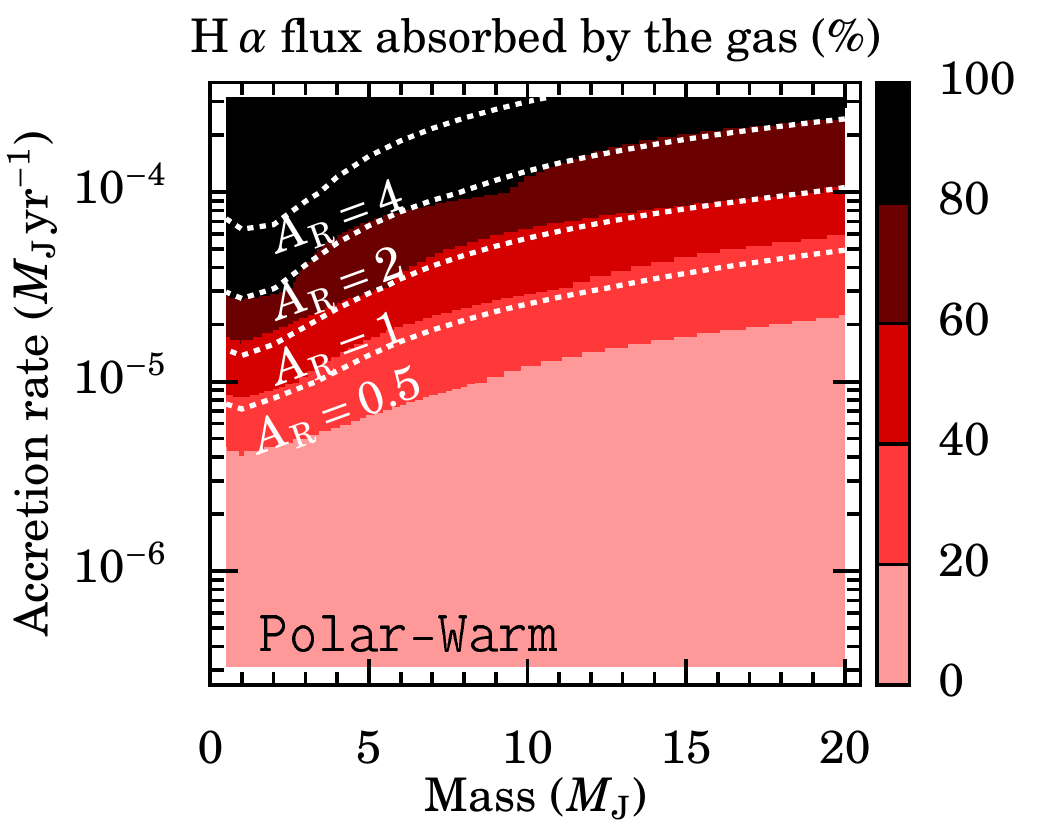}
 \includegraphics[width=0.47\textwidth]{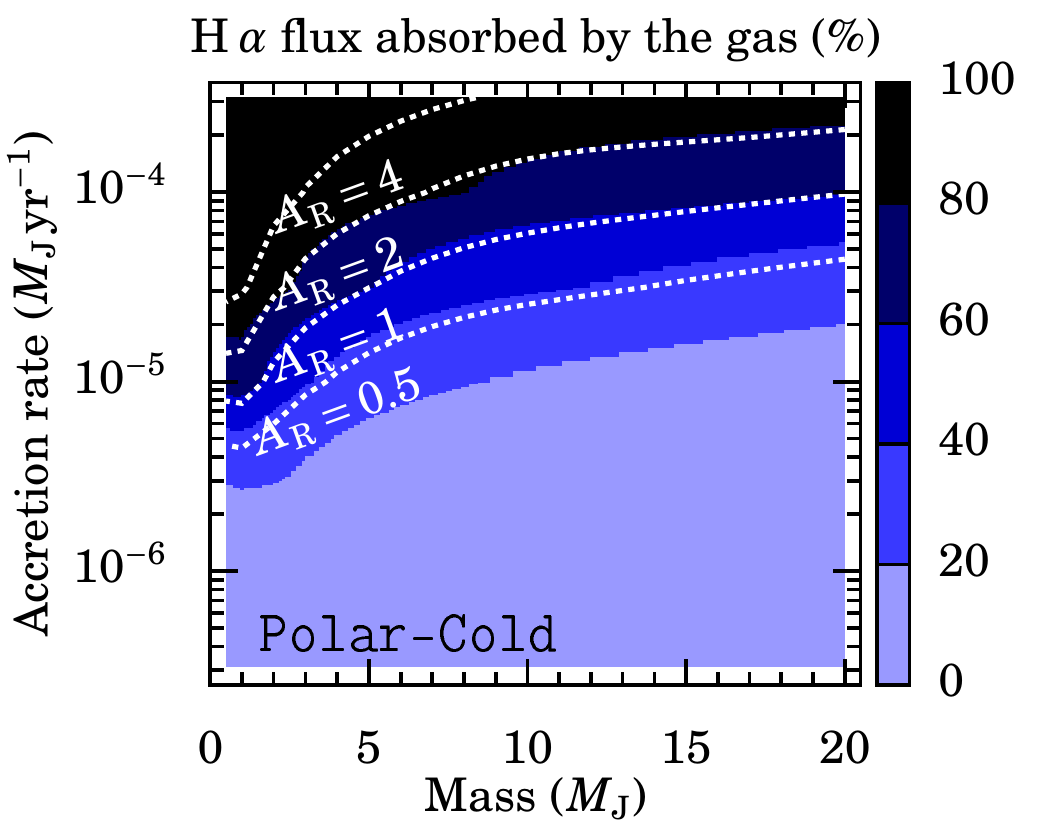}\\
 \includegraphics[width=0.47\textwidth]{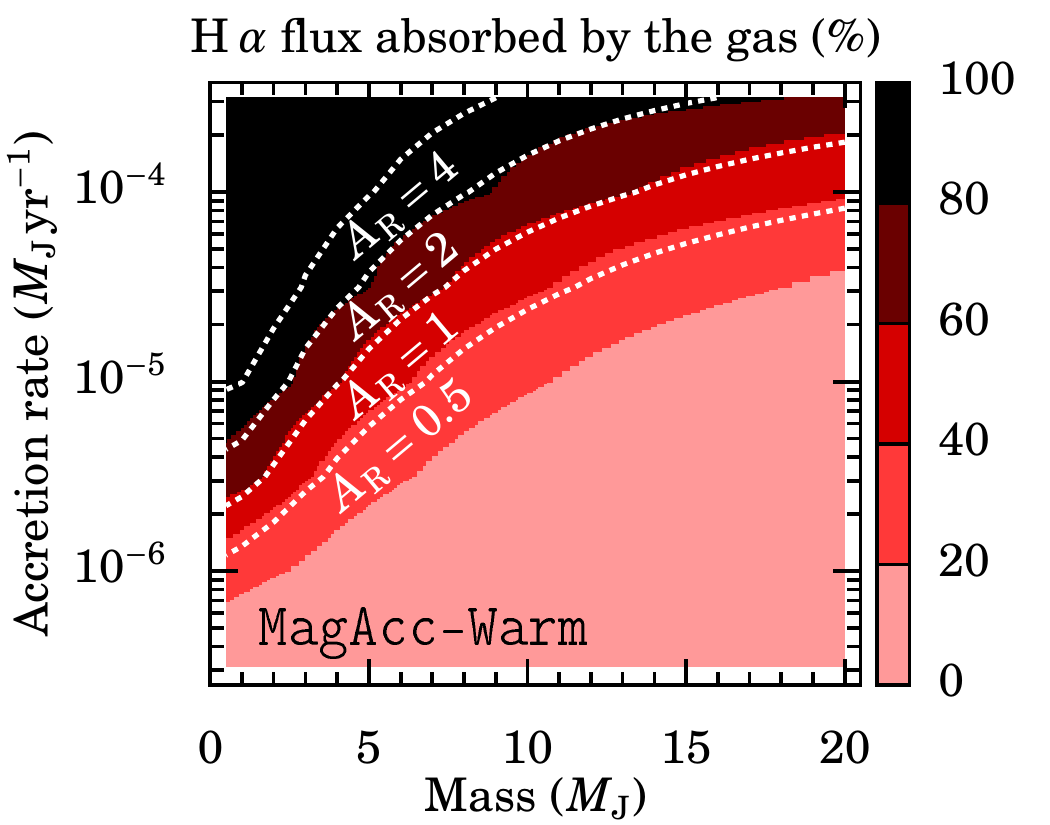}
 \includegraphics[width=0.47\textwidth]{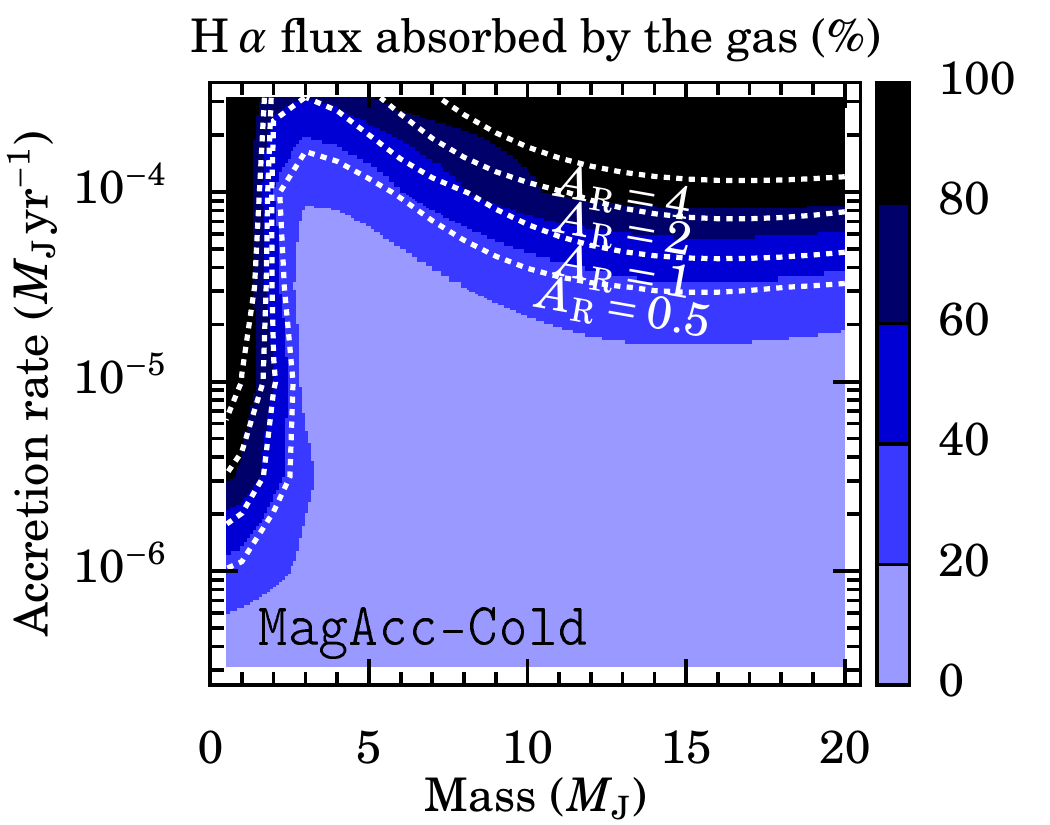}
\caption{
Percent reduction in the line-integrated flux for the \texttt{SpherAcc}, \texttt{Polar},
and \texttt{MagAcc} accretion geometries (\textit{top to bottom rows}) from the gas opacity only.
Shown are the results for the warm- (\textit{left column}) and cold-population radii
(right column).
The dotted lines highlight an extinction $\AR=0.5$, 1, 2, and~4~mag (bottom to top).
At a given accretion rate, high-mass planets suffer less from absorption. For moderate $\Mdot\lesssim10^{-5}~\MdotUJ$, the extinction is at most $\AR\approx0.5$~mag. For \texttt{SpherAcc} and \texttt{Polar}, both radius fits yield similar results.
}
\label{fig:DeltaF}
\end{figure*}

The first two rows of Figure~\ref{fig:DeltaF} display the relative drop in the flux $\DF$ across the accretion flow
for the \texttt{SpherAcc} and \texttt{Polar} geometries.
The drop is mainly an increasing function of $\Mdot/\ffill$ and ranges between $\DF=0~\%$ (no absorption)
and nearly $\DF=100$~\%\ over the parameter space considered.
Already for $\Mdot\approx3\times10^{-5}~\MdotUJ$, there is noticeable absorption
with $\DF\approx20$--50~\%\ depending on the mass, with more absorption at lower planet masses. 
This corresponds to an extinction $\AR\approx2$--4~mag.
The increase of extinction with decreasing planet mass (and thus, at a given $\Mdot$, with the gas column density) can be seen from
Equation~(\ref{eq:Siginf}).
Also, for the three highest $\Mdot$ values in Figure~\ref{fig:kappa gas Ha}, the opacity increases with decreasing temperature (due to the contribution from water; \citealp{m16Schock}) and increasing density, that is, decreasing mass.
For moderate $\Mdot\lesssim10^{-5}~\MdotUJ$, the extinction is at most $\AR\approx0.5$~mag.

Interestingly, the dependence of $\DF$ on the mass becomes larger
towards smaller filling factors.  %
The choice of warm- or cold-population radii barely changes the outcome but
the trend is as expected: $\DF$ is larger for the cold-population radii
since they are smaller, leading to higher preshock temperatures and densities ($\tau\propto\rho$), with the temperature peaking at a few thousand kelvin (see Figure~\ref{fig:kappa gas Ha}).

The \texttt{MagAcc} scenario, shown in the third row, leads to qualitatively similar but quantitatively different results.
The minimum accretion rate needed to have significant absorption ($\AR\gtrsim1$~mag) can be as low as
$\Mdot\sim3\times10^{-6}~\MdotUJ$, at low masses $\MP\approx1$--5~$\MJ$.
The mass dependence of $\DF$ is stronger for the warm-population radii,
and for the cold-population radii $\DF$ shows clear non-monotonic behaviour.
In particular, there is a ``window'' near $\MP=3$--7~$\MJ$ in which the absorption corresponds only to $\AR\lesssim0.5$~mag even for a colossal accretion rate $\Mdot=10^{-4}~\MdotUJ$; for only slightly smaller masses of $\MP\approx1$--2~$\MJ$, the extinction is $\AR\gg4$~mag. Thus these $\MP\approx5~\MJ$ planets could be particularly observable.
However, this depends
on the viewing geometry.

Overall, these results show that absorption can be significant (several magnitudes of extinction) at higher accretion rates or for low-mass planets in the \texttt{MagAcc} case, particularly for smaller (cold-population) radii. However, the extinction from the gas
is negligible
if $\Mdot\lesssim3\times10^{-6}~\MdotUJ$.
The outer integration limit for the absorption calculation
(i.e.\ in Equation~(\ref{eq:L pure abs})), $\rmax$,
is a poorly known quantity. 
However, one can argue that this is of little consequence:
Figure~\ref{fig:kappa2D} suggests that most of the absorption,
if there is any,
occurs at tens, not hundreds of Jupiter radii.
This is corroborated heuristically by Figure~\ref{fig:kappa gas Ha},
which shows that profiles with $\Mdot\gtrsim3\times10^{-6}~\MdotUJ$
cross the high-opacity region near the shock in density space,
and thus also in radial distance from the planet since $\rho$ is a monotonic function of radius for $r\ll\RAkk$ (Equation~(\ref{eq:rho})). %

\subsection{Line profiles}
 \label{sec:line profiles}

\begin{figure*}[ht] %
 \centering
 \includegraphics[width=0.97\textwidth]{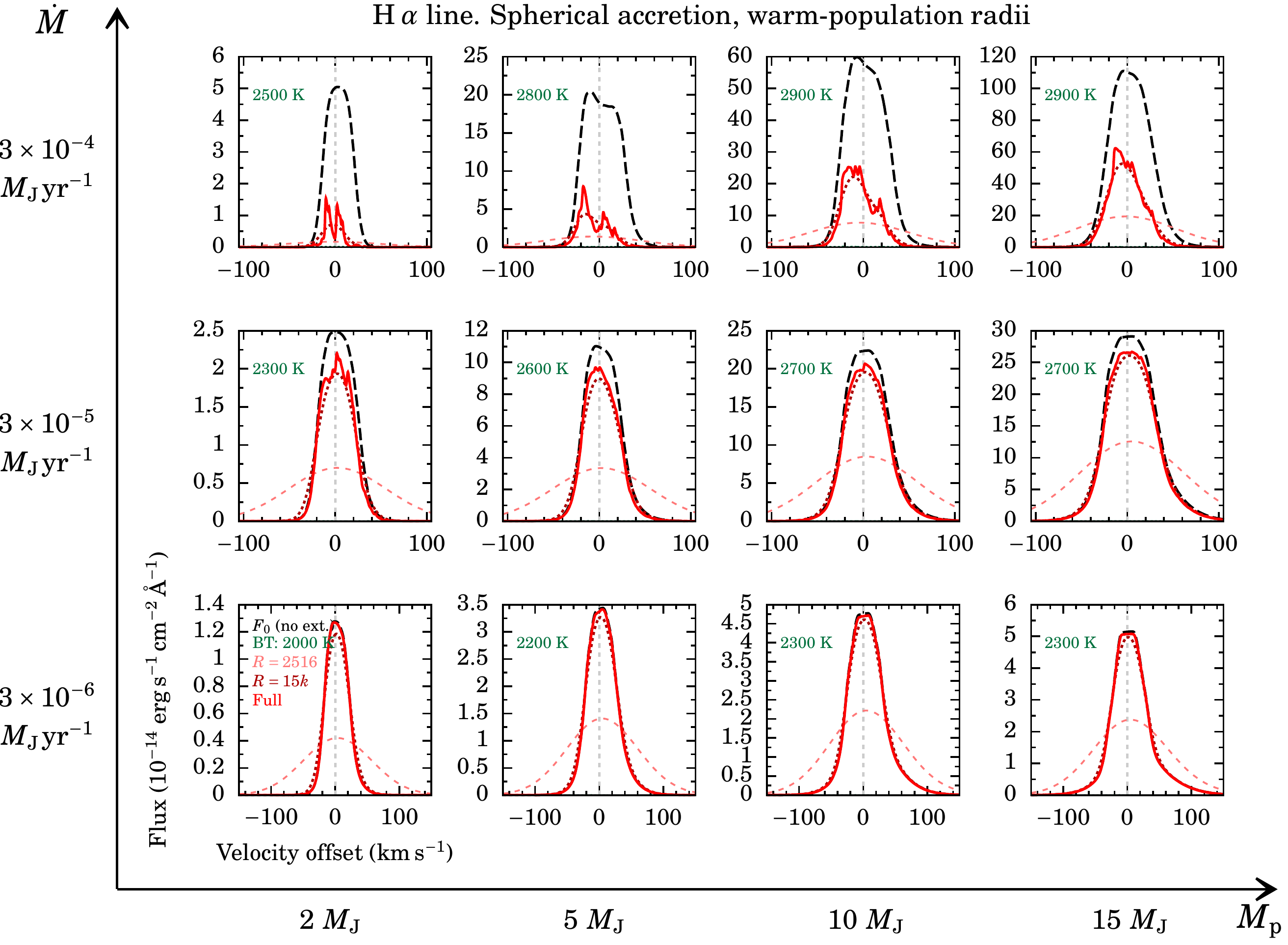}
\caption{
Effect of the extinction by the accreting gas on the \Ha profile.
Shown is the flux for sources at 150~pc as a function of planet mass and accretion rate
(\textit{outer axes}) for the \texttt{SpherAcc-Warm} case of Table~\ref{tab:scenarios}.
For each subpanel,
we plot the profile  %
without extinction
(black dashed line) 
and the profile after passing through the accreting material (red solid line). No ISM absorption is considered.
The observable profiles %
are also shown convolved with the resolution of MUSE ($R=2516$; dashed pale red line),
and of VIS-X ($R=15,000$; dashed dark red line).
The heated photosphere (BT-Settl model, with $\Teff$ from the fit of \citealp{Aoyama+2020}; green dotted line and label)
is too weak to be seen in any panel.
The horizontal axes are the velocity offset from the line centre.
The flux and velocity ranges differ from panel to panel.
}
\label{fig:profilesWARM100}
\end{figure*}

\begin{figure*}[ht] %
 \centering
 \includegraphics[width=0.97\textwidth]{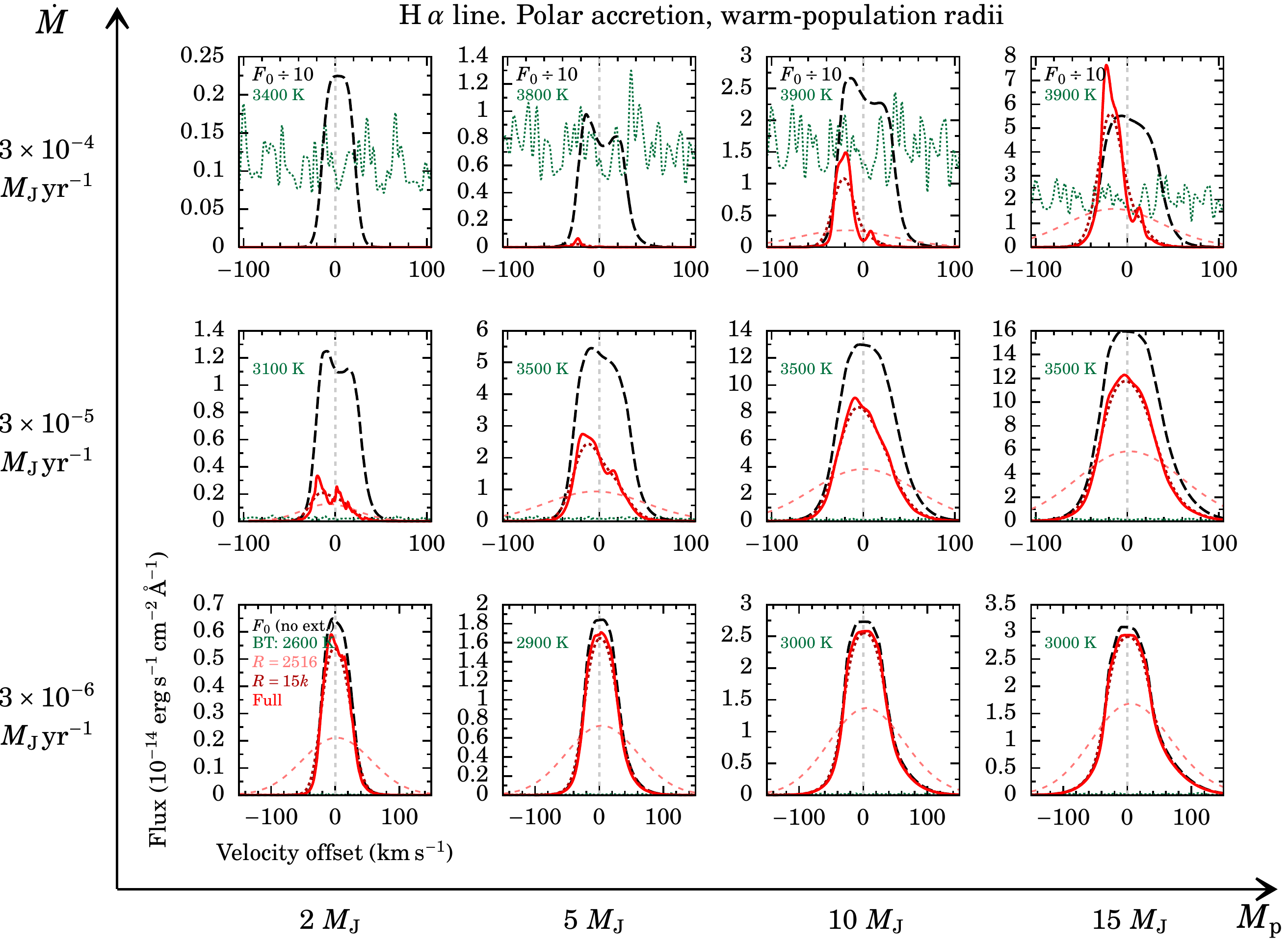}
\caption{
As in Figure~\ref{fig:profilesWARM100}, but for the \texttt{Polar-Warm} case.
The observer is looking into the accreting regions, along the accretion flow. In the top row, the flux 
without absorption
$F_0$ (black dashed line) has been reduced by a factor of ten to make the profiles with absorption visible.
}
\label{fig:profilesWARM30}
\end{figure*}

\begin{figure*}[ht] %
 \centering
 \includegraphics[width=0.47\textwidth]{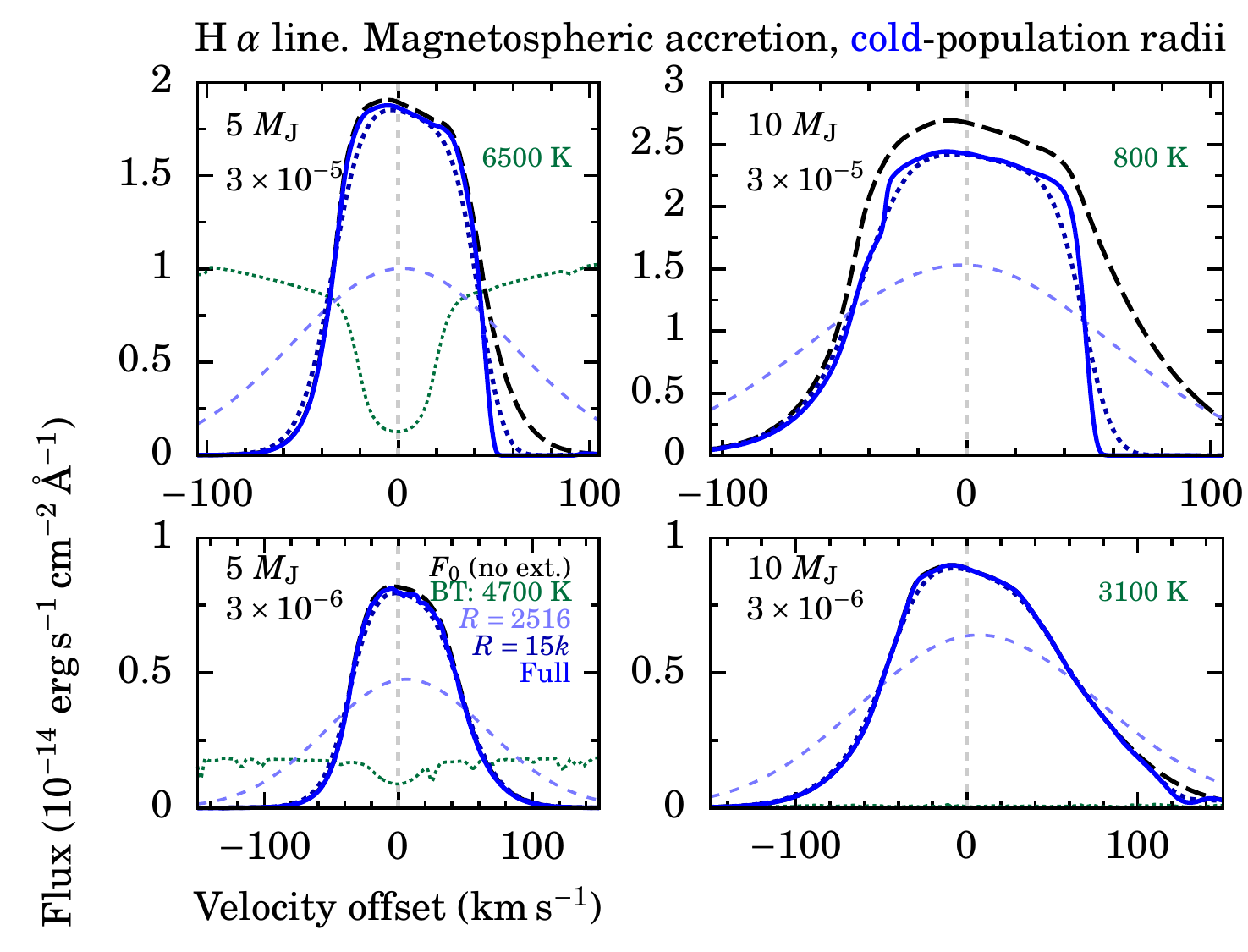}
 \includegraphics[width=0.47\textwidth]{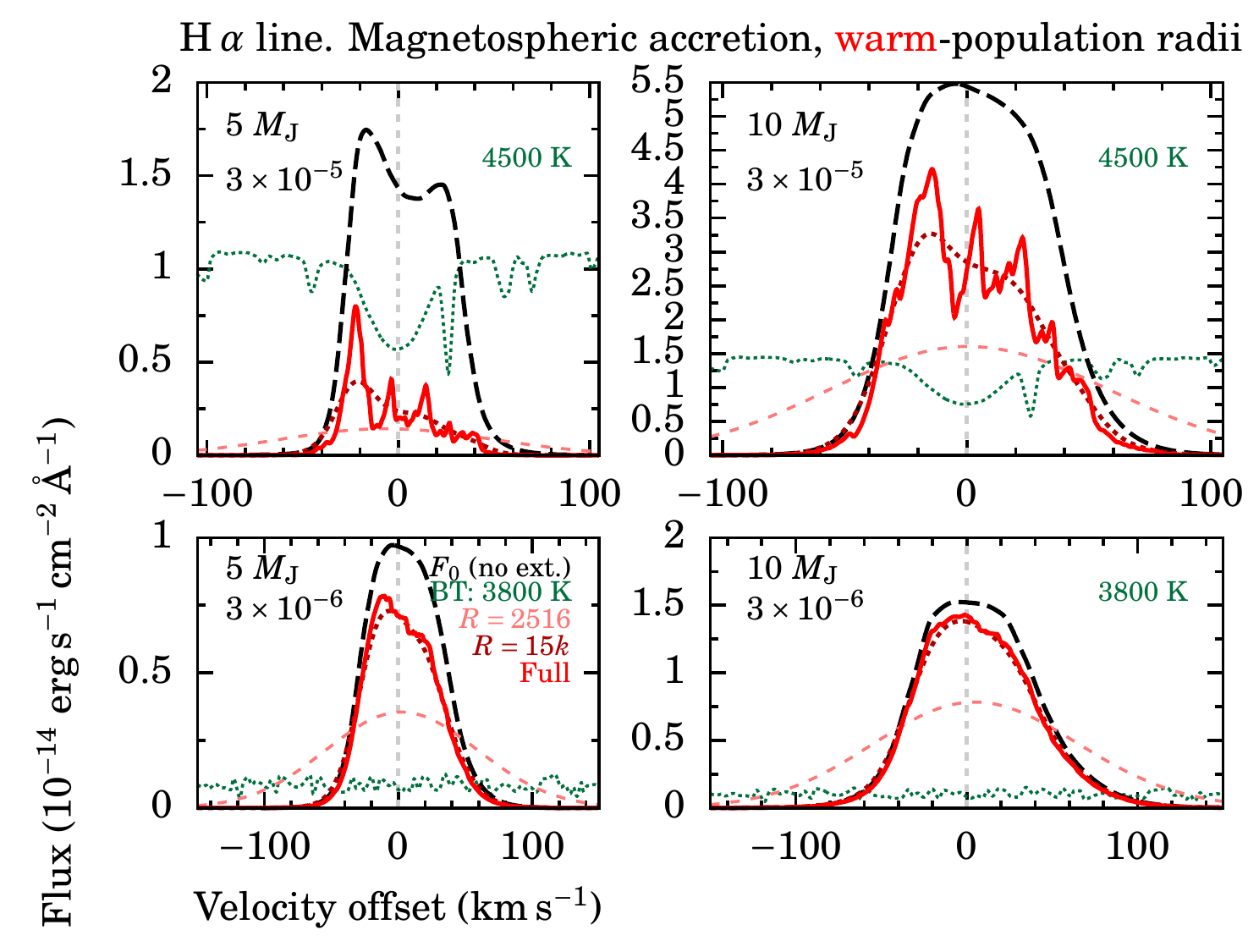}
\caption{
As in Figure~\ref{fig:profilesWARM100}, but for the \texttt{MagAcc} cases (cold and warm; left and right groups, respectively). The mass in $\MJ$ and accretion rate in $\MdotUJ$ are indicated in each subpanel. The spectra are for an observer looking into the accretion flow towards the base of the shock or more generally for radiation passing through only the layers closest to the planet and escaping (see Figure~\ref{fig:scenarios}).
Possible rotation and inclination (between the observer and the accretion column) are not taken into account. The strong photospheric absorption (green) in some cases may not be realistic; see text.
}
\label{fig:profilesMagAkk}
\end{figure*}

\subsubsection{Spherical and polar cases}

High-resolution ($R=1.5\times10^6$) line profiles are shown in Figures~\ref{fig:profilesWARM100} and~\ref{fig:profilesWARM30}
for several cases from Figure~\ref{fig:DeltaF} for the \texttt{SpherAcc} and \texttt{Polar} cases.
The absolute spectral densities are for sources at 150~pc
(relevant e.g.\ for Taurus, Lupus, or $\rho$~Oph--U~Sco)
and are mostly of order $F_\lambda\sim10^{-14}$--$10^{-13}$~erg\,s$^{-1}$\,cm$^{-2}$\,\AA$^{-1}$ for the range of $\Mdot$ and $\MP$ shown. We also show an estimate of the photospheric emission from the fit of \citet{Aoyama+2020}, to which we return in Section~\ref{sec:impcont}.

For a given $(\Mdot,\MP)$ and therefore, by assumption, radius, the line-integrated flux at the observer decreases when going from \texttt{SpherAcc} to \texttt{Polar}, that is, when reducing $\ffill$. This is due to the increased preshock density ($n_0\propto\rho\propto\Mdot/\ffill$), which leads to more self-absorption in the postshock region (\citealp{Aoyama+2020}; sketched in Figure~\ref{fig:Teresa}c), thereby reducing the luminosity in the \Ha line.

As discussed in \citet{Aoyama+Ikoma2019}, the line widths (at 10 and~50~\%\ of the maximum) of the profiles in Figures~\ref{fig:profilesWARM100} and~\ref{fig:profilesWARM30}
reflect both the temperature of the region at which most of the line is formed
as well as the absorption by the layers above this in the postshock region. We can now extend this statement by noting that the accretion flow too can change the width of the line, for $\Mdot>3\times10^{-5}$ ($3\times10^{-3}$)~$\MdotUJ$ in the \texttt{SpherAcc} (\texttt{Polar}) case.

The profiles as they emerge from
the accretion flow
reveal further information.
They exhibit, in part as expected from Figure~\ref{fig:DeltaF},
a range of qualitative outcomes: no absorption (low $\Mdot$),
moderate wavelength-independent (high $\Mdot$ and intermediate $\MP$),
moderate wavelength-dependent absorption (intermediate $\Mdot$ and low $\MP$),
and very strong absorption (high $\Mdot$ and low $\MP$).
Also, different line shapes are visible: %
roughly Gaussian (towards the bottom right); flattened (middle right);
or with self-absorption in the peak taking place in the postshock region (top and middle left).

In Figures~\ref{fig:profilesWARM100} and~\ref{fig:profilesWARM30},
the lines at (somewhat) high accretion rates $\Mdot\gtrsim3\times10^{-5}~\MdotUJ$
display some
asymmetry, as in the example shown in Figure~\ref{fig:kappa2D}.
In all examples considered, an asymmetry, if present,
is characterised by a peak shifted blueward by around 10 to~$20~\kms$,
which results from stronger absorption in the red wing.
This asymmetry is present in the opacity and comes at least in part from a slight slope in the continuum of the opacity, as can be seen in Figure~\ref{fig:kappa2D}. In that example, there happen to be several stronger features in the opacity between $\Delta v\approx-10$ and~$+150~\kms$ relative to the line centre. These features are physically unrelated to the \Ha\ line and might not be present for other elemental mixtures or atomic and molecular abundances (for instance due to disequilibrium chemistry).  %
The asymmetry
already present
in the input profile grows towards lower masses.
Only in a few cases is there a small dip at the central position of \Ha,
but this is not due to resonant absorption
because the latter is redshifted as discussed in Section~\ref{sec:oneex}.
For the highest accretion rates, especially at $\MP\lesssim5~\MJ$, the signal is reduced by orders of magnitude. This implies that a planet accreting at such high rates would be indistinguishable from the continuum, especially taking the emission from the accreting gas into account (see Figure~\ref{fig:fullI}b).

For the cold-population radii, the results (shown in Figures~\ref{fig:profilesCOLD100} and~\ref{fig:profilesCOLD30}) are similar,
with a moderate amount of extinction imprinting spectral structure at high accretion rates $\Mdot\gtrsim3\times10^{-5}~\MdotUJ$. However, the line fluxes (at the observer's position) are either very similar or smaller, with very few exceptions. In the \texttt{Cold} cases, the smaller radius leads to a higher accretion luminosity ($\Lacc\propto1/\RP$), but in the \texttt{Warm} cases the conversion of $\Lacc$ to $\LHa$ turns out to be more efficient by a factor greater than the ratio of the radii. Thus, overall, the flux is usually smaller in the \texttt{Cold} cases.

\subsubsection{Magnetospheric accretion}

Spectra for the \texttt{MagAcc} case are shown in Figure~\ref{fig:profilesMagAkk}.
In this geometry, we have assumed that the flux %
from the postshock region
passes through the accretion column out to a distance $\rmax\approx\RAkk/2$
(i.e.\ tangentially through the accretion arc; see Figure~\ref{fig:scenarios}).
The high spectral resolution $R\sim10^6$ of our calculations reveals
pronounced features for the warm-population radii
that are absent for the cold-population radii.
This is due to the different highest temperatures in the flow
(Equations~(\ref{eq:T} and~(\ref{eq:T0 not historical...})).
Indeed, the shock temperature (where the maximum is reached) is $T_0\approx 2000$--3000~K for the warm-population case
as opposed to $T_0\approx4000$--7000~K for the cold case, due to the smaller radii in the latter case.
For these higher temperatures in the cold-population case, close to the planet
molecules are absent and only a few atomic lines and the continuum leave a spectral imprint.

There are two consequences of the short integration length. %
One is that the temperature does not drop into a region where molecules are important.
The second consequence is that the features are sharper
because they are not blurred %
by a gradient in the Doppler shift;
in other words,
over the region of the strongest absorption, roughly $\RP$ to $\rmax\approx\RAkk/2=2.5\RP$, %
the change in the Doppler shift $\Delta(\Delta\lambda)\approx dv/dr\times(\rmax-\RP)\times\lambda_0/c\sim(3/4)(v_0/c)\lambda_0$ %
is smaller than the typical spacing between the spectral features.
Finally, in a few of the examples selected
(e.g.\ cold-population radii, $\Mdot=3\times10^{-5}~\MdotUJ$, $\MP=10$ and $5~\MJ$, $\ffill=10~\%$),
the spectral footprint of the \Ha\ resonance is clearly visible at $\Delta v\gtrsim+60~\kms$
as a strong absorption.
More generally, how smooth or ragged the line profile is,  %
is determined by how large the Doppler shift gradient in the flow (not the Doppler shift itself) is
over the length over which there are stronger spectral signatures,
which in turn is set by the temperature structure in the flow.
Very roughly speaking, the opacity is low both for $T\lesssim800$~K and at $T\approx3000$--6000~K (see Figure~\ref{fig:kappa gas Ha}) and does not depend strongly on density.
If the maximal temperature in the flow (at the shock) is above
or close to the upper edge of the opacity maximum,
a sufficiently large fraction of the flow will be in the high-opacity regime.
This matters especially because both the density (thus the optical depth) and the velocity gradient are stronger closer to the shock. A maximum temperature near the opacity maximum therefore leads to a blurring of the features by the Doppler shift (i.e.\ velocity) gradient.

\subsubsection{Importance of the continuum}
 \label{sec:impcont}

So far, we have effectively considered continuum-subtracted emission lines by calculating the radiative transfer only for the shock excess. However, currently used high-resolution spectral differential imaging (HRSDI) techniques
involve the subtraction of the continuum and of low-frequency spectral components (in \Ha: \citealp{Haffert+2019,xie20}; in the NIR: \citealp{snellen14,Hoeijmakers+2018,petrus21,cugno21}).
Therefore, we need to ensure that subtracting the photosphere, especially if it is somewhat mismatched, will still leave the line emission detectable. For spectrally resolved observations, what matters here is, to first order, not the ratio of the line peak to the continuum, but rather its ratio to the spread of the local (pseudo)continuum features, the ``photospheric noise''.  %

To assess whether the photospheric emission could hinder the line measurement, 
we plot in Figures~\ref{fig:profilesWARM100}--\ref{fig:profilesMagAkk} the approximate photospheric emission of the planet. This is to be compared to the line emission at the planet's surface. The more complete approach would require inputting the line together with the continuum to the radiative transfer including emission from the accretion flow (Appendix~\ref{sec:cffull}), but for our purposes the current estimate should suffice.

We use the fit\footnote{Available at \url{https://github.com/gabrielastro/St-Moritz}.} of \citet{Aoyama+2020} of $\Teff(\Mdot,\MP)$. These fits consider the downward-travelling radiative fluxes from the detailed shock models and ensure that the total emitted luminosity from the planet is equal to the sum of the interior and the incoming kinetic energy. Thus, for the photosphere, $\sigma\Teff^4$ is not simply given by $\Lacc/(4\pi\RP^2\ffill)$
because a part of this accretion energy is already emitted in the hydrogen lines and continua. 
Rather, the luminosity from the photosphere at $\Teff$ is $\Lphot=\Lacc+\Lint-\Lshock$, where $\Lshock$ is the total outward travelling luminosity from the shock models. This holds for the accreting region of fractional area $\ffill$, while the rest of the planet surface has $\Teff=\Tint$.

For the spectrum,
we use the high-resolution (at \Ha: $\Delta\lambda=0.01$--0.05~\AA, i.e.\ $R\approx130,000$--$660,000$) solar-metallicity BT-Settl/AGSS2009 models \citep{Allard+2012} obtained from the SVO Theory Server\footnote{See \url{http://svo2.cab.inta-csic.es/theory/newov2/}.} and round $\log g$ to the nearest 0.5~dex and $\Teff$ up (to be conservative) to the nearest multiple of 100~K. To plot against the Doppler distance $\Delta v$ from the line centre, we use as the central wavelength of the photospheric models $\lmbdHa=656.283$~nm in air (for the BT-Settl models on the SVO server; \citealp{Wiese+Fuhr2009}).

For the \texttt{SpherAcc-Warm} case (Figure~\ref{fig:profilesWARM100}), the continuum is never visible on the linear scale of the line peak and would need to be $\sim10$ stronger to be even barely noticeable. Thus the photospheric emission is entirely negligible.
In the \texttt{Polar-Warm} case (Figure~\ref{fig:profilesWARM30})
for $\Mdot\lesssim3\times10^{-4}~\MdotUJ$, the continuum noise is not important,
but for the highest accretion rate it makes the \Ha line undetectable, except at 15~$\MJ$.
In the \texttt{SpherAcc-Cold} and \texttt{Polar-Cold} cases (Figures~\ref{fig:profilesCOLD100} and~\ref{fig:profilesCOLD30}), the photospheric noise is
mostly very low, with some exceptions at the highest accretion rate. %

In the \texttt{MagAcc-Warm} or \texttt{-Cold} cases shown in Figure~\ref{fig:profilesMagAkk}, the photospheric noise
(i.e.\ the amplitude of the features) is small to insignificant for most cases shown. The exception is for the combination ($\MP=5~\MJ$, $\Mdot=3\times10^{-5}~\MdotUJ$), whether in the \texttt{Warm} or in the \texttt{Cold} scenario,
in which
the \Ha line is in absorption in the atmospheric spectrum, with the other features smaller. This is maybe not realistic
because the BT-Settl models were calculated with the pressure--temperature structure of an isolated, not accreting, atmosphere.
From Equation~(\ref{eq:Pram}), the ram pressure for the cases in Figure~\ref{fig:profilesMagAkk} is $\Pram\approx3\times10^{-3}$--$3\times10^{-2}$~bar.
These pressures are likely neither very far up in the atmosphere nor very deep, in which cases the shock would, respectively, not or very much change the structure. Therefore, without detailed calculations of the resulting pressure--temperature structure of the atmosphere and the radiation transport, it is difficult to estimate the effect of the heating by the shock on the line emerging from the atmosphere below the shock.

\subsubsection{Observability}

Line-integrated luminosities of our models are presented in \citet{Aoyama+2020}, and the line-integrated extinction in Section~\ref{sec:line fluxes}. Up to now, the few known accreting low-mass objects have been observed in \Ha filters with a width of order of 1--2~nm (\citealp{wagner18}; \citealp{Zhou+2021}) or with MUSE at a resolution $\Delta\lambda\approx0.3$~nm \citep{Haffert+2019,Eriksson+2020}.
However, as seen in Section~\ref{sec:instr},
much more powerful instruments are underway or expected soon.
For example,
with $R\geqslant130,000$, %
the resolution of RISTRETTO
is comparable to the effective resolution %
of the model curves
given the native width of the spectral features
and the slight smear coming from the velocity gradient.
Therefore,
we discuss in this section the %
observability at high resolution of accreting planets
whose line profiles are in part shaped by the accreting gas.

In Figures~\ref{fig:profilesWARM100}, \ref{fig:profilesWARM30}, and~\ref{fig:profilesMagAkk}, we show 
line profiles $F_\lambda(\lambda)$ for our grid of models
convolved to $R=2516$ (for MUSE and slightly broader than HARMONI)
and to $R=15,000$ (for VIS-X).  %
The full-resolution curve (for RISTRETTO) was discussed above,
and Subaru+SCExAO/RHEA %
with $R=60,000$ is in an intermediate range.
The flux densities are for an observer on Earth for a source $d=150$~pc away, typical of several star-forming regions.

The resolution of MUSE corresponds to $\Delta v=120~\kms$, which is much larger than the spectral features. Therefore, they are not recognisable at this resolution. In some cases, very slight asymmetries between the red and blue wings are visible but they would be undetectable given any amount of noise and given the finite number of pixels per resolution element. Unfortunately, the higher resolution of HARMONI with $\Delta v=100~\kms$ will not yield a significant improvement. This emphasises the need for very-high-resolution spectrographs.

Even at $R=15,000$ ($\Delta v=20~\kms$), the finest spectral features are not distinguishable. Nevertheless, some asymmetries are preserved, for instance in the \texttt{Polar} case at $(\Mdot=3\times10^{-5}~\MdotUJ,\MP=2$--$5~\MJ)$, or more clearly at the highest \Mdot and \MP. This leaves the line centre shifted by approximately $-20~\kms$. Such large offsets are not expected from other effects such as planetary spin or orbital motion (see Section~\ref{sec:lineasymm}). Therefore, a line displaced from the theoretical value by a large amount could be a tell-tale sign of extinction from the surrounding gas.

In summary, the %
spectral resolving power of planned instruments such as RISTRETTO and VIS-X will make it possible in the next few years to start studying in detail the line profiles of low-mass companions, revealing the spectral imprint of the absorbing gas in the  accretion flow.

\section{Absorption by the dust}
 \label{sec:cont abs dust}

So far, we have neglected the contribution from the dust in our radiative transfer calculation.
Since dust exists only at low temperatures and these temperatures tend to be reached at larger distances from the planet,
the outer parts of the accretion flow can be important for setting the total optical depth.
In the case of accretion directly onto the planet, it seems likely that the last stages, that is, close to the planet surface, are radial and supersonic.
This leaves only the filling factor as the main free parameter of the accretion geometry relevant for the gas.
In the case of the dust, however, the fact that parts further out can contribute makes the analysis more tentative.

Given these uncertainties, we estimate what dust opacity is needed to have a significant optical depth in the accretion flow (Section~\ref{sec:dustoptdepth}). We then derive a range of plausible values for the dust opacity based on the relevant recent literature (Section~\ref{sec:dustopac}),
and in Section~\ref{sec:compdustkap} compare this to other recent estimates.
We derive the total dust optical depth in Section~\ref{sec:dustabsorbcombined}.

There is only a small range of temperatures in which
(i)~gas--magnetic field coupling is possible thanks to sufficient ionisation ($T\gtrsim1000$~K; \citealp{desch15})
and (ii)~at the same time the dust has not been evaporated ($T\lesssim\TZerst\approx1500$~K,
$\TZerst$ being the temperature at which the last component of the dust evaporates; \citealp{poll94,semenov03}).
In other words, in the \texttt{MagAcc} scenario the temperature is usually too high for dust to survive.
Therefore, we will put %
\texttt{MagAcc} %
aside for this section.

\subsection{Calculating the optical depth}
 \label{sec:dustoptdepth}

We first calculate the optical depth of the dust in the accretion flow.
Given 
the large uncertainties on the dust opacity (see Section~\ref{sec:dustopac}), we take a crude approach and let the dust opacity
be independent of density and temperature in the region where dust is present ($T<\TZerst$, or $r>\RZerst$)\footnote{A test with the weak powerlaw dependence $\kappa\propto T^{1/2}$ appropriate for the ``metal grains''
regime of \citet{bl94} indeed did not lead to a significantly different optical depth.}.
The contribution of the dust to the optical depth through the accreting matter is given by
\begin{equation}
 \label{eq:taudHa general}
 \taudHa = \int_\RZerst^\rmax \fpg\kappaStbint \,\rho \,dr,
\end{equation}
where we remind that $\kappaStbint$ %
is the opacity of the dust at \Ha\ as  %
a cross-section per unit mass of dust.
We have defined in Section~\ref{sec:oneex} the dust opacity as a cross-section per gram of gas as $\kapStbfpg\equiv\fpg\kappaStbint$ and will use this below.
The dust-to-gas mass ratio is essentially zero at $r<\RZerst$,
which is why the lower limit of the integration is the ``dust destruction front'' %
$\RZerst$ \citep{stahlerI}.
From Equation~(\ref{eq:T}), this is:
\begin{equation}
 \label{eq:RZerst}
 \RZerst = \left(\frac{T_0}{\TZerst}\right)^2\RP
\end{equation}
if %
the shock temperature is larger than $\TZerst$.
At low $\Mdot$ there is no destruction radius in the classical sense because the dust survives down to the shock (see Figure~\ref{fig:kappa gas Ha} and Section~6.2 of \citealt{m18Schock}).
In that case the lower limit of the integral in Equation~(\ref{eq:taudHa general}) is $\RP$ instead of $\RZerst$.
For simplicity, we do not take the weak density dependence of $\TZerst$ \citep{isella05} into account. The dust destruction front is illustrated in Figure~\ref{fig:Teresa}.

Since $\rho\propto r^{-3/2}$ (Equation~(\ref{eq:rhob}) when $\zeta=1$), the integral in Equation~(\ref{eq:taudHa general}) will converge, as $\rmax$ tends to infinity,
if $\kapStbfpg$ is constant or does not increase faster than $r^{1/2}$. However, this also assumes that the accretion radius $\RAkk$ is sufficiently large; if not, the outer parts of the flow contribute significantly because the density $\rho\propto v$ is higher there due to the smaller velocity $v\propto(1/r-1/\RAkk)$.

If $\RZerst\geqslant\RP$ and in the limit $\rmax\gg\RZerst$ and $\RAkk\gg\RP$, the optical depth (Equation~(\ref{eq:taudHa general})) becomes
\begin{subequations}
\label{eq:taudHa simple beides}
\begin{align}
  \taudHa =&~\fpg\kappaStbint\TZerst \left(\frac{\sigSB\Mdot^3\RP}{4\pi^3 \ffill^3 G^3\MP^3}\right)^{1/4}
           \label{eq:taudHa simple}\\  %
        = &~0.2\,
        \left(\frac{\fpg}{10^{-4}}\right)
        \left(\frac{\kappaStbint}{10^4~\mathrm{cm}^{2}\,\mathrm{g}_\mathrm{dust}^{-1}}\right)
                   \left(\frac{\TZerst}{1500~\mathrm{K}}\right) \notag \\
                  &  \times \left(\frac{\ffill}{0.30}\right)^{-3/4} \left(\frac{\Mdot}{\upmu\MdotUJ}\right)^{3/4}
                          \left(\frac{\MP}{5~\MJ}\right)^{-3/4}\notag \\
                  & \times \left(\frac{\RP}{2~\RJ}\right)^{1/4},\label{eq:taudHa simple num}
\end{align}
\end{subequations}
where the subscript ``dust'' in the units of $\kappaStbint$ indicates that
this is the cross-section per gram of dust, that is, the intrinsic (material) cross-section.
Equation~(\ref{eq:taudHa simple num}) was written with $\RP$ as an independent quantity
while in our fits it depends on $\Mdot$ and $\MP$.
However, $\taudHa$ does not depend strongly on $\RP$
(only as the fourth root).
The $\ffill$ factor on the denominator in these expressions comes from the fact that
$\taudHa$ is the optical depth through the accretion flow, with $\rho\propto\Mdot/\ffill$
(see Equation~(\ref{eq:rho})).
Thus the optical depth should increase with increasing accretion rate
(all the more since $\RP$ also grows with $\Mdot$),
whereas it should decrease somewhat with mass
(in part counterbalanced by the growth of $\RP$ with $\MP$, but only weakly because $\taudHa\propto\RP^{1/4}$).
Unfortunately, $\taudHa$ has a non-negligible dependence (linear) on
the very uncertain quantities $\fpg$ and $\kappaStbint$. In this light, the uncertainty on $\TZerst$ and its weak density dependence
do not matter. %

The prefactor in Equation~(\ref{eq:taudHa simple num}) implies that
for the nominal values in that equation, the dust is not able
to absorb much \Ha\ emission, with a flux decrease $\DF = 1-\exp(-0.2)\approx20$~\%. The parameter space is large, however, and we explore it more systematically in this and the following sections.

One can ask what the minimum average dust opacity
(as a cross-section per unit mass of gas) is needed to have $\taudHa\sim1$.
Equation~(\ref{eq:taudHa general}) can be written as
\begin{equation}
 \label{eq:taudHa}
  \taudHa = \Sigwhered \langle\kapStbfpg\rangle,
\end{equation}
where
\begin{equation}
\label{eq:Sigwhered}
\Sigwhered\equiv\int_\RZerst^\rmax\rho\,dr
\end{equation}
is the gas column density of the accretion flow where the dust is present (as opposed to the full column density)
and $\langle\kapStbfpg\rangle=\langle\fpg\kappaStbint\rangle$ is the dust absorption cross-section per unit gas mass averaged over that part of the accretion flow. Since we assumed the dust material opacity to be constant where it is non-zero, and if we also take $\fpg$ constant in the accretion flow, it holds simply that $\langle\kapStbfpg\rangle=\kapStbfpg$. Equation~(\ref{eq:taudHa}) implies that
$\Sigwhered^{-1}$ is equal to the minimum dust opacity
(as a cross-section per unit mass of gas) needed to have $\taudHa\approx1$ and thus contribute noticeably to the absorption. Over the narrow width of the line ($\Delta\lambda\lesssim0.001~\upmu$m), the dust opacity is %
constant, so that it can affect only the total flux %
but not the line shape.

\begin{figure*}[!ht] %
 \centering
 \includegraphics[width=            0.47 \textwidth]{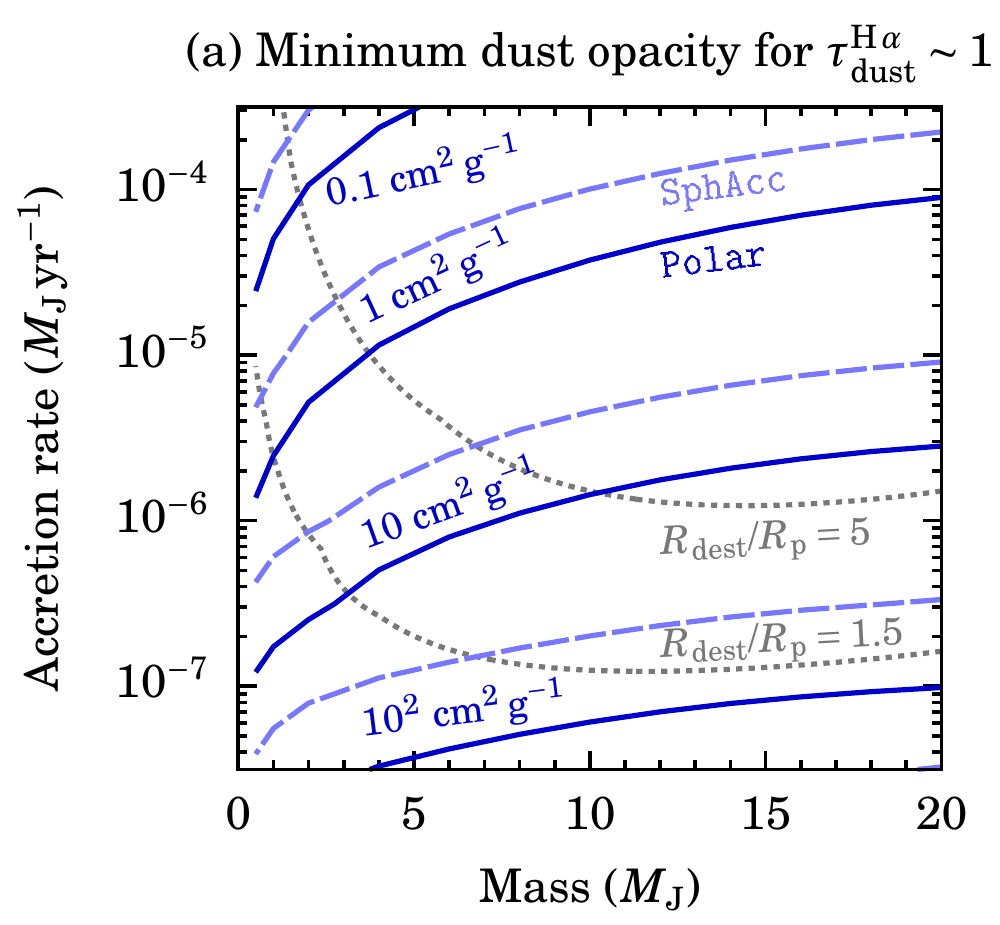}
 \includegraphics[width=            0.47 \textwidth]{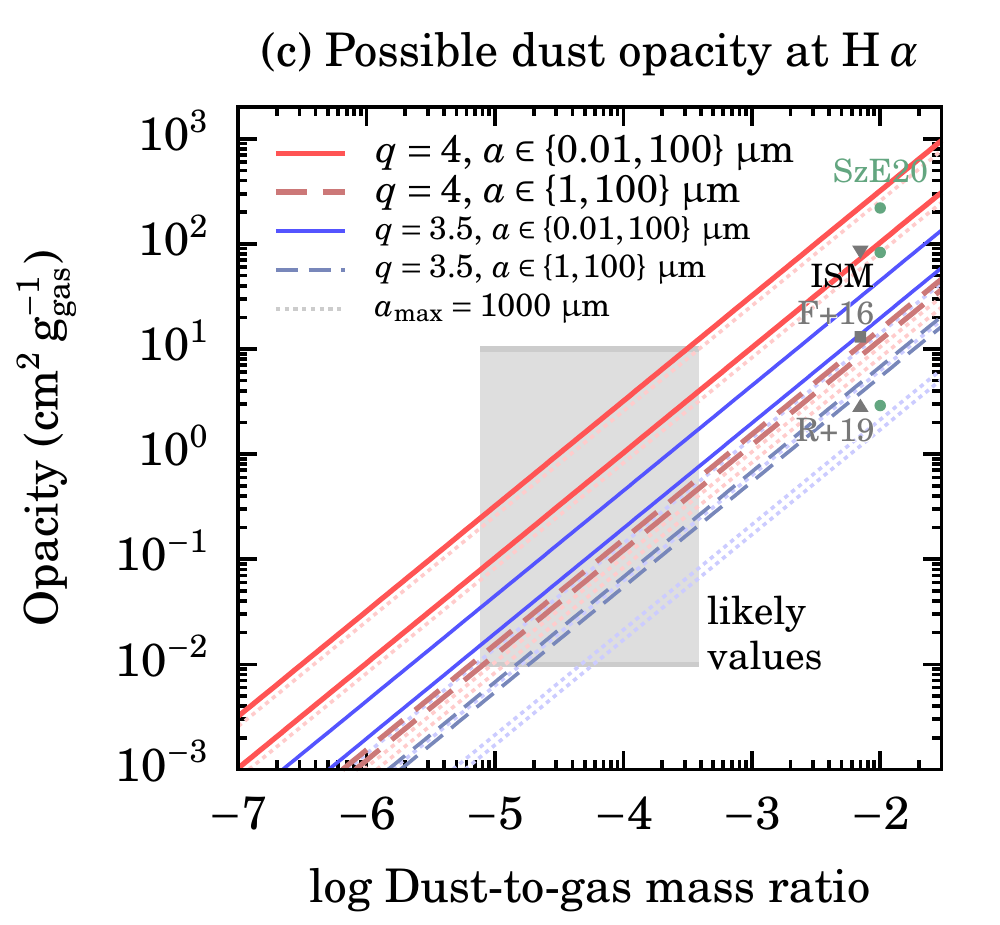}\\
 \includegraphics[width=0.94\textwidth]{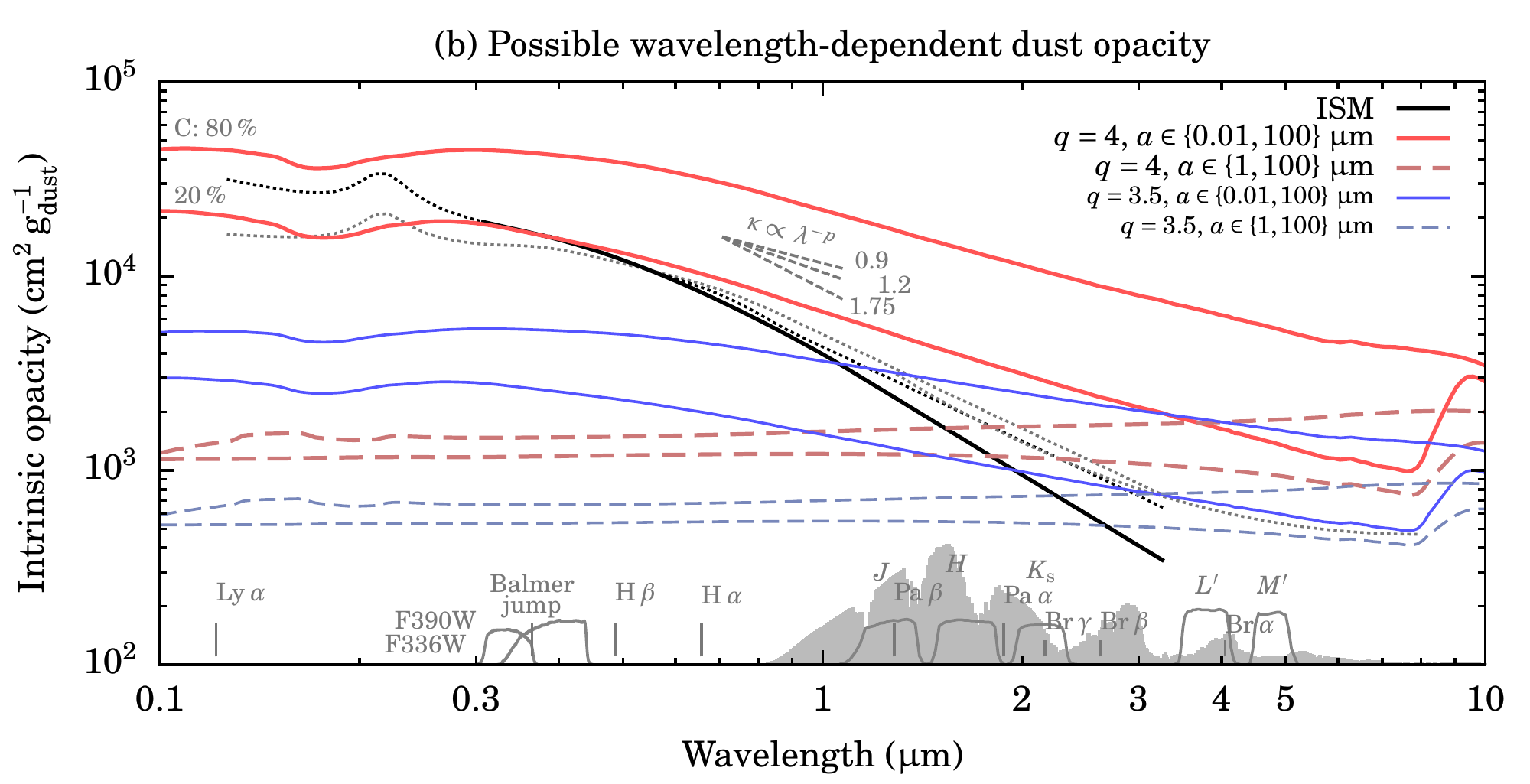}
\caption{
Estimate of the dust absorption in the accretion flow.
\textit{(a)}~Minimum dust opacity at \Ha,  %
 $\fpg\kappaStbint$ %
(per unit gas mass; contour labels), %
needed to have $\taudHa\approx1$ in the accretion flow onto a growing gas giant (Equation~(\ref{eq:taudHa})) in the \texttt{Polar} (solid lines)
and \texttt{SpherAcc}
(dashed lines)
cases.
We take the cold-population radii and fix $\TZerst=1500$~K.
Grey dotted lines show where $\RZerst/\RP=1.5$ and~5 (Equation~(\ref{eq:RZerst})).
\textit{(b)}~Material %
opacity  %
$\kappaStbint$.
We show the ISM fits of \citet{cardelli89} for $\RV=3.1$ (dotted black)  %
and~5.5 (dotted grey), with the absolute scale from \citet{g09}; %
\citet{wang19} (solid black);
and \citet{chiar06} (dotted grey, up to 8~$\upmu$m). %
Red and blue curves are for size distributions set by the slope $q$ and minimum size (see legend).
Curves for each model are for 20 or 80\,\%\ carbon %
(bottom to top), with silicate completing.
At the bottom, hydrogen lines, two HST filters, and NACO IR filters are shown. The grey area is a BT-Settl model with $\Teff=1200$~K and $\log g=4$.
\textit{(c)}~Estimate of the dust opacity $\kapStbfpg=\fpg\kappaStbint$ in the accretion flow.
Pale dotted lines are for $\amax=1$~mm instead of $0.1$~mm.
Red curves represent %
recent simulation results (see text),
with a low dust abundance $\fpg\sim10^{-5}$--$10^{-4}$,
implying $\kapStbfpg\sim0.3$~cm$^2$\,g$^{-1}$ with a half-spread $\sigma=1.5$~dex (grey shaded region).  %
The opacity of \citet[][``F+16'']{flock16}, \citet[][``R+19'']{rab19}, and \citet[][``ISM'']{Sanchis+2020} is shown for $\fpg=0.01$ %
(grey symbols, shifted left) and the pure-graphite, ``mixture'', and pure-silicate opacity of \citet[][``SzE20'']{szul20} is shown (green circles; top to bottom). All but \citet{flock16} assume $\fpg=0.01$ in their work.
}
\label{fig:SigmadkapStb}
\end{figure*}

Figure~\ref{fig:SigmadkapStb}a displays the minimum average dust opacity required to have an optical depth of unity.
We focus on the \texttt{SpherAcc} and \texttt{Polar} cases.
We find that over the parameter range considered but restricting ourselves to $\Mdot\lesssim3\times10^{-5}~\MdotUJ$,
the minimum dust opacity required in order for the dust to absorb a significant fraction of the \Ha\ signal
is of order $\Sigwhered^{-1}\gtrsim1$--100~cm$^2$\,g$^{-1}$.  %
The results of
Figure~\ref{fig:SigmadkapStb}a are only moderately sensitive to the planet mass and scale roughly as $\Sigmad\sim\Mdot^{1.5}$.
The choice of the cold- or warm-population radii is not crucial (not shown),
and the filling factor only has a moderate effect (solid vs.\ dashed lines).
For most of parameter space, the shock is hot enough that the dust destruction radius
is at several times the planet radius, which typically corresponds to $\RZerst\approx10$--30~$\RJ$ (see grey dotted lines). %
For increasing accretion rate, the size of the dust-free inner region relative to the planet radius, $\RZerst/\RP$, grows. Nevertheless, the dust column density increases with $\Mdot^{1.5}$, so that the minimum opacity required becomes smaller.

We have assumed that the dust opacity has a constant value throughout the accretion flow at $r>\RZerst$ and is zero within the dust destruction front.
If it is not too far out (i.e.\ if $\RZerst\ll\rmax$), and assuming that $\RAkk\gg\RP$, the dust optical depth $\taudHa$ can be expressed analytically (Equation~(\ref{eq:taudHa simple beides})).
In this case, but also in general, $\taudHa$ depends linearly on the dust opacity.
We now turn to the task of estimating its value.%

\subsection{Realistic estimates of dust opacity}
 \label{sec:dustopac}

\subsubsection{General considerations about dust parameters}

The dust monochromatic opacity is set amongst others by the composition, shape, space- and time-dependent size distribution, and dynamics of dust particles in CSDs \citep{andrews20} and specifically  %
in the vicinity of a growing and %
migrating gas giant (e.g.\ \citealp{poll94}).
In particular, the dust-to-gas mass ratio $\fpg$ and---to a lesser extent \citep{chachan21}---the minimum and maximum grain sizes $\amin$ and $\amax$ impact the opacity (e.g.\ \citealp{cuzzi14,kataoka14,woitke16,krapp21}).
In turn, these properties are set by %
many processes
(e.g.\ \citealp{flock16,birnstiel18,vorobyov18a,dr19}; see review in the latter work).

To obtain accurate dust opacities would require global-disc radiation-hydrodynamic simulations following the dust growth, drift, and evaporation with sufficiently high resolution both in space and in the dust size, while covering the full range of grain sizes that set the opacity, and taking the feedback of the grains on the disc structure into account. This is not yet computationally feasible. However, different studies have looked at some of these aspects (e.g.\ \citealp{dr19,savvidou20,chachan21,binkert21,szul21,krapp21}),
with some properties emerging. We highlight briefly four of them.

Firstly,
since the density in the accretion flow decreases outwards, the layers in the dusty part of the flow (where $T\lesssim1500$~K) that are closest to the planet will likely contribute the most.
For $T\gtrsim700$~K, %
only
silicate along with carbon, iron, or troilite %
remain
\citep{semenov03,woitke16}.
Secondly,
in 2D simulations,
the pressure perturbation of the planet keeps the large grains outside of its orbit (``filters them out''; e.g.\ \citealp{paardekooper04,rice06,zhu12,bae19}).
For example,  %
\citet{dr19} found a maximum grain size $\amax\sim0.03$~cm around the planet instead of $\amax\sim3$~cm at larger orbital radii in theirglobal-disc hydrodynamical simulations combined with a dust evolution model.
Also the size distribution near the planet is different, with
$n(a)\propto a^{-q}$ with $q\approx4$, where $n(a)$ is the number density of grains of size $a$ (J.~Dr\k{a}\.zkowska 2020, priv.\ comm.). This is steeper than the commonly used \citet{mrn77} ISM distribution with $q=3.5$.

In 3D, meridional circulation could bring large grains towards a forming planet \citep{bi21,szul21},
bridging the pressure bump.
Realistically, this %
however depends on the amount of turbulence and settling \citep{dullemond04} and the strength, for instance, of the vertical shear instability (VSI; \citealp{flock20}), or the viscosity and the gap depth \citep{kanagawa18b}.   %
Thirdly,
a range of different minimum grain sizes $\amin$ is used in the literature:
$\amin=0.1$--1~$\upmu$m (e.g.\ \citealp{okuzumi12,kataoka14,bae19,stammler19}; see brief review in \citealp{xiang20});  %
$\amin=1~\upmu$m in \citet{dr19} but with some pile-up %
near $\amin$ in the resulting distribution;
\citet{flock16} used a smaller value: $\amin=5~$nm\footnote{%
Not $\amin=5~\upmu$m as their Appendix~A states (M.\ Flock 2020, priv.\ comm.).
That their opacity curve has features at $\lambda\ll5$~$\upmu$m suggests this, since at
$\lambda\ll2\pi\amin$ the geometric, $\lambda$-independent limit for the cross-section per particle $\sigma=2\pi a^2$ must hold (e.g.\ \citealp{morda14b}).\label{fn:aminfl16}}.

Fourthly,
concerning the dust abundance,
the simulations of \citet{pinilla12} or \citet{dr19} find depletions
by a factor of 
$10^3$--$10^4$ relative to the global disc abundance.
In a sample of seven discs, \citet{powell19} found with a method independent of a tracer-to-mass conversion a global\footnote{Although $\fpg$ likely varies on global disc scales \citep{soon19}.}
dust-to-gas ratio of order $\fpg\sim10^{-4}$--$10^{-3}$.
Recent global-disc simulations of grain growth and drift (e.g.\ \citealp{savvidou20,chachan21}) lend support to this.
Finally, if planets form early (e.g.\ \citep{manara18}), the dust particles may be locked up in macroscopic objects (pebbles, planetesimals, planetary cores), also reducing $\fpg$.

We assume that
the dust flowing onto the planet
has the same size distribution as in the gap around the planet on scales of $\RHill$.
Based on the previous discussion,
to estimate the dust opacity we consider the following parameter values:
$\amin\in\{0.01,1\}~\upmu$m, $\amax\in\{0.1,1\}$~mm, $q\in\{3.5,4\}$, the fraction of (amorphous) carbon $\fC\in\{0.2,0.8\}$ by mass, with the rest made up of the usual astrophysical silicate (pyroxene: Mg$_{0.7}$Fe$_{0.3}$SiO$_3$), and the porosity $P\in\{0.1,0.2,0.3\}$, the vacuum fraction by volume (and not by mass).  %

As shown in \citet{woitke16}, $\kappaStbint$ increases with increasing $q$ (more small grains) and $\fC$ (more carbon relative to silicate) or with decreasing $\amin$ (smaller grains included) or $\amax$ (larger grains excluded). The dependence of the opacity on the size can be understood from the fact that small grains have a larger cross-section-to-mass (area-to-volume) ratio than large grains. For the parameter values here, the opacity is almost insensitive to the porosity $P$, varying at most (at the highest carbon fraction) by 30~\%\ for the range of $P$ covered here. Therefore, we fix $P=0.2$ hereafter.

\subsubsection{Resulting monochromatic dust opacity}

To calculate the dust opacity, we use
with the convenient \texttt{OpTool}\footnote{Available at \url{https://github.com/cdominik/optool}.} \citep{dominik21},
which uses \citet{toon81} and, amongst others, the ``distribution of hollow spheres'' \citep{min05}.
Optical constants are from \citet{zubko96} and \citet{dorschner95} for the carbon and silicates, respectively.
Other parameters are as preset. 

We plot in Figure~\ref{fig:SigmadkapStb}b the opacity from the UV to the mid-infrared (MIR) for different dust models.
We label the curves by $q$, $\amin$, and $\amax$.
At \Ha, this leads to a similar or larger range of opacities compared to including other materials such as iron (troilite) or water (ice) or even CHON organic material, and varying their abundance. Silicate-rich grains ($\fC=0.2$) lead to  strong absorption at 10~$\upmu$m (the ``ten-micron bump'').

For these opacity curves,
the logarithmic average opacity slope %
$\pUV=\Delta\log\kappa/\Delta\log\lambda$
between \Hb ($\lmbdHb=486$~nm) and \Ha
is
roughly $\pUV\approx-0.5$ to $-1$.
Larger dust grain size distribution powerlaw exponent values %
$q$ and smaller values of $\amin$ make the slope steeper, that is, $\pUV$ more negative,
while 
$\amax$ essentially does not change the slope much (see Figure~3 of \citealp{woitke16}) and only decreases the absolute amount of extinction.
For comparison,
the extinction law of \citet{cardelli89}, valid at \Ha and \Hb, has
$\pUV=-1.2$ ($\pUV=-0.9$) for $\RV=3.1$ ($\RV=5$).
\citet{wang19} recently adjusted the \citet{cardelli89} extinction law with $\RV=3.1$ and obtained $\pUV=-1.4$.
We note that the fundamental reason why our curves are flatter than %
the empirical ISM curve is not yet understood. %

For reference, we indicate in Figure~\ref{fig:SigmadkapStb}b the position of some hydrogen lines and of the Balmer jump (H\,$\infty$ at 3646~\AA) as well as the flanking $U$-band F336W and F390W filters of the \textit{Hubble Space Telescope} (HST).
This is motivated by the recent detection of \PDSb by \citet{Zhou+2021} at F336W, a first for an accreting planet.
We also diplay NIR and MIR filters from VLT/NACO ($J$, $H$, \Ks, $L'$, $M'$).
Finally, we also show a scaled spectrum with $\Teff=1200$~K and $\log g=4$ from BT-Settl, which approximately fits the data for \PDSb \citep{Stolker+20b,wang21vlti}.

Figure~\ref{fig:SigmadkapStb}c displays the opacity $\kapStbfpg$ as a function of $\fpg$. 
We consider that a range of $\fpg\sim10^{-5}$--$3\times10^{-4}$ is plausible given the relative depletion by a factor of 100--1000 found by \citet{dr19} and spatial variations (on scales larger than planets' Hill spheres, as the results of \citealp{soon19} suggest).

We obtain that the material opacity of the dust can be anywhere between $\kappaStbint\sim300$ and~$3\times10^{4}$~cm$^2$\,g$^{-1}_{\mathrm{dust}}$, which translates into $\kapStbfpg\sim10^{-2}$--$10$~cm$^2$\,g$^{-1}_{\textrm{gas}}$ at $\fpg\sim10^{-5}$--$3\times10^{-4}$. Uncertainties in both the size distribution of the grains and the material properties (including composition and porosity) lead to this spread of three orders of magnitude.
The large uncertainty in $\kappaStbint$ justifies our simplification that it is constant in the accretion flow.

\subsection{Comparison with other estimates of dust opacity}
 \label{sec:compdustkap}

We compare to the opacity used by \citet{Sanchis+2020}. (See also the discussion of their work in Section~\ref{sec:cfSanchisSzul}.) They took the ISM opacity in the $V$~band
$\kappa_V=107$~cm$^2$\,g$^{-1}_{\textrm{gas}}$
$\kappa_V=107$~cm$^2$\,g(gas)$^{-1}$
from \citet{g09}\footnote{For comparison, the commonly used law of \citet{bohlin78} gives an opacity only 10~\%\ smaller than in \citet{g09}.%
} assuming $\fpg=0.01$ and a mean molecular weigth $\mu=2.353$, and used the ISM dust extinction law from \citet{cardelli89} with $\RV=3.1$ (E.~Sanchis 2020, priv.\ comm.) to scale to other bands up to $K$, and the one from \citet{chiar06} for longer wavelengths. Doing this for \Ha, we obtain $\kapStbfpg =88$~cm$^2$\,g$^{-1}_{\textrm{gas}}$.
This is much higher than the upper range of our suggested values. Nevertheless, the intrinsic opacity (cross-section per gram of dust) is similar to that of some of our curves.

In Figure~\ref{fig:SigmadkapStb}c we also include for reference the dust opacities used in \citet{szul20} at their assumed $\fpg=0.01$. We computed them with \texttt{OpTool} for their pure-silicate, silicate--water--carbon (``mixture''), and pure-graphite cases (respectively from weakest to strongest).
The lowest dust opacity of \citet{szul20}, their ``silicate'' case ($\kapStbfpg=2.9$~cm$^2$\,g$^{-1}_{\textrm{gas}}$), is consistent with our opacities, whereas the two other cases are 10--20~times higher than ours.
Here again, however, the range of intrinsic opacities is almost identical to ours, with the real difference only in the adopted dust-to-gass mass ratio. %

\citet{flock16}, who assumed
$\fC=37.5$~\%\ and silicate for the rest (see above for the other parameters), used the \texttt{MieX} code \citep{wolf04} to obtain $\kappaStbint=1300$~cm$^2$\,g$_\mathrm{dust}^{-1}$. With their reference global mass ratio $\fpg=10^{-2}$, their opacity as a cross-section per gram of gas fits within our range. The same holds for the opacity of \citet{rab19} (see above), which we computed with \texttt{OpTool} to be $\kappaStbint=280$~cm$^2$\,g$_\mathrm{dust}^{-1}$. It is however at the lower edge of the range of intrinsic opacities because of their large maximal grain size ($\amax=3$~mm). Such large grains are not expected in our case, in the accretion flow onto a growing planet.

\subsection{Final estimate of the absorption by the dust}
 \label{sec:dustabsorbcombined}

Comparing the estimated dust opacity with the minimum opacity for absorption to be important (Figure~\ref{fig:SigmadkapStb}),  %
we find that the dust optical depth in the accretion flow is likely less than unity ($\AR<1$~mag) for planets more massive than a few~$\MJ$ accreting at $\Mdot\lesssim3\times10^{-6}~\MdotUJ$ and even at a very high rate $\Mdot\lesssim3\times10^{-5}~\MdotUJ$ if we consider the middle of the opacity range ($\kapStbfpg\approx0.3$~cm$^2$\,g$_\mathrm{gas}^{-1}$). This does not depend much on the geometry (\texttt{SpherAcc} or \texttt{Polar}), and even less on the choice of the hot- or cold-population radii (not shown). For the highest accretion rates that we consider this would lead to some absorption (optical depth of a few), similar to the contribution from the gas to the absorption (Figure~\ref{fig:DeltaF}). If the dust opacity is at the lower end of the estimated range ($\kapStbfpg\approx10^{-2}$~cm$^2$\,g$_\mathrm{gas}^{-1}$), there is no accretion rate for which the dust can reduce the \Ha\ flux by more than a percent.

At low masses ($\MP\lesssim3~\MJ$) and for opacity values at the high end ($\kapStbfpg\sim10$~cm$^2$\,g$_\mathrm{gas}^{-1}$), however,
there can be a large amount of extinction reaching $\AHa\sim10$~mag for $\Mdot\sim10^{-5}~\MdotUJ$.

In Section~\ref{sec:cfSanchisSzul},
we compare these results to the work of \citet{Sanchis+2020} and \citet{szul20},
whose density structures are very different from ours due to their smoothing of the gravitational potential.

\section{Observational consequences and example applications}
 \label{sec:obscons}

Here, we %
present 
an absorption-modified $\LHa$--$\Mdot$ relationship (Section~\ref{sec:MdotfromHa}), and apply the models to \Dlrmb and \PDSb (Section~\ref{sec:app}).

\subsection[Accretion rate from H alpha luminosity]{Accretion rate from \Ha\ luminosity}
 \label{sec:MdotfromHa}

\begin{figure*} %
 \centering
 \includegraphics[width=0.47\textwidth]{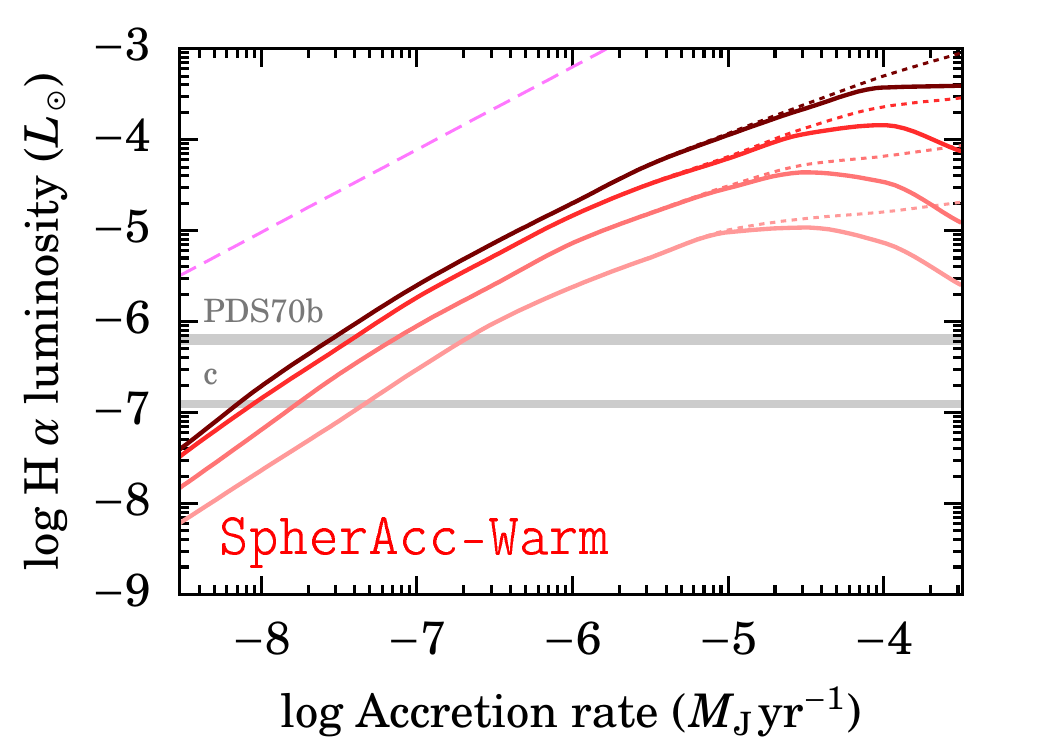}
 \includegraphics[width=0.47\textwidth]{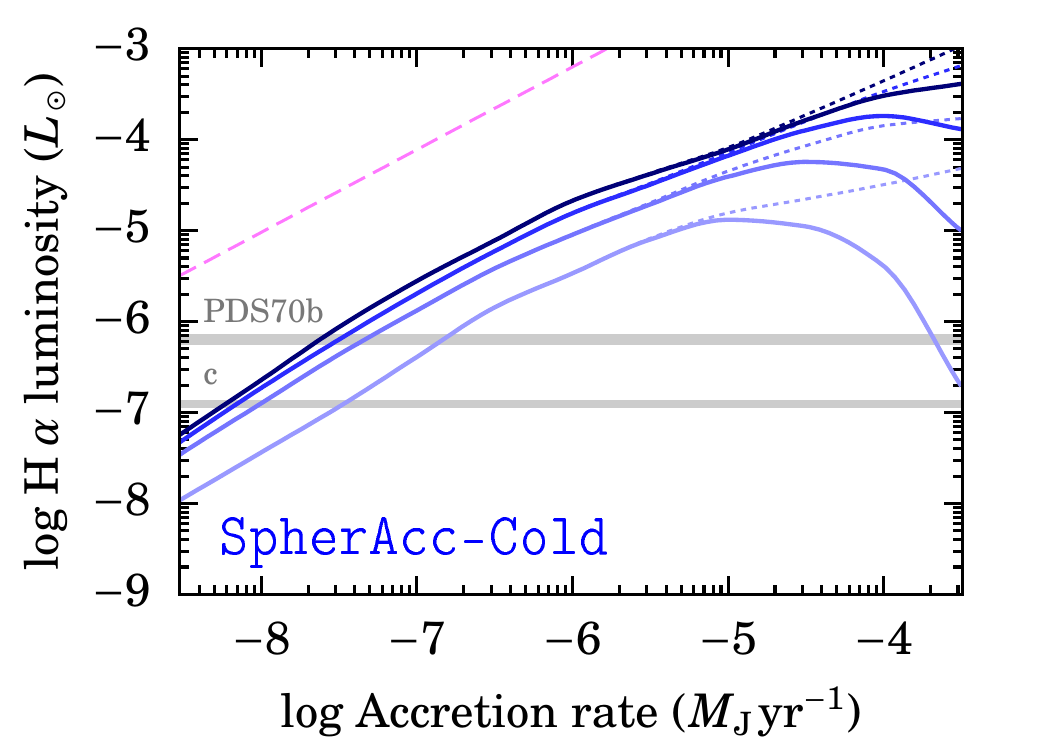}\\
 \includegraphics[width=0.47\textwidth]{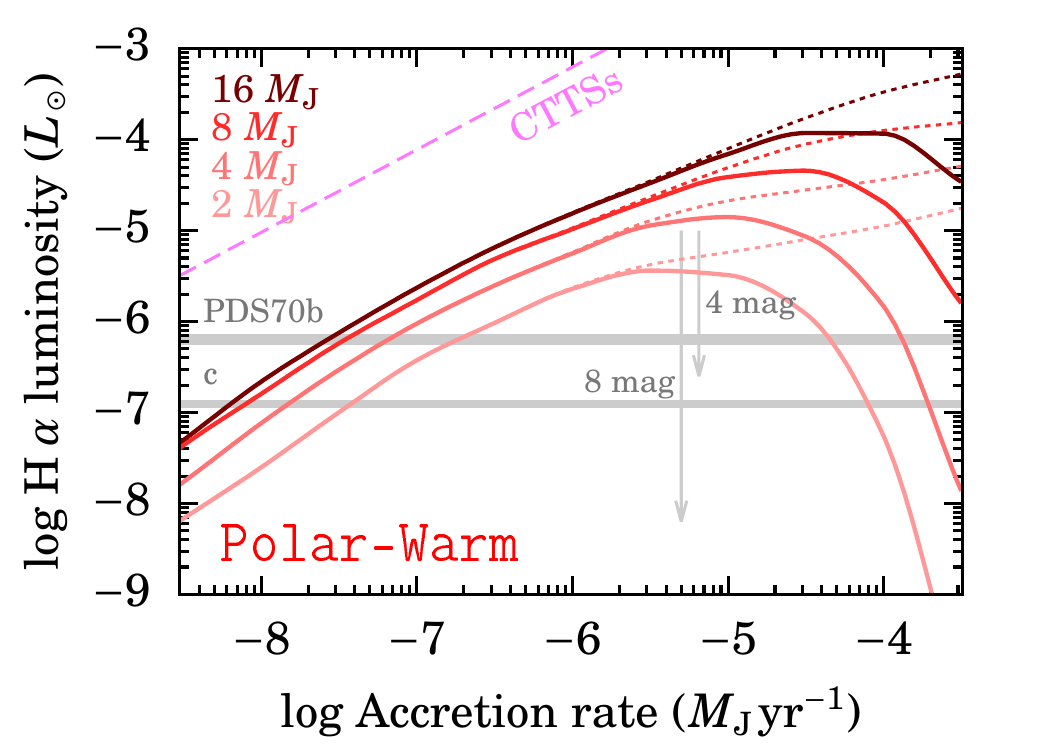}
 \includegraphics[width=0.47\textwidth]{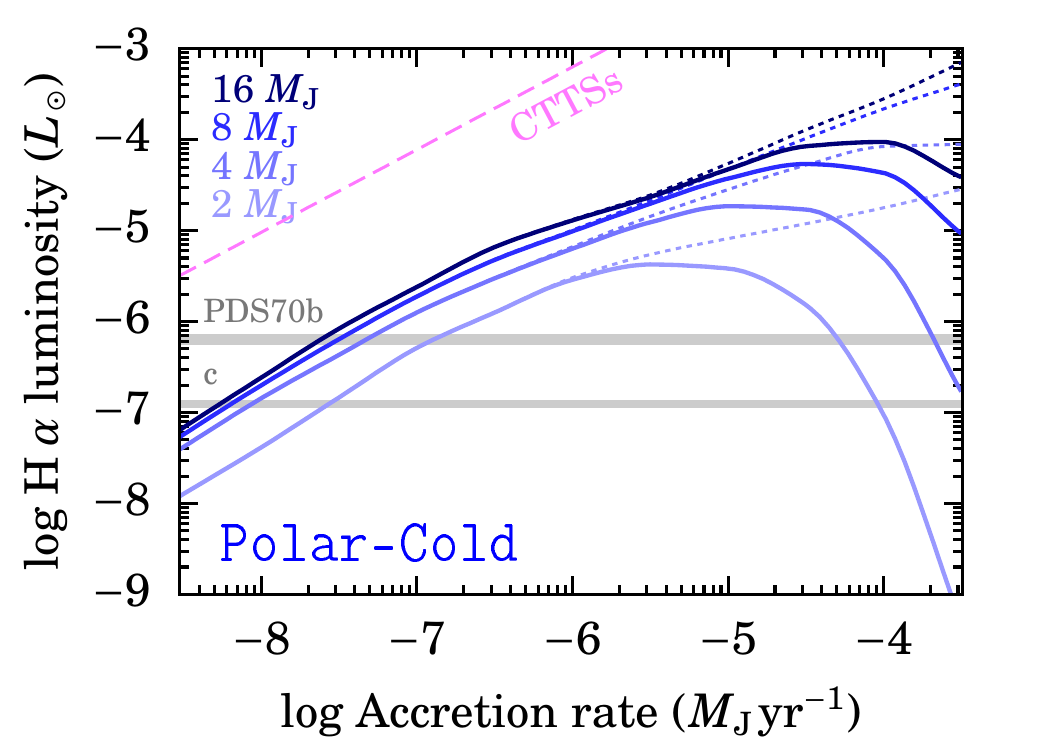}\\
 \includegraphics[width=0.47\textwidth]{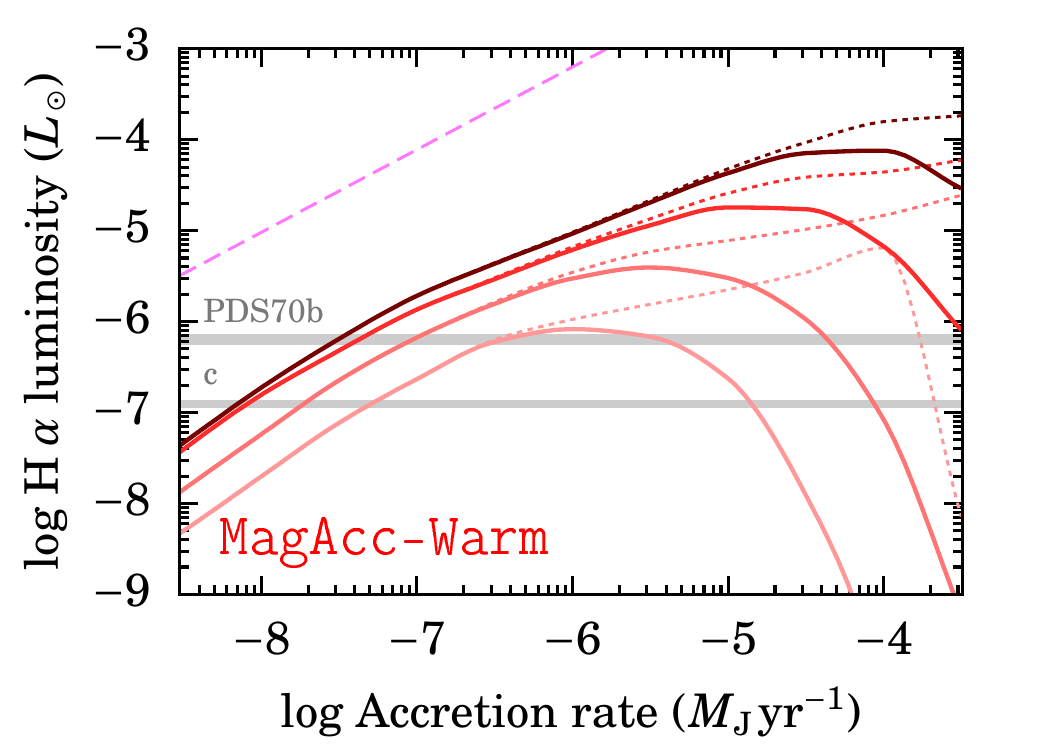}
 \includegraphics[width=0.47\textwidth]{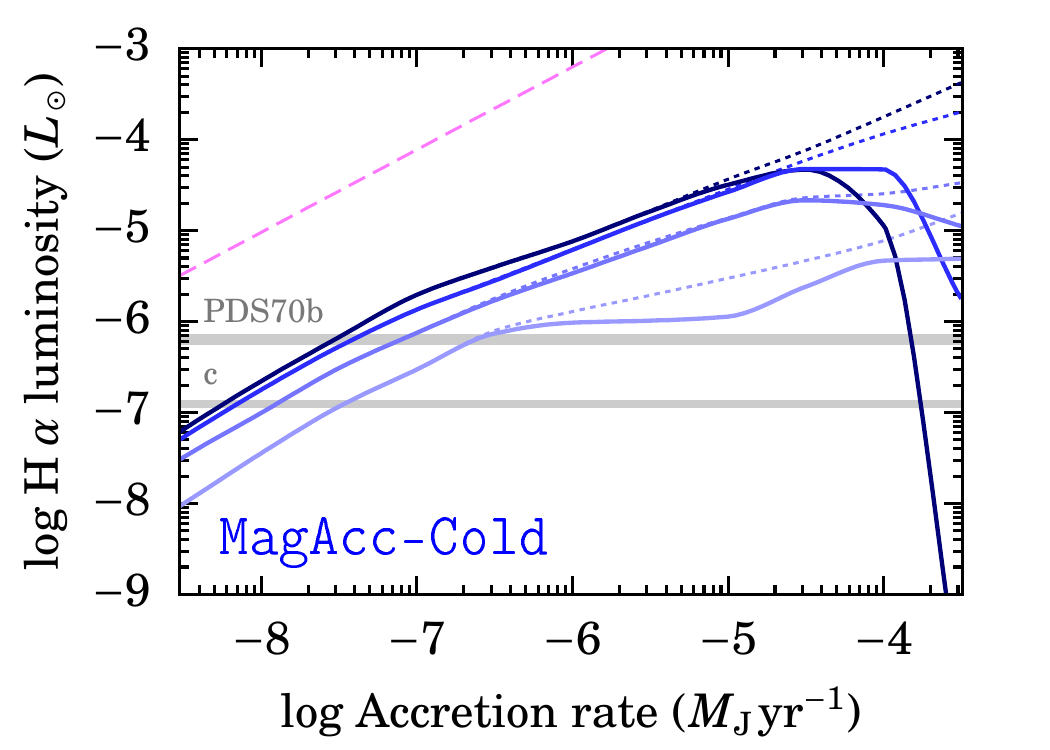}\\
\caption{
Absorption-modified relationship (\textit{solid curves}) between %
\Ha luminosity
and accretion rate for the warm- (left panel) and cold-population (right panel)
radius fits for the three accretion geometries (rows; see labels).
We consider only absorption by the gas,and also show the case of no absorption (\textit{dotted curves}).
Curves are for masses of 2--16~$\MJ$ (bottom to top, but with an inversion in \texttt{MagAcc-Cold} at very high $\Mdot$).
Horizontal bands are for the luminosity %
of \PDSb\ \citep{Zhou+2021} %
and \PDSc\  
\citep{Hashimoto+2020}.
The fit of \citet{Ingleby+2013} for CTTSs is also shown (\textit{pink dashed line}).
For reference, one panel shows extinction arrows for $\AHa=4$ and 8~mag.
}
\label{fig:LHaMdot}
\end{figure*}

Several authors have studied the empirical correlation between \Ha luminosity and accretion rate for CTTSs (e.g.\ \citealp{Natta+2004,Rigliaco+2012,Ingleby+2013}). Recently, this has been extended theoretically to a mass of 6~$\MJ$ for the scenario in which the \Ha is generated by magnetospheric accretion columns \citep{Thanathibodee+2019}.
In \citet{AMIM21L}, we presented the $\Mdot$--$\LHa$ correlation
for our models for accreting planets, looking also at other hydrogen lines.
Here, we briefly present this correlation again but including the effects of absorption.
Given the large uncertainty about the dust opacity (Section~\ref{sec:dustopac}), we consider only the gas opacity here.

\citet{Aoyama+2020} explained that self-absorption, which occurs in the postshock region (depicted in Figure~\ref{fig:Teresa}c),
lets the scaling of $\LHa$ with $\Mdot$ become sub-linear.
We find that absorption by the matter flowing onto the planet strengthens this trend
and, depending on $\ffill$, can lead to a maximum (``saturation'') luminosity,
that is, a flattening and turning over of $\LHa$ as a function of $\Mdot$.
At low \Mdot, the \Ha luminosity is intrinsically low, while a high \Mdot leads both to a higher \Ha luminosity and to a stronger extinction, with the second effect dominating.

Our $\Mdot$--$\LHa$ correlation is shown in Figure~\ref{fig:LHaMdot} for $\MP=2$ to 16~$\MJ$ for all geometries and both radius fits.
For objects with $\MP\lesssim15~\MJ$, the maximum line-integrated luminosity is $\LHamax\approx3\times10^{-4}~\LSun$ in the \texttt{SpherAcc} geometry
and $\LHamax\approx10^{-4}~\LSun$ in the \texttt{Polar} and \texttt{MagAcc} geometries,
reached over a range of accretion rates around $\Mdot\gtrsim3\times10^{-5}~\MdotUJ$.
This is relatively insensitive to the choice of the radius fit.
Including the extinction by the dust would only 
lead to a stronger downturn
because its importance increases with \Mdot (see Figure~\ref{fig:SigmadkapStb}).

Interestingly, this implies that for a range of \LHa values, especially in non-spherically symmetric geometries, there are two solutions to explain a given \LHa observation: a low \Mdot without extinction, or a high \Mdot with extinction. The luminosities of \PDSb and~c fall precisely in this range of \LHa. We return to this in detail in Section~\ref{sec:app}. The high-\Mdot solution might be statistically not preferred because observing a planet in a (presumably short) phase of massive accretion is unlikely.
Assuming additional amounts of absorption by dust implies accretion rates intermediate between the low and the high solutions.

Could this maximum on the luminosity be responsible for the non-detections of dedicated recent surveys \citep{Cugno+2019,Zurlo+2020,xie20}? Roughly, these surveys typically reached sensitivities down to $\LHa\sim10^{-7}$--$10^{-6}~\LSun$ beyond 15~au.
For most extinction geometries (see Figure~\ref{fig:LHaMdot}), this is a few to several orders of magnitude below the peak luminosity. Only for low masses and small filling factors could extinction push a planet into non-detectability, at high accretion rates (in Figure~\ref{fig:LHaMdot}, where the dotted lines differ from the solid ones).
Closer in to the surveyed stars, where planets are expected to be more numerous (e.g.\ \citealp{bowler16,fernandes19,nielsen19,vigan21}), the sensitivity curves become quickly much less constraining; that is, the \LHa upper limits are higher. In that case, extinction affects only the higher-mass planets, but less than lower-mass planets. The low-mass planets\footnote{For example $\MP\lesssim2~\MJ$, but the steepness of the sensitivity curves implies that no single value is representative.} are not detectable in any case.
Altogether, extinction by the accreting gas does not appear to be the reason why so few accreting planets have been observed.
Other, more likely explanations are discussed in %
\citet{Aoyama+2020} and include the intrinsic rarity of giant planets at the large separations from their host stars to which observations have been sensitive up to now. Depending on the value of the dust opacity (Figure~\ref{fig:SigmadkapStb}), however, the dust could also contribute to hiding accretors. Finally, in general the circumstellar disc could (also) be a source of extinction, to which we return %
at the end of Section~\ref{sec:cfSanchisSzul}.

We also briefly compare our absorption-modified \Mdot--$\LHa$ relationship to the fit of \citet{Ingleby+2013} for CTTSs.
A detailed comparison, including a discussion of the physical differences, is given in \citet{AMIM21L}.
Up to $\Mdot\sim10^{-5.5}~\MdotUJ$ for \texttt{SpherAcc} and \texttt{Polar}, and up to $\Mdot\sim10^{-6.5}~\MdotUJ$ for \texttt{MagAcc}, the slopes are similar between the linear fit and our curves, but with a significant offset of approximately 1.5~dex (several times the spread $\sigma\approx0.5$~dex of the fit of \citealt{Ingleby+2013}). At large accretion rates, the difference grows dramatically. This implies that a given observed \Ha luminosity requires much more vigorous accretion in planets than in stars.

\subsection{Application to known accreting planets}
 \label{sec:app}

Robust detections of \Ha emission from planetary-mass objects exist only for the \PDS\xspace companions \citep{wagner18,Haffert+2019} and for \Dlrmb \citep{Eriksson+2020}.  
The latter, a circumbinary low-mass companion, exhibits hydrogen- and helium-line emission, which is naturally explained by an accretion  scenario\footnote{%
There are several arguments against the alternative origin, chromospheric activity; see \citet{Eriksson+2020}.
}.
The inferred accretion rate onto \Dlrmb is $\Mdot\approx10^{-9.5}$ or $\approx10^{-8}~\MdotUJ$, depending on the model used to convert the line luminosities into $\Mdot$. %
Even at %
$\Mdot\approx10^{-8}~\MdotUJ$,
extinction by either the gas or the dust (Figures~\ref{fig:DeltaF} or~\ref{fig:SigmadkapStb}, respectively) is entirely negligible. %

\begin{figure}
 \centering
 \includegraphics[width=0.47\textwidth]{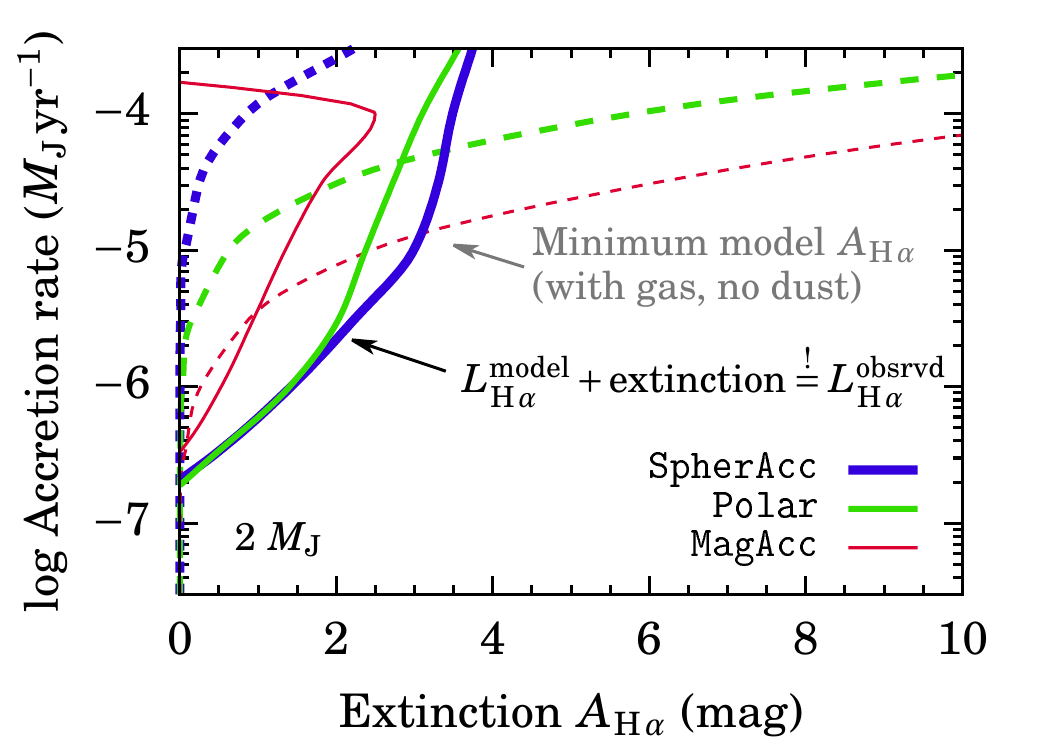}
 \caption{%
Constraints on the accretion rate onto and total extinction towards \PDSb ($\MP\approx2~\MJ$).
For a given geometry (line thickness or colour), possible values of \AHa are rightward of the dashed line (minimum extinction as a function of $\Mdot$) and along the solid curve %
(implied by our emission model and \LHa from \citealt{Zhou+2021}).
Results barely depend on the radius fit (\texttt{Warm} is used here).
}
\label{fig:MPktAHa}
\end{figure}

We combine the measured luminosity of \PDSb with our models to derive joint constraints on the accretion rate and the total amount of extinction.
We consider a low mass ($\MP=2~\MJ$ for definiteness) as suggested by, for instance, \citet{Stolker+20b} and \citet{wang21vlti},
while acknowledging that in $\log(\MP)$, the latter study suggests similar probabilities for low and high (logarithmic) masses. %

We derive the $\Mdot$--\AHa constraints in two steps.
First, we use
the measured value\footnote{%
This assumes uniform redistribution over a sphere of radius $113.4$~pc. We ignore the errorbars on \LHa for the analysis here.%
} of $\LHa=(6.5\pm0.9)\times10^{-7}~\LSun$
 \citep{Zhou+2021}
and draw in Figure~\ref{fig:MPktAHa}, for each geometry, the contour of constant extincted \LHa equal to the observed value (solid lines).
This reveals a minimum accretion rate $\Mdot\approx2\times10^{-7}~\MdotUJ$.

Then, we show the extinction $\AHa(\Mdot)$ calculated in this work
coming only from the gas and without dust,
again for each geometry (dashed lines).
This is independent of the measured luminosity.
The total extinction will therefore be on the line or rightward of it.

Comparing the solid and the dashed line of Figure~\ref{fig:MPktAHa} for each geometry, there is thus only a range of possible accretion rates and extinctions. The maximum $\Mdot$, set by the crossing of the curves,
is near $\Mdot\approx3\times10^{-6}~\MdotUJ$ for \texttt{MagAcc},
near $\Mdot\approx3\times10^{-5}~\MdotUJ$ for \texttt{Polar},
and at higher $\Mdot$ (slightly beyond the plotted range) for \texttt{SpherAcc}.

In turn, the total extinction (of gas and dust taken together) must be
$\AHa\lesssim1$~mag for \texttt{MagAcc},
$\lesssim3$~mag for \texttt{Polar}, and
$\lesssim4$~mag for \texttt{SpherAcc}.
Stronger extinction would leave less flux than observed for any $\Mdot$.
This is broadly consistent with modelling of the
NIR SED, which suggests a significant amount of extinction ($\AHa\approx2$--10~mag or more; \citealp{Hashimoto+2020,wang21vlti,cugno21}). The highest values ($\approx10$~mag) might not be consistent with our findings ($\lesssim3$~mag),
but this could due to our simplifications in the radiative transfer geometry.%

Therefore, the accretion rate onto \PDSb is likely $\Mdot\approx2\times10^{-7}$--$10^{-4}~\MdotUJ$. This large spread folds in the uncertainties on the accretion geometry. The results are very similar for $\MP=3~\MJ$ or the \texttt{Cold} radius function instead (not shown).
This range of \Mdot is higher than $\Mdot=10^{-8}~\MdotUJ$ as derived by \citet{Haffert+2019}, but that was based on a low-mass extrapolation of empirical correlations for stars.
\citet{AMIM21L} found that it is invalid at low masses,
as can also be seen in Figure~\ref{fig:LHaMdot}
from the discrepancy between the \citet{Ingleby+2013} correlation for CTTSs and our curves.

\K{  %

\section{Application to the PDS~70 planets}
 \label{sec:PDS70}

Robust detections of \Ha emission at planet masses exist only for the \PDS objects \citep{wagner18,Haffert+2019} and for \Dlrmb \citep{Eriksson+2020}.  
The latter, a circumbinary low-mass companion exhibits hydrogen- and helium-line emission, which is naturally explained by %
accretion%
\footnote{%
There are several arguments against the alternative origin, chromospheric activity; see \citet{Eriksson+2020}.%
}.
The inferred accretion rate onto \Dlrmb is $\Mdot\approx10^{-9.5}$ or $\approx10^{-8}~\MdotUJ$, depending on the model used to convert the line luminosities into $\Mdot$. %
Even at %
$\Mdot\approx10^{-8}~\MdotUJ$,
extinction by either the gas or the dust (Figures~\ref{fig:DeltaF} or~\ref{fig:SigmadkapStb}, respectively) is entirely negligible. %

Therefore, in this section we apply our estimates of the absorption by the gas and by the dust  
to the other secure detections of accreting planets, \PDSb and~c.
We take however a different approach: instead of calculating the extinction for $(\Mdot,\MP)$ values from the literature, we use the additional piece of information available, the upper limit on the \Hb flux from \citet{Hashimoto+2020}.
Assume a mass. Do what for Mdot?
This lets us derive a minimum extinction at \Ha.
From this one gets a filling factor.
The corresponding mass is however not consistent with the assumed one.
Therefore:
Do not use the (n0,v0) whose width is consistent with the MUSE data because unresolved.
Use only the line-integrated fluxes.
Straightforward:
Can we obtain this extinction (from flux ratio) at this Mdot and Mp?
Yes/no/stretching the dust opacity.

\subsection[Deriving the accretion rate from the H beta upper limit]{Deriving the accretion rate from the \Hb upper limit}
 \label{sec:getMdotlowlim}

\citet{Hashimoto+2020} re-analysed the MUSE data of \citet{Haffert+2019} and placed upper limits on $\FHb$, the flux at \Hb, at the position of \PDSb\ and \PDSc.
The 3-$\sigma$ upper limit on the flux ratio is $\fluxratobs\equiv\FHb/\FHa\lesssim0.28$ for \PDSb\ and $\fluxratobs\lesssim0.52$ for \PDSc.
They identified in the theoretical shock models of \citet{Aoyama+2018,Aoyama+2020}
the combinations of $(n_0,v_0)$ that match the measured 10~\%\ and 50~\%\ width of the \Ha line. (Since these are microphysical and therefore local models, this is independent of whether the shock is on the planet's or the CPD's surface.) Those models however predict that at the surface of the object, $\FHb$ should be $\fluxratth\approx1$--1.1~times as strong as $\FHa$, the \Ha flux. %
This may seem surprisingly high given that \Hb is a higher-energy transition than \Ha and thus in principle weaker, but $\FHb\approx\FHa$ is due to the saturation of \Ha in the postshock region at high densities.
This theoretical ratio $\fluxratth$ is larger than 
the observed upper limits.

One possible way to reconcile the predicted flux ratio with the upper limit is to invoke the finite spectral resolution of MUSE \citep{uyama21}. If the intrinsic line width is narrower than observed, the matching $n_0$ and $v_0$ would be smaller than currently inferred. At those values, also $\fluxratth$ %
is smaller (see Figure~3 of \citealt{Hashimoto+2020}). Therefore, it could be possible to find a self-consistent solution.

However, differential absorption between \Ha and \Hb could also explain the discrepancy by a factor $\fluxratth/\fluxratobs\gtrsim2$--3. Neglecting emission from the accreting matter as we have done so far (see justification in Appendix~\ref{sec:cffull}), the measured upper limit on the flux ratio $\fluxratobs$ can be converted to a lower limit on $\AHa$, the extinction at \Ha, through
\begin{equation}
\label{eq:AvonRundp}
 \AHa > 2.5~\mathrm{mag}\times\frac{\log_{10}\left(\fluxratth/\fluxratobs\right)}{\left(\lmbdHa/\lmbdHb\right)^{-\pUV}-1},
\end{equation}
where $\pUV=\Delta\log\kappa/\Delta\log\lambda$ is the logarithmic average opacity slope (the extinction law) between \Hb ($\lmbdHb=486$~nm) and \Ha, %
a UV opacity index. Equation~(\ref{eq:AvonRundp}) follows exactly and straightforwardly from %
the pure-absorption solution $F=F_0\exp(-\Delta\tau)$ (Equation~(\ref{eq:L pure abs}), with the radial dependence cancelling out) of the radiative transfer equation. Equation~(\ref{eq:AvonRundp})
is independent of the source (gas or dust) of the opacity
and %
of $\fpg$ or the absolute opacity; only the slope of the extinction law matters.
The shallower the absorption slope (the smaller $|\pUV|$ is), the larger the extinction has to be for the differential extinction to hide the signal at \Hb, and for $\pUV=0$ or positive, there is no solution.

As long as some sub-micron dust particles are present (compare the dashed to the solid lines in Figure~\ref{fig:SigmadkapStb}), the extinction increases with decreasing wavelength,
such that $\pUV<0$.
This is what \citet{Hashimoto+2020} assumed,
and took
$\pUV=-1.75$ from \citet{draine89},
leading to
$\AHa>2.0$~mag %
for \PDSb,
and $\AHa>1.0$~mag for \PDSc, as they report and Equation~(\ref{eq:AvonRundp}) shows.

As \citet{Hashimoto+2020} noted, however, the \citet{draine89} fit
holds only down to 700~nm,  %
and the slope between \Hb and \Ha\xspace is  %
gentler (i.e.\ $|\pUV|$ smaller).
Indeed, the extinction law of \citet{cardelli89}, valid at \Ha and \Hb, has
$\pUV=-1.2$ ($\pUV=-0.9$) for $\RV=3.1$ ($\RV=5$), instead of $-1.75$.
\citet{wang19} recently adjusted the \citet{cardelli89} extinction law with $\RV=3.1$ and obtained $\pUV=-1.4$
(see Figure~\ref{fig:SigmadkapStb}b), which is also flatter.
For the opacity models considered in this work (Section~\ref{sec:dustopac}), we find roughly $\pUV\approx-0.5$ to $-1$.
Larger dust grain size distribution powerlaw exponent values %
$q$ and smaller values of $\amin$ make the slope steeper, that is, $\pUV$ more negative,
while 
$\amax$ essentially does not change the slope much (see Figure~3 of \citealp{woitke16}) and only decreases the absolute amount of extinction.
We note that the fundamental reason for the difference with the empirical ISM curve is not obvious. %
Using our opacity\footnote{We neglect the gas opacity, which will be justified below.} slope
values $\pUV$ raises the minimum $\AHa$, as Equation~(\ref{eq:AvonRundp}) shows,
to $\AHa\gtrsim4$--8~mag for \PDSb,  %
and to $\AHa\gtrsim2$--4~mag for \PDSc.  %
This is higher than in \citet{Hashimoto+2020}, who took $\pUV$ from
\citet{draine89},
The highest values of $\AHa$ might be unlikely because they could lead to absorption of the \Ha\ line to the point that it is indistinguishable compared to the surrounding (also extincted) continuum, whereas a line is unambiguously detected \citep{Haffert+2019}. However, whether a high extinction makes the line blend into the continuum or not depends on the (three-dimensional) radiative structure of the accretion flow (Appendix~\ref{sec:cffull}). Then, %
the line could remain detectable %
despite a high extinction.

A higher extinction than in \citet{Hashimoto+2020} implies that a larger filling factor of the accretion shock on the planetary surface is sufficient to explain the observed luminosity.
From $4\pi d^2 \FHa = 4\pi\RP^2\ffill \FHamod10^{-0.4\AHa}$, which leads to their Equation~(2),
\begin{subequations}
\begin{align}
 \log_{10}\ffill &= \frac{\AHa}{2.5~\mathrm{mag}} - Q,\\
  Q & \equiv -\log_{10}\left(\frac{d^2}{\RP^2}\frac{\FHa}{\FHamod}\right),  %
\end{align}
\end{subequations}
where $\FHa$ is the observed flux (energy per time per detector area) and $\FHamod$ is the surface flux from the model with a 10~\%\ and 50~\%\ line width consistent with the MUSE spectrum, $d=113$~pc is the distance to \PDS\ \citep{pecaut16}, and $\RP$ is the planetary radius, set here to $2~\RJ$ as in \citet{Hashimoto+2020} to simplify the present discussion.
They report   %
$Q=2.7$ ($Q=3.0$) for \PDSb\ (\PDSc), so that with our ranges of $\AHa$ inferred above we find
$\ffill\gtrsim0.1$
($\ffill\gtrsim0.007$--0.04).
For \PDSb, formally the minimum $\ffill$ can be larger than unity for the heaviest extinctions. This means only that the radius would need to be larger than %
assumed, %
which would still agree with models of forming planets (\citealp{morda12_I}).
\citet{Hashimoto+2020} derived $\ffill\gtrsim0.01$ ($\ffill\gtrsim0.003$) for \PDSb\ (\PDSc), which implies that our lower bounds have a greater constraining power.

The consequence is that the lower limits on the accretion rates are now higher than inferred in \citet{Hashimoto+2020} since $\Mdot$ is directly proportional to $\ffill$ given the other constraints and assumptions (see Equation~(\ref{eq:rho})). Thus the minimum accretion rates deduced from the upper limit on $\FHb$ are approximately a factor of ten higher:
\begin{subequations}
\label{eq:Mdot}
\begin{align}
\Mdot\gtrsim5\times10^{-6}~\MdotUJ~\mbox{for \PDSb},\\
\Mdot\gtrsim1\times10^{-6}~\MdotUJ~\mbox{for \PDSc}.
\end{align}
\end{subequations}
This is within the range of accretion rates typically considered in formation calculations (e.g.\ \citealp{boden00,Tanigawa+Tanaka2016,Emsenhuber+2020a}).

Are such large extinctions consistent with the planet properties?
From Figure~\ref{fig:DeltaF}, at accretion rates $\Mdot\sim10^{-6}$--$10^{-5}~\MdotUJ$, the contribution of the gas to the extinction
is $\AHa\lesssim0.1$~mag for \texttt{SpherAcc}, $\AHa\approx0$--1~mag for \texttt{Polar}, and $\AHa\approx0.5$--4~mag for \texttt{MagAcc} towards low masses of \PDSb, for which there are indications from different approaches \citep{bae19,Stolker+20b,wang21vlti}.
Except for the extreme value in the \texttt{MagAcc} case, %
this is clearly smaller than the $\AHa$ we derived. However, the minimum dust opacity to have $\taudHa\approx\AHa\approx1$ is approximately $\kapStbfpg\approx3$--30~cm$^2$\,g$^{-1}_{\textrm{gas}}$
(Figure~\ref{fig:SigmadkapStb}a), implying that $\kapStbfpg\approx10$--300~cm$^2$\,g$^{-1}_{\textrm{gas}}$ (10--100~cm$^2$\,g$^{-1}_{\textrm{gas}}$) is needed to obtain our inferred extinctions for \PDSb\ (\PDSc).
Looking at Figure~\ref{fig:SigmadkapStb}c, this is outside of the nominal estimated range of opacities given the highlighted $\fpg$. However, a dust abundance $\fpg\sim10^{-3}$--$10^{-2}$ would be sufficient to reach the calculated amount of extinction. Thus there is some tension but the required values for $\fpg$ remain in a reasonable range.

We note finally that the upper limits on $\FHb$ that \citet{Hashimoto+2020} reported were derived from one spectral channel,
which is 1.25~\AA\ wide, corresponding to only $\Delta v =77~\kms$.
Thus a fraction of the line flux could be missing, especially if the signal does not peak at the channel centre at $\lambda=4860.96$~\AA, which is blueward by $|\Delta v|=23~\kms$ of the central wavelength in air of $\lambda=4861.34$~\AA.
(Figure~\ref{fig:kappa2D} provides a counterexample with a line that does peak at $\Delta v\approx-20~\kms$.)
As for the \Ha\ fluxes, they came from three channels ($\Delta v=230~\kms$), which is more than sufficient to cover the whole flux from the shock-emitted line.
Thus the true limit on the integrated $\FHb$ flux could be higher,
increasing the $\fluxratobs$ upper limit and thus reducing the required extinction. Further data re-reductions or new measurements could be useful.

\subsection[Non-detection at Br alpha of a shock excess for PDS 70 b]{Non-detection at \Bra\ of a shock excess for \PDSb}

\citet{Stolker+20b} recovered for the first time \PDSb\ in the narrow filter NB4.05 of VLT/NACO centred at \Bra
($\lambda_{\mathrm{c}}=4.05~\upmu$m, effective width\footnote{From the SVO at \url{http://svo2.cab.inta-csic.es/theory/fps3/index.php?id=Paranal/NACO.NB405}. The width of 0.02~$\upmu$m mentioned by several (e.g.\ \citealp{janson08,quanz10,meshkat14fomal,kervella14,Stolker+20a,Stolker+20b})
is not correct. However, this does not change their results. The incorrect value was from \url{http://www.eso.org/sci/facilities/paranal/decommissioned/naco/inst/filters.html} (up to 2021 January), which nevertheless provides the right filter transmission profile.} $\Delta\lambda=0.0616~\upmu$m)  %
as part of the MIRACLES survey \citep{Stolker+20a}.
Using also their detection of \PDSb in the M$'$ filter, \citet{Stolker+20b} argued that the spectrum of \PDSb\ from 1~to 5~$\upmu$m is overall consistent with a blackbody.
Comparing the NB4.05 photometry with the best-fit spectrum, there is no evidence for a shock excess at NB4.05, where the thermal emission from the atmosphere is significant, contrary to the case at \Ha.

We can verify whether this is consistent with the shock models and the inferred extinction.
Indeed, the extinction is less strong at \Bra than at \Ha; could it then be that, based on the emission at \Ha, the extincted shock emission at \Bra would be expected to be visible?
We follow the approach of ``forced photometry'' (\citealp{lang16}; see also the discussion in \citealt{samland17}) as \citet{Stolker+20b} did for the ALMA measurement at 855~$\upmu$m. We take the 1-$\sigma$ uncertainty on the NB4.05 integrated flux as the upper limit on the shock emission line flux,
$\FBra\leqslant2.7\times10^{-16}$~erg\,s$^{-1}$\,cm$^{-2}$,
leading to
$\fluxratobs'\equiv\FBra/\FHa<1.0$ with $\FHa=8.1\times10^{-16}$~erg\,s$^{-1}$\,cm$^{-2}$ from \citet{Hashimoto+2020}.
Defining $\fluxratth'$ similarly to $\fluxratobs'$ and applying Equation~(\ref{eq:AvonRundp}) now between \Ha\ and \Bra, a lower limit on $\AHa$ predicts a reddened flux ratio $\fluxratthred'$
\begin{equation}
 \label{eq:fluxratobsshould}
 \fluxratthred' < \fluxratth' \times\exp{\left(   %
    \frac{\AHa\ln 10}{2.5~\mathrm{mag}} \left[1-\left(\frac{\lmbdHa}{\lmbdBra}\right)^{-\pNIR}\right]\right)},
\end{equation}
where $\pNIR$ is the opacity index between \Ha\ and \Bra\ in the
NIR%
, specifically at $\lmbdBra=4.052~\upmu$m.
From \citet{Aoyama+2020}, the ratio of fluxes at the object's surface is $\fluxratth'\approx10^{-2}$. %
Using $\pNIR=-1.75$ from \citet{draine89} and $\AHa=2.0$~mag (see above), Equation~(\ref{eq:fluxratobsshould}) predicts
$\fluxratthred'\lesssim 6\times10^{-2}$. %
In this case, indeed $\fluxratobs'$ and $\fluxratthred'$ are consistent: the reddened \Bra\ from the shock is not predicted to exceed the observed upper limit.
Using instead $\pNIR$ from the adopted dust models leads to the same conclusion:
because they have a $\pNIR$ shallower than $-1.5$ but 
a steeper opacity curve in the NIR than in the UV, %
the dust models discussed above only yield larger (i.e.\ less constraining) upper limits on $\fluxratthred'$. 

Thus, the predicted reddened flux ratio between \Bra\ and \Ha\ from the \citet{Aoyama+2018,Aoyama+2020} shock model is fully consistent with the observed upper limit. Put differently, the photometry at NB4.05 does not constrain the shock emission in the \Bra line, neither providing evidence for nor excluding it. Observations with higher spectral resolution at \Bra should be used to place more stringent constraints.

\subsection{Discussion of the results on the PDS~70 planets}
 \label{sec:discPDS70}

In summary, differential absorption between \Hb and \Ha can be explained by a
set of accretion and extinction parameters for the \PDS planets.
The extinction is derived to be $\AHa\gtrsim4$--8~mag for \PDSb\ and $\AHa\gtrsim2$--4~mag for \PDSc.
This %
is higher than the extinction of at most $\AR=3$~mag that \citet{wagner18} considered for \PDSb, and the upper end might be inconsistent with the fact that a line is clearly detected relative to the continuum.
This applies even more to the detection at F336W ($U$ band) by \citet{Zhou+2021} since $\AU\approx2\AHa$ according to \citet{cardelli89} or $\AU\approx1.3\AHa$ using our estimates.
Detailed radiative transfer would however be needed to assess this, while taking the emission from the gas and dust around the planet into account could lead to a lower requirement on the optical depth to explain the upper limit on \Hb.

The required dust opacity is tolerably higher than our estimate of realistic dust opacities and could be explained by a dust-to-gas ratio closer to the ISM value than we have argued and/or a greater relative proportion of small grains. Also, a line intrinsically narrower than observed by MUSE (for which \citealp{uyama21} provide support) would help alleviate the tension by implying a smaller amount of extinction.

Thanks to their new $K$-band medium-resolution ($R\approx500$; \citealp{lacour21}) spectroscopy from VLTI/GRAVITY combined with the existing data, \citet{wang21vlti}
provided evidence that a dusty atmosphere (a DRIFT-PHOENIX model on the one hand, or an extincted BT-Settl or Exo-REM atmosphere on the other; \citealp{helling08,charnay18}) matches the data clearly better than a blackbody. They find %
support for heavy extinction ($\AV\approx4$--8~mag for \PDSb, $\AV\approx15$--20~mag for \PDSc, although with a much larger uncertainty), which is in line with our conclusions.

Also, the non-detection of a shock excess in \Brg at 2.166~$\upmu$m by \citet{Christiaens+2019b} with VLT/SINFONI and \citet{wang21vlti}, as well as in other NIR lines (see \citealp{uyama21} for \Pab), is entirely consistent with the models of \citet{Aoyama+2020}, which predict that these lines should be much lower than the photospheric emission. The moderate resolution of the instruments lets any shock excess be spread into the continuum. %

\subsection{Discussion: Other possible sources of extinction}

\citet{wagner18} mentioned the ISM and the circumstellar and circumplanetary discs (CSD and CPD) as other possible sources of extinction. \citet{mueller18} found that the contribution of the ISM is very modest in the visual, which translates to only $\AR=0.04$~mag (Section~\ref{sec:oneex}). Similarly, to $3\sigma$, \citet{Wang+2020} constrained $\AV<0.15$~mag from the ISM. 

Concerning the CSD, \citet{bae19} and \citet{toci20} have  %
an azimuthally averaged gas column density of $\Sigma\sim10^{-3}$~cm$^2$\,g$^{-1}$ at the location of \PDSb.
From Figure~\ref{fig:SigmadkapStb}c, $\AR\approx\kapStbfpg\Sigma\sim10^{-4}$--0.1~mag.  %
This suggests that the CSD should not lead to any significant extinction.
The increased column density at the position angle of the planet is effectively considered in the calculation of the accretion flow.

Given the inclination of $i=52\degr$ \citep{keppler19}, the emission from the planet passes through a significant fraction of the column density only if the CPD has an aspect ratio greater than approximately $\arctan(i)\approx0.7$. %
\citet{gressel13} found such thick discs for masses $\MP\sim0.3~\MJ$ and \citet{ab09b} found thinner discs with an aspect ratio $\lesssim0.4$, decreasing with increasing mass.
Simulations with a higher resolution than available up to now and without potential smoothing would be required to make these results more robust, but parameters such as the strength of the viscosity also play an crucial role. Nevertheless, it seems likely, if $\MP\gtrsim\MJ$, that the CPD is sufficiently thin to not absorb significantly.%

If the \Ha\ is generated on the CPD surface, the inclination will lead to a higher optical depth by a factor of only $\approx1/\cos i=1.6$ since the disc is not too edge-on.
Thus, it seems realistic that the CPD and CSD material does not contribute to the absorption, especially for \PDSb, which is found in a gap. Only for \PDSc (close to the gap rim) could there be an effect, especially at short wavelengths such as \Ha and \Hb.

As mentioned by \citet{Aoyama+Ikoma2019},
absorption could in general be due 
to a disc wind,
with a sizeable effect due to the inclination.
This wind could be dusty, with the radiation pressure playing a key role and the dust porosity and size evolving significantly while being transported by the wind \citep{vinkov20}.
\citet{Thanathibodee+2020} find that the mass-loss rate in the wind is $\Mdot_{\mathrm{wind}}\sim10^{-11}~\MdotUS\sim10^{-8}~\MdotUJ$, which might lead to a very small surface density given the large area over which it is spread.
However, estimating the absorption by a disc wind is well beyond the scope of this work.
It is not clear which explanation(s) for the extinction are the most relevant, if any is present.
The situation is further complicated by the presence of sub-mm emission offset from the position of \PDSb\ \citep{Isella+2019}.
Dedicated simulations of the growth (and material properties) of dust for the \PDS\ system, continuing in the direction of \citet{bae19} and \citet{toci20}, could help %
understand this system that is unique amongst the ones discovered so far.

}  %

\section{Discussion}
 \label{sec:disc}

As \citet{Aoyama+Ikoma2019} point out, the models of \citet{Thanathibodee+2019},
which are an application of those of \citet{Hartmann+1994} to the planetary regime,
are premised on the assumption that, as for CTTSs, the \Ha\ emission
originates in the column(s) of gas accreting onto the planet.
Because the heating mechanism of these columns is highly unknown \citep{muzerolle01}, these models are parametrised by a maximal temperature $\Tmax$ in the column. \citet{Thanathibodee+2019} do not consider emission from the post-shock region, which is appropriate for the stellar case.
Thus in \citealt{Aoyama+2018}, \citet{Aoyama+Ikoma2019}, \citet{Aoyama+2020}, and this work, we are exploring a complementary approach. Namely, we are calculating the emission of \Ha\ and other lines by the postshock gas,
showing that it can be a detectable source,
and investigating here the absorption by the accreting material.

Fortunately, the structure of the line-emitting, cooling postshock region is also 
less uncertain than that of the accretion columns. It would be interesting to deal simultaneously and self-consistently with the emission from the shock itself as well as from the gas and dust accreting onto the planet.

In the following we discuss different aspects of this study or beyond it.
We 
comment on $\ffill$ values (Section~\ref{sec:ffill}),
discuss the time variability of the \Ha line (Section~\ref{sec:timevar}),
take a critical look at the temperature structure in the accretion flow (Section~\ref{sec:Tstructr}),
comment on asymmetries in the line profiles (Section~\ref{sec:lineasymm}),
discuss other sources of extinction (Section~\ref{sec:otherabs}),
and relate this work to other recent efforts dealing with extinction (Section~\ref{sec:cfSanchisSzul}).
In Appendix~\ref{sec:mit anderen Linien}, we discuss briefly what data exist on other lines for the \PDS objects, and what can be learned from this.

\subsection{Filling factor values}
 \label{sec:ffill}
With no dedicated hydrodynamical simulations of magnetospheric accretion onto planets,
we take as rough guidance the values for the filling factor inferred for CTTSs.
In the sample by \citet{Ingleby+2013}, $\approx80~\%$ of the stars
have an estimated global $\ffill < 10~\%$,
and often amounting to a few percents or fractions of a percent.
The same order-of-magnitude estimates have been reported in \citet{calvetgull98},
with there $\ffill$ typically less than 10~\%, using spectroscopic data.
More recently, \citet{robinson19} similarly found filling factors of tens of percent for a few objects from HST low-resolution observations.
Estimates derived from multi-wavelength time-series photometry can be less precise,
but they also indicate typically $\ffill<10~\%$--20~\%\ for the bulk of CTTSs
dominated by hot accretion spots (e.g.\ \citealp{bouvier93,bouvier95,venuti15}).

In this study, we have assumed that the accretion occurs only over one or several region(s) with $\Mdot\neq0$, while $\Mdot=0$ outside of this. More realistically, there will be a distibution of local accretion rates covering different fractions of the surface as for CTTSs (e.g.\ \citealp{bouvier07PPV,Ingleby+2013}). The resulting line profile in this case could be a linear combination of the profiles for each hot spot, but geometrical effects (not all spots being radially directly towards the observer) will complicate this picture.

Using the \citet{Hartmann+1994} model,
\citet{Thanathibodee+2019} describe the magnetospheric accretion flow
as originating from the CPD between the inner (or truncation) radius $R_\mathrm{i}$ and the outer radius $R_\mathrm{o}=R_\mathrm{i}+W_r$, where $W_r$ is the width of the flow at the launching point.
With these definitions,
the axisymmetric angles on the planet covered by accretion
are given by $\sin^2\theta_{\mathrm{i},\mathrm{o}}=\RP/R_{\mathrm{i},\mathrm{o}}$ \citep{Hartmann+1994}.
From Equation~(\ref{eq:ffilltheta}), the filling factor is %
\begin{equation}
 \ffill = \sqrt{1-\frac{\RP}{R_\mathrm{i}+W_\mathrm{r}}} - \sqrt{1-\frac{\RP}{R_\mathrm{i}}}.
\end{equation}
\citet{Thanathibodee+2019} consider a range of $R_i=(2$--8$)~\RP$ and $W_r=(1$--$6)~\RP$, which
corresponds to
$\ffill=1$~\%--20~\%.
They find a roughly flat distribution of $R_\mathrm{i}$ and $W_\mathrm{r}$ matching the observed fluxes
for \PDSb\ and \PDSc.
This corresponds also to the range we consider in this work.

\subsection{Time variability}
 \label{sec:timevar}
Time variability in the \Ha\ flux (and in other lines and filters as well)
is a well-known phenomenon for CTTSs (e.g.\ \citealp{herbst94,siwak18}).
Monitoring campaigns of their accretion variability showed that this is typically dominated
by the timescale of rotational modulation of the accretion features (e.g.\ \citealp{costigan14}),
that is usually of the order of 0.5--2~weeks (e.g.\ \citealp{roquette17}).
If we exclude the most ``extreme'' cases (for instance, unstable accretors),
which can exhibit prominent variations over timescales of hours,
the amplitude of the $\Mdot$ variability on CTTSs is typically observed to increase
from timescales of hours to timescales of days,
and then flatten out and remain approximately constant over timescales as long as years \citep{costigan14,sergison20}.
This indicates that the global structure of accretion persists over hundreds of rotational cycles,
albeit with smaller-scale variations on shorter timescales (e.g.\ \citealp{grankin07}).
The typical $\Mdot$ variability measured on week-long timescales
is $\approx0.4$--0.5~dex \citep{costigan14,venuti14},
out of which $\approx70~\%$ can be explained in terms of geometric modulation of the accretion shock.
This implies that the intrinsic $\Mdot$ variability on such timescales typically amounts to only $\approx0.15$~dex.
Also, radiation-magnetohydrodynamical simulations of CTTSs (e.g.\ \citealp{kurosawa13}) indeed find that both stable and unstable accretors have stochastically changing line profiles, with paradoxically a more stationary appearance of the line profile for unstable accretors.
The estimate of \citet{Thanathibodee+2020} for the variability of $\Mdot$ onto the PDS 70 star,
0.56~dex over the rotation cycle, is indeed consistent with these typical estimates
of accretion variability for CTTSs.  %
Time variability has also been seen in low-mass brown dwarfs,
for example in the $\approx47~\MJ$,
$\approx1$-Myr-old brown dwarf candidate DENIS~1538--1038 \citep{nguyenthanh20}.

For the planetary regime studied here, in
the case of magnetospheric accretion,
time variability in the strength of the accretion tracer
can come from (at least) the following:
\begin{itemize}
 \item[(i)] from variations of the accretion rate onto the planet, which might be on the keplerian timescale around it, %
 $P\sim2~\mathrm{d}\times[R/(10~\RJ)]^{3/2}[\MP/(3~\MJ)]^{-1/2}$ at cylindrical radius $R$;
 \item[(ii)] from variations of the magnetic field topology, with a timescale linked to the one over which the field-generating gas in the planet's interior moves;
 \item[(iii)] from variations of the viewing angle onto the hot spot (through the column or not), over the course of a planetary rotation ($\sim10$~h; \citealp{snellen14,bryan18,bryan20,wang21hr8799}); and
 \item[(iv)] from variations of the optical depth along the column, for example from fluctuations in the temperature structure. If the preshock region is transmissive (``optically thin''), its thermal time should set the timescale \citep{malygin17}; otherwise, the diffusion time is the relevant quantity. Simple estimates of their value is however challenging, but they are likely fast compared to the other, dynamical processes.
\end{itemize}
Concerning point~(iii):
it is not clear whether the light emitted at the shock needs to pass through
the accretion column, as we have assumed here;
this might be the case only at certain phases.

\citet{Zhou+2021} recently detected \PDSb at \Ha and the UV continuum ($U$ band) with HST at several epochs over five months, separated typically by weeks, and found no statistical evidence for variability in \Ha beyond the $\sim10$\,\%\ level.   %
These timescales correspond to the orbital period at tens of $\RJ$ from the planet (see item~(i) above). Constraints on variability of the signal on the timescale of a possible planet rotation would be a valuable extension.

It is not clear %
how to disentangle the variation in the signal
coming from optical-depth effects and from accretion-rate variations.
Variability in the accretion rate (item~(i) above) is expected over a wide range of timescales but it decreases towards short timescales \citep{g21}. Therefore, the accretion rate might show only small variations over a rotational period, but both high-resolution simulations and monitoring campaigns will be needed to assess this in detail.
Also, if the accretion rate and mass are such that extinction
is absent or negligible, variations in the \Ha signal would be due
only to variations in the accretion rate, for instance due to episodic accretion \citep{lubow12,brittain20,martin21}, which might not be periodic,
or to rotational modulation of the accretion features (e.g.\ hot spots),
which would be periodic. Ideally,
this could help distinguish the source of the variability while
providing constraints on the spin rate of young objects
\citep{bryan18,ginzburg20,bryan20}.

\subsection{Temperature structure}
 \label{sec:Tstructr}

Our assumption that the temperature decreases away from the accreting object
finds support in the empirical determination by \citet{petrov14} for a CTTS
using line ratios probing different sections of the accretion column,
and agrees with the self-consistent model of \citet{Martin1996}.
This temperature distribution contrasts with the one assumed in the models of \citet{Hartmann+1994} and \citet{muzerolle98a,muzerolle01} (see brief review in \citealt{bouvier07PPV}), which \citet{Thanathibodee+2019} use, and their more recent developments (e.g.\ \citealp{lima10}). There, the temperature profile scales everywhere inversely  with the density, and the maximum temperature is a free parameter.  %
\citet{hartmann16} suggest that magnetic effects could play a role in setting the temperature structure but this remains unknown. Thus our approximation is a possible one but its validity is currently difficult to assess.

\subsection{Line asymmetries}
 \label{sec:lineasymm}
When the line profile becomes asymmetric because of non-uniform absorption
(see Figures~\ref{fig:profilesWARM100}, \ref{fig:profilesWARM30}, and~\ref{fig:profilesMagAkk}),
a shift of the line peak ensues, generally to the blue side.
This is typically around $|\Delta v|\approx20~\kms$.
While this is only a fraction of the resolution of MUSE (FWHM of 120~$\kms$),
it is possible to determine the line centroid better than a resolution element.
Indeed,
\citet{Haffert+2019} reported shifts of
$\Delta v=25\pm8$ and~$30\pm9$~$\kms$ %
for \PDSb and~c relative to the star,
towards the red side. 
It seems to have the opposite sign relative to the models.
However, this was with respect to the stellar \Ha, which has an asymmetric line profile. Thus the signal itself could still be overall blue-shifted.

If there is any, the redshifted resonant absorption should be at a clearly larger velocity offset (see Figure~\ref{fig:profilesMagAkk}) than the Keplerian speed
of the planet on its orbit around its host, which is $\vK= 9.4~\kms\times\sqrt{\MSternen/a_{10}}$
for a circular orbit, where $\MSternen\equiv\MStern/(1~\MSun)$ and $a_{10}\equiv a/(10~\mathrm{au})$.
Thus the orbital motion will not be able to shift significantly the absorption at the central rest wavelength of \Ha.
Similarly, the spin broadening should not be important for the line shape as a whole since young planets have equatorial spin velocities of order $v\approx10~\kms$ \citep{snellen14,bryan18,bryan20,wang21hr8799}, which is much narrower than the line. However, the fine spectral features seen for the warm-population radii in the \texttt{Polar} or the \texttt{MagAcc} case (Figures~\ref{fig:profilesWARM30} and~\ref{fig:profilesMagAkk}b, respectively) would possibly be more challenging to distinguish.
On the other hand, this depends on the latitude from which the line is emitted.

Note that, puzzlingly, the line profiles of \citet{Thanathibodee+2019} do not display a redshift
despite the free-fall velocities that they should be obtaining where the emission is maximal,
judging from the equivalent problem for CTTSs in \citet{Hartmann+1994}.

\subsection{Other possible sources of extinction}
 \label{sec:otherabs}

In this work, we have dealt only with the extinction due to the material in the accretion flow within the Hill sphere of the planet.
The contribution from the ISM is in principle easy to estimate from the stellar SED.
In this section, we discuss other sources, looking at \PDSb as an example.

The \PDS\ planets have been found inside a large cavity (whose size depends on wavelength; \citealp{hashimoto15,keppler18,long18,Isella+2019}) that seems to be devoid of gas and dust,
except for some small amounts of various optically thin molecules \citep{facchini21}.
To model the CSD of \PDS,
\citet{bae19} and \citet{toci20} have  %
an azimuthally averaged gas column density of $\Sigma\sim10^{-3}$~cm$^2$\,g$^{-1}$ at the location of \PDSb.
From Figure~\ref{fig:SigmadkapStb}c, $\AR\approx\kapStbfpg\Sigma\sim10^{-4}$--0.1~mag.  %
This suggests that the CSD should not lead to any significant extinction.
Finally, \PDSc could be affected more severely by extinction because of its proximity to the gap edge \citep{Haffert+2019}. However, the extinction at ALMA wavelengths is not strong enough as to prevent the detection of a CPD around \PDSc \citep{benisty21}.

Given the inclination of $i=52\degr$ \citep{keppler19}, the emission from the planet passes through a significant fraction of the column density only if the CPD has an aspect ratio greater than approximately $\arctan(i)\approx0.7$. %
\citet{gressel13} found such thick discs for masses $\MP\sim0.3~\MJ$ and \citet{ab09b} found thinner discs with an aspect ratio $\lesssim0.4$, decreasing with increasing mass.
Given that \PDSb cannot be too light in order to still emit \Ha, its
CPD is likely sufficiently thin to not absorb significantly.

If the \Ha\ is generated on the CPD surface, the inclination will lead to a higher optical depth by a factor of only $\approx1/\cos i=1.6$ since the disc is not too edge-on.
Thus, it seems realistic that the CPD and CSD material does not contribute to the absorption, especially for \PDSb, which is found in a gap. Only for \PDSc (close to the gap rim) could there be an effect, especially at short wavelengths such as \Ha and \Hb.

As mentioned by \citet{Aoyama+Ikoma2019},
absorption could in general be due 
to a disc wind.
This wind could be dusty, with the radiation pressure playing a key role and the dust porosity and size evolving significantly while being transported by the wind \citep{vinkov20}.
\citet{Thanathibodee+2020} find that the mass-loss rate in the wind of \PDS is $\Mdot_{\mathrm{wind}}\sim10^{-11}~\MdotUS\sim10^{-8}~\MdotUJ$, which might lead to a very small surface density given the large area over which it is spread.
However, estimating the absorption by a disc wind in detail is well beyond the scope of this work.

Finally,
the reduction in the surface density at the location of the planet depends on the viscosity, which is still a mostly unconstrained parameter of CSDs. Thus it possible for the disc column density above the planet to be negligible for a given accretion rate.
We note that a non-extinguishing gas and dust surface density is one of the arguments made in the ``massive accreting gap planet'' model of \citet{Close2020}.
In summary, the case of no absorption by the CSD is a useful limit.

\subsection{Relation to other theoretical work dealing with extinction}
 \label{sec:cfSanchisSzul}

The recent theoretical study of \citet{Sanchis+2020} also deals with the absorption of the flux from a planet by material in the system.
Our approaches are similar but distinct:
they use interstellar medium (ISM) dust opacity values (we attempt to estimate them specifically for accretion onto planets, and consider also absorption by the gas) to address the extinction in different infrared filters (we deal with one accretion line).
The flow geometry is handled differently (in 3D in their case, albeit with a non-zero smoothing length),
the parameter space is surveyed to a different extent (more broadly in our case, and with \Mdot and $\MP$ independent), and the thermodynamics are treated differently (they assume isothermal gas, while we take the radiation transport into account), while the radiative transfer is similar in both.
Thus in many respects the study of \citet{Sanchis+2020} and ours are complementary.

\citet{Sanchis+2020} performed hydrodynamics simulations for specific stellar and planetary parameters, but being isothermal, their simulations can be rescaled (approximately; see their Section 3.2.3) to apply to other values.
For \PDSb, using a gas surface density  %
from \citet{keppler18} and scaling to the stellar mass of $\MStern\approx0.8~\MSun$, they considered masses of $\MP=0.5$--2.4~$\MJ$ and accretion rates $\Mdot=(2$--$3)\times10^{-8}~\MdotUJ$. %
This \Mdot is dictated by the rescaling of their simulations through (E.\ Sanchis 2021, pers.\ comm.)
\begin{equation}
 \label{eq:rscl}
\Mdot' = \Mdot \left(\frac{a}{a'}\right)^3
\frac{\Sigma_1'}{\Sigma_1}
\sqrt{\frac{\MStern'}{\MStern}},
\end{equation}
where non-primed quantities are the ones used in the actual simulations and primed quantities the rescaled ones, with $a$ the semimajor axis of the planet and $\Sigma_1$ the surface density at a fixed reference semimajor axis, for instance 1~au. For $(a,a')=(5.2,22)$~au, $(\Sigma_1,\Sigma_1')=(290,12.5)$~g\,cm$^{-2}$ (at 1~au), and $(\MStern,\MStern')=(1.6,0.76)~\MSun$ as appropriate for their parameter choices for \PDSb at 22~au, Equation~(\ref{eq:rscl}) yields $\Mdot'/\Mdot = 3.92\times10^{-4}$, that is, $\Mdot'=(2$--$3)\times10^{-8}~\MdotUJ$.
This is a somewhat low, but reasonable value.

We can derive the \Ha extinction that \citet{Sanchis+2020} would predict
and compare this to our results.
They obtained $K$-band extinctions $\AK=0.74$, 0.22, and 0.0030~mag
for $\MP=0.48$, 0.95, and 2.38~$\MJ$ respectively (E.\ Sanchis 2021, pers.\ comm.). %
With the opacity law they used \citep{cardelli89}, 
this translates to $\AHa\approx5.3$, 1.6, and 0.021~mag respectively.

The dependence on the planet mass is thus very strong:
the extinction varies from $\AHa\approx5$~to 0.02~mag
despite a change of only a factor of $\approx2.5$ in mass, for essentially the same accretion rate.
This seemingly does not agree with our results,
in which the extinction decreases much more slowly with mass: $\AHa\propto\Sigma$, and $\Sigma\propto\MP^{-0.5}$ approximately (Equation~(\ref{eq:Siginf})\footnote{This equation applies because at such small $\Mdot$, the shock temperature is low ($T_0\approx600$~K 
at most) and thus the dust does not sublimate down to the shock. Otherwise, Equation~(\ref{eq:taudHa simple num}) would apply.}).
However, this holds for a fixed accretion geometry,
and in their case higher-mass planets open a deeper gap, with a surface density above the planet reduced by a factor of $\approx2$ between the 0.48- and 0.95-$\MJ$ simulations ($\Sigma\sim\MP^{-1}$), and of ten between the 0.95- and 2.38-$\MJ$ simulations ($\Sigma\sim\MP^{-2.5}$).
The velocity is certainly not larger than the smoothing-free free-fall velocity from infinity we assume and, in fact, because of the gravitational smoothing, is likely much smaller. Therefore, to have the same $\Mdot$ despite a reduced density above the planet, the gas must be accreting nearer to the equator in their case. Thus our assumption that the shock radiation passes entirely through the accretion flow (Figure~\ref{fig:scenarios}) does not apply to their simulations and indeed provides an upper limit on the amount of absorption.
This is an important aspect of the quantitative differences.

Another aspect is that \citet{Sanchis+2020} calculate the extinguishing column density from a distance $r=0.03$--$0.1~\RHill$ outwards from the planet, due to the smoothing.
Given the rescaling, for \PDSb at 22~au this corresponds to $r=80$, 340, and $\approx200~\RJ$ as a starting radius for the integration for the $\MP=0.48$, 0.95, and 2.48~$\MJ$ simulations. Thus, their column densities are smaller than what higher-resolution simulations would yield.
Nevertheless,
the qualitative result
of a decreasing extinction with increasing mass should be robust.

The work of \citet{szul20} deals with the emission of hydrogen lines
by accreting planets
and the absorption of these lines by the surrounding gas.
We
discuss their approach to calculating the emission in \citet{AMIM21L}
and 
assess in \citet{Aoyama+2020} the applicability of \citet{Storey+Hummer1995}, which \citet{szul20} used, to the planetary accretion shock.

Here, we comment briefly on the relation between \citet{szul20} and this work.
As for the comparison to \citet{Sanchis+2020},
our approaches are in principle similar,
with the major advantage of \citet{szul20} that they too consider an intrinsically 3D gas structure, from
the three-dimensional simulations of \citet{szul16}.
As we have shown in Figure~\ref{fig:SigmadkapStb}, the dust opacities\footnote{%
Their ``gas-only'' opacity case
seems
to include the %
photoionisation of atoms \citep{draine03} but
no 
molecular-line opacities.
This would severely underestimate the opacity.%
}
of \citet{szul20} per gram of dust are consistent with the range that we consider as more likely, with a different net result because of our much lower assumed dust-to-gas mass ratio ($\fpg\sim10^{-5}$--$10^{-3.5}$ instead of $\fpg=10^{-2}$ in their case).

The main difference between \citet{szul20} and this work is again in the density structure, which is affected by the need to use a non-zero gravitational smoothing length ($\sim$ fraction of $\RHill$) and large grid cells ($\sim\RJ$), due to the fact that these are three-dimensional, global-disc and thus numerically %
expensive simulations. Judging by the density structures seen in their Figure~4, there is likely also a considerable amount of variability. Consequently, %
estimating
the mean influx rate would require dozens of snapshots.
Therefore, the extinction might be calculated more accurately in our work but our density structure is simplified and thus approximate, albeit in a different way.
The reader is refered to Appendix~B of \citet{Aoyama+2020} %
and also \citet{AMIM21L}
for the more important discussion of how the emission was computed in \citet{szul20}.

Altogether, in \citet{Sanchis+2020} or \citet{szul20}
the flow is not correctly captured from small scales downwards in their simulations (at best a few tens of $\RJ$; corresponding to medium scales downwards in our model). This implies that those studies cannot study reliably the contribution of the accretion flow to the extinction as we have done here.
\section{Summary and conclusions}
 \label{sec:summ}
Motivated by recent detections of accreting planets and by the non-detections from surveys at \Ha,
we have studied to what extent the gas and dust accreting onto a forming planet
can absorb the \Ha emission coming from the shock at the planet surface,
and what the spectral signatures of this extinction %
might be.

We surveyed the large parameter space of planet properties
and considered three accretion geometries:
spherical, polar, and magnetospheric (Section~\ref{sec:geo}).
For each parameter combination, we integrated the equation of radiative transfer along the accretion flow.
This integration is needed because of the strong wavelength, density, and temperature dependence of the gas opacity (Figure~\ref{fig:kappa gas Ha}).

The main take-aways of this study are the following:
\begin{enumerate}
\item
The accreting gas absorbs the line-integrated flux with 0.5~mag or more only for accretion rates $\Mdot\gtrsim3\times10^{-5}~\MdotUJ$
in spherical symmetry (\texttt{SpherAcc})
or $\Mdot\gtrsim3\times10^{-6}~\MdotUJ$ for polar accretion,
with a weak mass dependence (Figure~\ref{fig:DeltaF}).
In the \texttt{MagAcc} case the mass dependence is stronger.
Except for \texttt{MagAcc-Cold},  %
the amount of extinction increases towards smaller masses.
This is partly due to the increase in density with decreasing planet mass at a given \Mdot (see Equation~(\ref{eq:rho})). Another reason is that the opacity happens mostly to increase with decreasing mass (see Figure~\ref{fig:kappa gas Ha}).
Up to $\Mdot\approx3\times10^{-4}~\MdotUJ$ and for masses {above} $\MP=10~\MJ$, the gas absorbs with at most $\AHa\lesssim4$~mag in the \texttt{SpherAcc} and \texttt{Polar} geometries.
\item
At low \Mdot, the observed line profile is the same as the one leaving the planet, but moderate to high $\Mdot$ values lead to clear spectral features. There is no systematic trend in these features because they are due to several molecular and atomic lines. Larger amounts of extinction tend to be associated with a ``spiky'' spectrum showing features visible at a resolution of 10--20~$\kms$ ($R\gtrsim15,000$; Figure~\ref{fig:profilesMagAkk}) and suggest relatively localised absorption, as could occur in a magnetospheric accretion geometry.
\item

 Larger-scale ($\Delta v\sim100~\kms$) slopes in the gas opacity at \Ha\ introduce asymmetries in the line shape. This is often 
 more important than
 asymmetries in the line emerging from the shock.
\item
Based on the recent literature, we estimated that the accreting dust has an extinction curve shallower than in the ISM \citep{cardelli89,wang19}.
A conservative range of %
opacity values at \Ha is 
$\kapStbfpg\sim10^{-2}$--$10$~cm$^2$\,g$^{-1}_{\textrm{gas}}$ at $\fpg\sim10^{-5}$--$3\times10^{-4}$ (Figure~\ref{fig:SigmadkapStb}),
which is 10--$10^4$ times lower than in the ISM.
Our dust opacity implies that for most parameter combinations the absorption by the dust is negligible to small (Section~\ref{sec:dustabsorbcombined}).
\item
For a given planet mass there is a maximal \Ha luminosity because absorption increases more strongly with $\Mdot$ than $\LHa$ does (Figure~\ref{fig:LHaMdot}). Considering only the gas opacity, this maximum is $\LHa\approx10^{-4}~\LSun$ for $\MP\sim10~\MJ$. Lower masses peak at lower \LHa.
Therefore, for certain values, 
an %
\LHa measurement can be interpreted as two different accretion rates, a low or a high value.
Also considering some dust absorption implies \Mdot values between the two extremes.

\item
The current computational capacities impose a coarse spatial resolution of the flow near the planet in multidimensional studies. This affects dramatically their estimate of the flow's contribution to the absorption (Section~\ref{sec:cfSanchisSzul}). This highlights the complementarity of our approach, which assumes a simplified geometry but more realistic densities, temperatures, and velocities in the accretion flow and at the shock.
\item
The accretion rate onto \Dlrmb is much too low for %
absorption by gas or dust in the accretion flow to matter (Section~\ref{sec:app}).
For \PDSb, we used the indication of a low mass and the observed luminosity to derive from our models a range of possible accretion rates and extinctions. We found $10^{-7} \lesssim (\Mdot/\MdotUJ) \lesssim 10^{-4}$ and $\AHa\lesssim4$~mag ($\AV\lesssim5$~mag).

\end{enumerate}

Thus, for planets found in gaps (i.e.\ when the surface density of the circumstellar disc is negligible), it may not be necessary to correct for any extinction at \Ha within the system.
It also suggests that the paucity of detected planets might not be mainly due to heavy extinction by the accreting material but rather to a less efficient conversion of the accretion energy to \Ha than for CTTSs \citep{AMIM21L}.

We note that for a shock on a CPD, compared to the shock on a the planet surface, at a given total accretion rate the mass flux is spread over a much larger area, so that the absorption will likely be essentially zero. Thus the \Ha line should be smooth and shaped only by the postshock region.%

The complexity of the problem,
due to uncertainties in the accretion geometry and %
the dust opacity, %
will make detailed modelling of individual objects enlightening.
Obtaining high-resolution spectra will be an important breakthrough to overcome the limitations of MUSE, from which only upper limits on the line width can be set. With high-resolution spectra, our grid %
of line shapes %
could be used to perfom fits to the observed line profile,
using the total flux also as a fit criterion,
and ideally fitting at the same time other accretion tracers
such as \Hb, \Pab, \Paa, \Brg, \Bra, or metal lines.
\begin{acknowledgements}   %
It is a pleasure to
thank
V.~Suleimanov,   %
P.~Molli\`ere,   %
W.~Kley,         %
W.~B\'ethune,    %
M.~Flock,        %
J.~Dr\k{a}\.zkowska,   %
P.~Pinilla,      %
Ch.~Rab,         %
S.~Facchini,     %
T.~Stolker,      %
A.~Marconi,      %
N.~Choksi,       %
and M.~Keppler
for useful comments and for illuminating discussions on different aspects of this work.
Detailed answers
by and interesting discussions with
E.~Sanchis     %
are gratefully acknowledged.
We thank warmly Thomas M\"uller (MPIA/Haus der Astronomie) for producing Figure~\ref{fig:scenarios}, and
A.~Ziampras %
for his patient help
and for providing data for the disc of Figure~\ref{fig:scenarios}.
We thank Carsten Dominik for making available the versatile and powerful yet easy-to-use \texttt{OpTool} (\url{https://github.com/cdominik/optool}).
G-DM and RK acknowledge the support of the DFG priority program SPP 1992 ``Exploring the Diversity of Extrasolar Planets'' (KU~2849/7-1 and MA~9185/1-1).
G-DM also acknowledges the support from the Swiss National Science Foundation under grant BSSGI0\_155816 ``PlanetsInTime''.
Parts of this work have been carried out within the framework of
the NCCR PlanetS supported by the Swiss National Science Foundation.
RK acknowledges financial support within the Emmy Noether research group on
``Accretion Flows and Feedback in Realistic Models of Massive Star Formation''
funded by the German Research Foundation (DFG) under grants no.~KU 2849/3-1 and KU 2849/3-2.
GC thanks the Swiss National Science Foundation for financial support under grant number 200021\_169131.
This work was carried out in part at the Jet Propulsion Laboratory, California Institute of Technology, under a contract with NASA and with the support of Exoplanets Research Program grant 17-XRP17\_2-0081.
CFM acknowledges an ESO fellowship.
This project has received funding from the European Union's Horizon 2020 research and innovation programme
under the Marie Sk\l{}odowska-Curie grant agreement no.~823823 (DUSTBUSTERS).
This work was partly supported by the Deutsche Forschungsgemeinschaft (DFG, German Research Foundation),
Ref.\ no.~FOR 2634/1 TE 1024/1-1.
Support for this work was provided by NASA through the NASA Hubble Fellowship grant \#HST-HF2-51436.001-A awarded by the Space Telescope Science Institute, which is operated by the Association of Universities for Research in Astronomy, Incorporated, under NASA contract NAS5-2655.
DK acknowledges partial financial
support from the Center for Space and Habitability (CSH), the PlanetS National Center of Competence in Research (NCCR).
Support for this work was provided by NASA through the NASA Hubble Fellowship grant HST-HF2-51472.001-A awarded by the Space Telescope Science Institute, which is operated by the Association of Universities for Research in Astronomy, Incorporated, under NASA contract NAS5-26555. 
The results reported herein benefited from collaborations and/or information exchange within NASA's Nexus for Exoplanet System Science (NExSS) research coordination network sponsored by NASA’s Science Mission Directorate.
LV acknowledges support by an appointment to the NASA Postdoctoral Program at the NASA Ames Research Center, administered by Universities Space Research Association under contract with NASA.
MJ gratefully acknowledges support from the Knut and Alice Wallenberg Foundation.
We acknowledge support by the High Performance and Cloud Computing Group at the Zentrum f\"ur Datenverarbeitung of the Universit\"at T\"ubingen, the state of Baden-W\"urttemberg through bwHPC and the German Research Foundation (DFG) through grant no.~INST 37/935-1 FUGG.
This publication makes use of VOSA, developed under the Spanish Virtual Observatory project supported by the Spanish MINECO through grant AyA2017-84089 \citep{bayo08}. VOSA has been partially updated by using funding from the European Union's Horizon 2020 Research and Innovation Programme, under Grant Agreement Nr.~776403 (EXOPLANETS-A).
This research has made use of NASA's Astrophysics Data System Bibliographic Services.
All figures except Figure~\ref{fig:scenarios} were produced using \texttt{gnuplot}'s terminal \texttt{epslatex} with the font package \texttt{fouriernc}.
\end{acknowledgements}

\bibliographystyle{aa}

\begin{appendix}

\section{Including the emission from the infalling matter}
 \label{sec:cffull}

We assess to what extent the emission by the accreting gas and dust can be neglected.
Keeping the emission term, the formal solution of the radiative transfer equation (Equation~(\ref{eq:dIds})) reads at a given wavelength (omitting the $\lambda$ subscript)
\begin{equation}
 I(\tau)=I(\taumax)e^{-\taumax-\tau} + \int_\tau^\taumax S(\tau')e^{-(\tau'-\tau)}d\tau',
\end{equation}
where $S$ is the source function, $I$ is the outward intensity, and $\tau$ is the inward optical depth with $\tau=0$ formally at $r=\infty$ (at the observer) and $\tau=\taumax$ at the shock (the surface of the planet). Therefore, the intensity at the observer is, for a purely radial ray,
\begin{equation}
 \label{eq:Iform}
 I(\tau=0) = I(\taumax)e^{-\taumax} + \int_0^\taumax S(\tau')e^{-\tau'} d\tau',
\end{equation}
so that, neglecting limb darkening, the luminosity at the observer is given simply by $L=4\pi\RP^2 \times\pi I(\tau=0)$.
The emerging intensity is made up of the attenuated intensity from the shock (first term) and the weighted integral of the source function (second term).
Assuming LTE, $S=B$ is a known function if the density, opacity, and temperature profiles are given, as we do assume here, making it possible to compute the integral in Equation~(\ref{eq:Iform}).
For simplicity in this work, we have kept only the first term in Equation~(\ref{eq:Iform}) (compare to Equation~(\ref{eq:L pure abs})). However, for large total optical depths $\taumax$, it can become negligible relative to the emission term.

\begin{figure*} %
 \centering
 \includegraphics[width=0.49\textwidth]{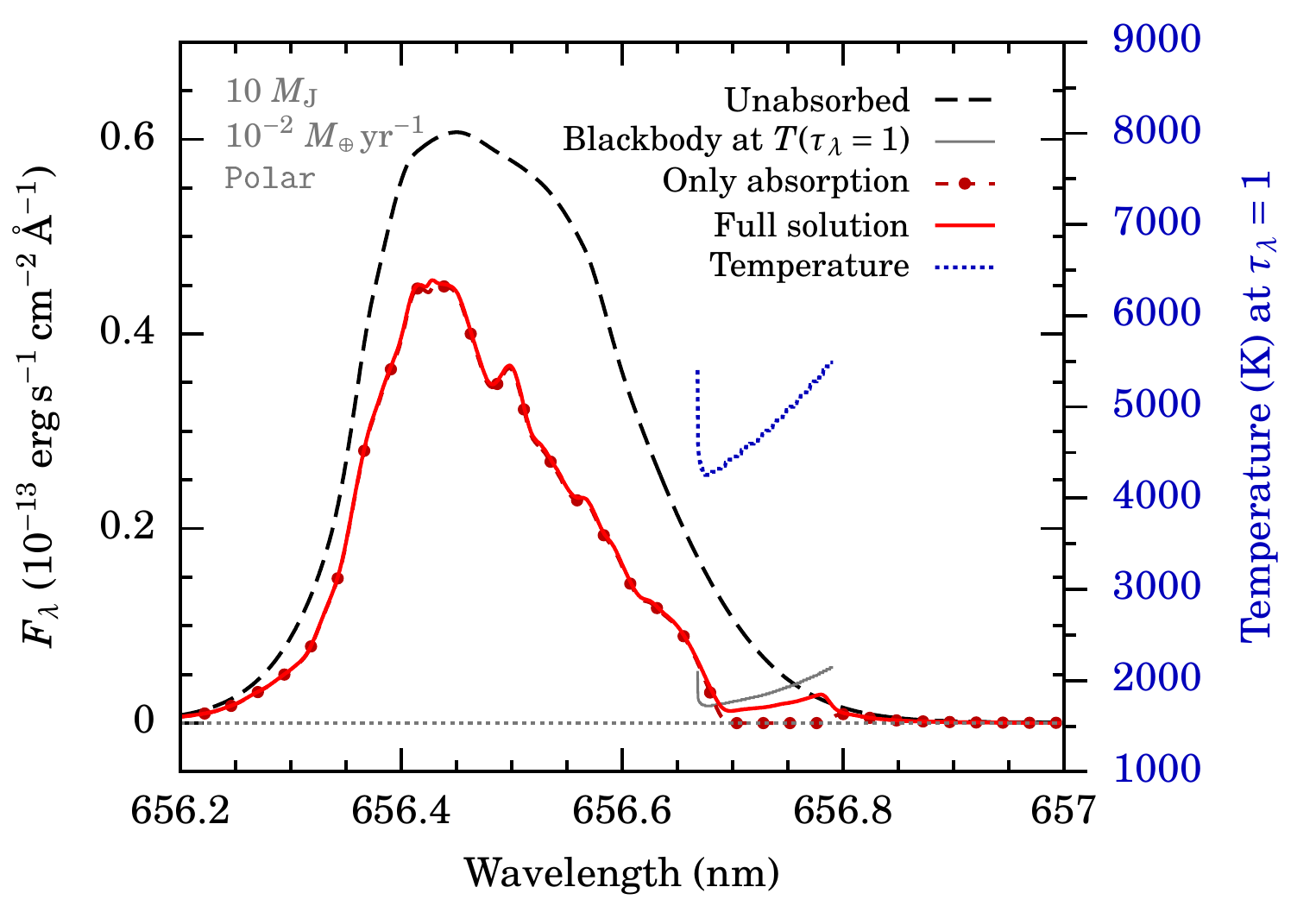}
 \includegraphics[width=0.49\textwidth]{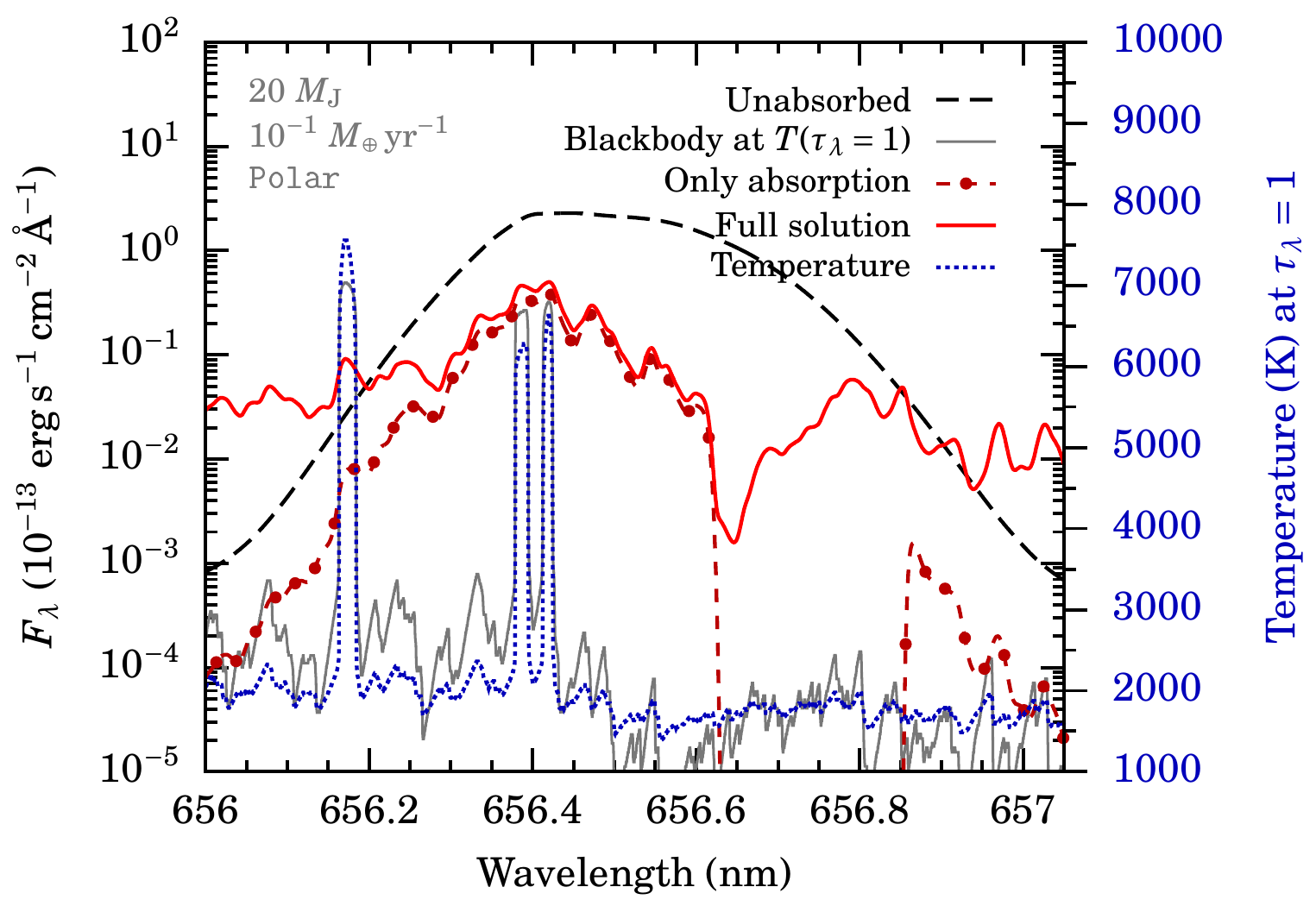}
\caption{
Full solution to the radiative transfer approximation (Equation~(\ref{eq:dIds})) including the emission term (see Equation~(\ref{eq:Iform})) compared with the approximate, absorption-only solution (Equation~(\ref{eq:L pure abs})).
We show the flux leaving the planet (dashed black line), the solution taking only absorption into account (dark red dashed line with dots), and the full solution (pale red full line), as well as the source function (blackbody intensity) at the location where $\tau_\lambda=1$ radially inwards (grey line), for those wavelengths where $\tau_\lambda=1$ is reached. The temperature of the blackbody there is shown against the right axis (blue dotted line).
\textit{Left panel:} Case of Figure~\ref{fig:kappa2D}. The strong preshock absorption at the Doppler-shifted \Ha\ resonance near the planet is filled in somewhat by the emission in the accretion flow, almost to the level of the blackbody at $\tau_\lambda=1$, roughly according to the Milne--Barbier--Uns\"old relationship.
\textit{Right panel:} Case of polar accretion for $\Mdot=3\times10^{-4}~\MdotUJ$, $\MP=20~\MJ$ for the cold-population radius fit ($\RP=3.4~\RJ$). Outside of the line centre, the outcoming intensity is much higher than the source function at $\tau_\lambda=1$ and thus the full solution is not roughly the sum of the absorption-only curve and the source function at $\tau_\lambda=1$.%
}
\label{fig:fullI}
\end{figure*}

\begin{figure} %
 \centering
 \includegraphics[width=0.49\textwidth]{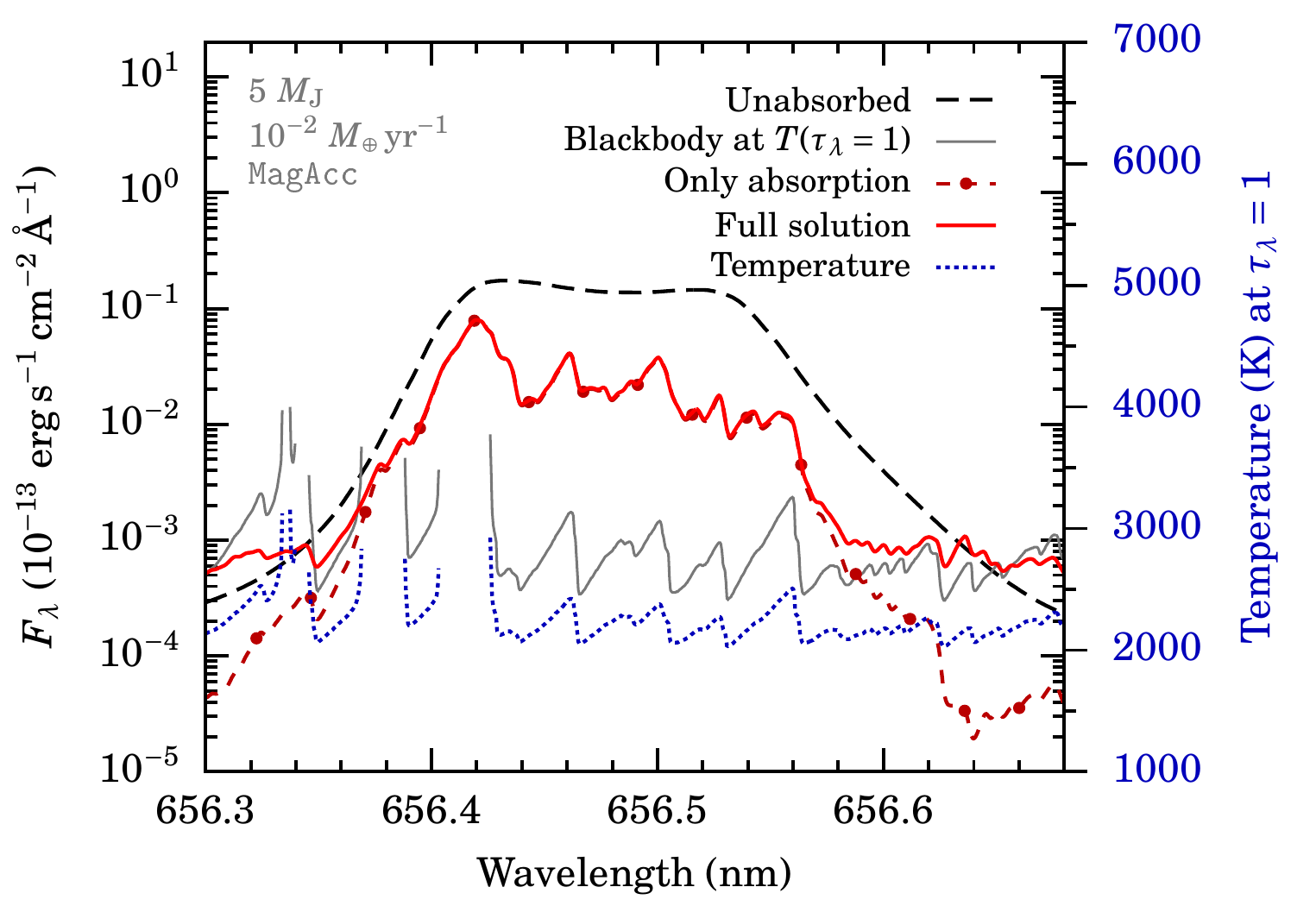}
\caption{
As in Figure~\ref{fig:fullI}, but for a case from \texttt{MagAcc} (top left in Figure~\ref{fig:profilesMagAkk}b). The emission from the accreting material region is important only in the far wings.
}
\label{fig:fullIMagAkk}
\end{figure}

We show in Figure~\ref{fig:fullI} the result of including the emission term for two examples. This is compared to the correction provided by the Barbier--Eddington (perhaps more accurately ``Milne--Barbier--Uns\"old''; see a rewiew of its history in \citealt{paletou18}) relationship. It states that the emerging intensity, less the first term in Equation~(\ref{eq:Iform}), is equal to the value of the source function at a depth $\tau=1$. Importantly, the derivation assumes that the source function is a linear function of the optical depth: $S=a+b\tau$, for constant $a$ and $b$.

In the first case (Figure~\ref{fig:fullI}a), which uses the parameters of Figure~\ref{fig:kappa2D}, the full solution is practically identical with the absorption-only solution, except at the strong absorption caused by the Doppler-shifted \Ha\ resonance near the planet at $\lambda\approx656.7$--656.8~nm (see the solid yellow region in Figure~\ref{fig:kappa2D}). The Barbier--Eddington or Milne--Barbier--Uns\"old correction (MBU; grey line) is non-zero only there because the total optical depth at the other wavelengths is less than unity. This correction is only some 20--70~\%\ away from the full solution and thus yields a reasonable approximation. The full solution is lower than predicted by the MBU relationship, implying that the observable intensity effectively comes from a region of lower temperature and therefore lower optical depth. This is because the source function, a blackbody, is a monotonic function of the temperature and the temperature increases with increasing optical depth (decreasing radius).

In the second example (Figure~\ref{fig:fullI}b), the absorption-only curve profile is 1--2~dex fainter than the flux at the planet's surface (i.e.\ $\tau_\lambda\approx2$--4) outside of the resonance, where the flux is effectively zero. For most of the range shown, the source function at $\tau_\lambda=1$ is orders of magnitude smaller than the absorption-only curve. However, in the full solution the flux can be as high as the absorption-only line (near the centre) or even much higher (in the wings, including the Doppler-shifted resonance). Where it is much higher, the MBU relationship can be seen not to hold. The reason is that the source function is a strongly non-linear function of the optical depth: it---or more relevantly the emissivity $j_\lambda = \alpha_\lambda S_\lambda=\alpha_\lambda B_\lambda$---is nearly zero from $\tau_\lambda=0$ to an optical depth $\tau_\lambda>1$, where it rises sharply. Thus the emission mostly comes from a depth with a high blackbody temperature.
For example, at $\lambda=656.0$~nm, this transition occurs at $\tau_\lambda=1.8$, and the source function at $\tau_\lambda=1.89$ is equal to the outcoming intensity.

Since in Figure~\ref{fig:fullI}b the extinction structure of the flow does not depend too much on wavelength, the resulting line shape is relatively flat, within a factor of ten. The only exceptions are near the line peak, where the attenuated shock flux (first term of Equation~(\ref{eq:Iform})) is important, and in the blue part of the Doppler-shifted \Ha\ resonance, where the emission is noticeably less strong. Also, in the narrow opacity window at $\lambda=656.17$~nm the outcoming flux is smaller by 1~dex than that coming from a blackbody at $T(\tau_\lambda=1)\approx7500$~K, whereas near 656.37~and 656.42~nm the observable flux is indeed approximately equal to the sum of the extincted flux from the planet surface and that of a blackbody at $T(\tau_\lambda=1)$.

A further example is provided in Figure~\ref{fig:fullIMagAkk}, now for $\Mdot=3\times10^{-5}~\MdotUJ$, $\MP=5~\MJ$ in the warm population ($\RP=4.5~\RJ$).
The absorption-only line matches the full solution everywhere except in the outer parts of the line. This is however visible only because the fluxes are displayed on a logarithmic scale; in reality it is only a very small difference.

The example of Figure~\ref{fig:fullI}b is an extreme one (very high accretion rate and mass), and in most cases the emission by the gas accreting onto the planet is negligible. When the optical depth to the planet is not very large, the extincted shock flux will likely dominate over the emission from the accretion flow. Near the line wings, emission from the accretion flow emission or the planet's can become important, leading to an apparently narrower line, but this should be generally only a small correction because the shock (peak or total) flux is large.

\section{Line profiles for the cold-population radii}
 \label{sec:line profiles COLD}

\begin{figure*}[ht] %
 \centering
 \includegraphics[width=0.8\textwidth]{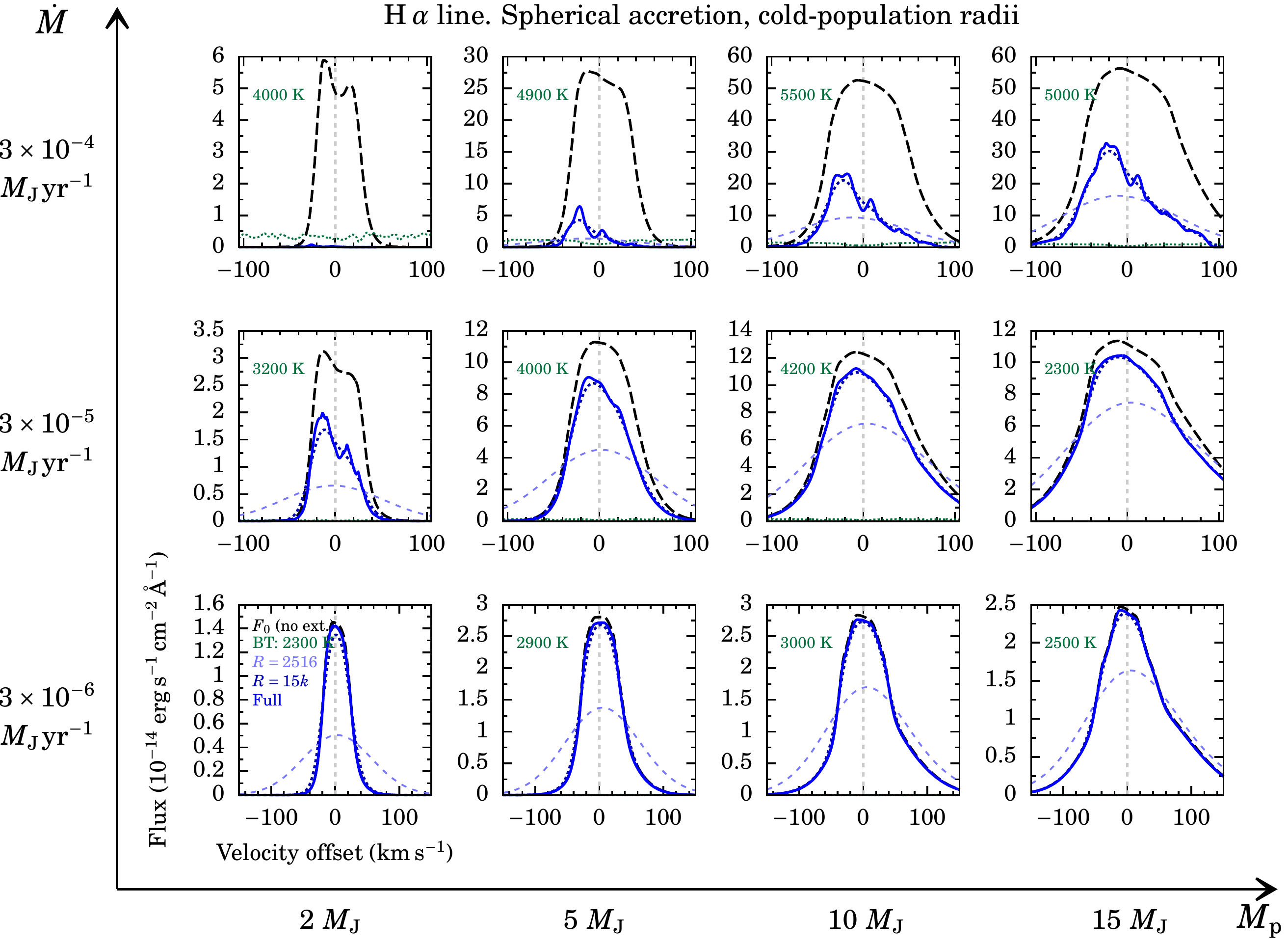}
\caption{
As in Figure~\ref{fig:profilesWARM100}, but for \texttt{SpherAcc-Cold}.
}
\label{fig:profilesCOLD100}
\end{figure*}

\begin{figure*}[ht] %
 \centering
 \includegraphics[width=0.8\textwidth]{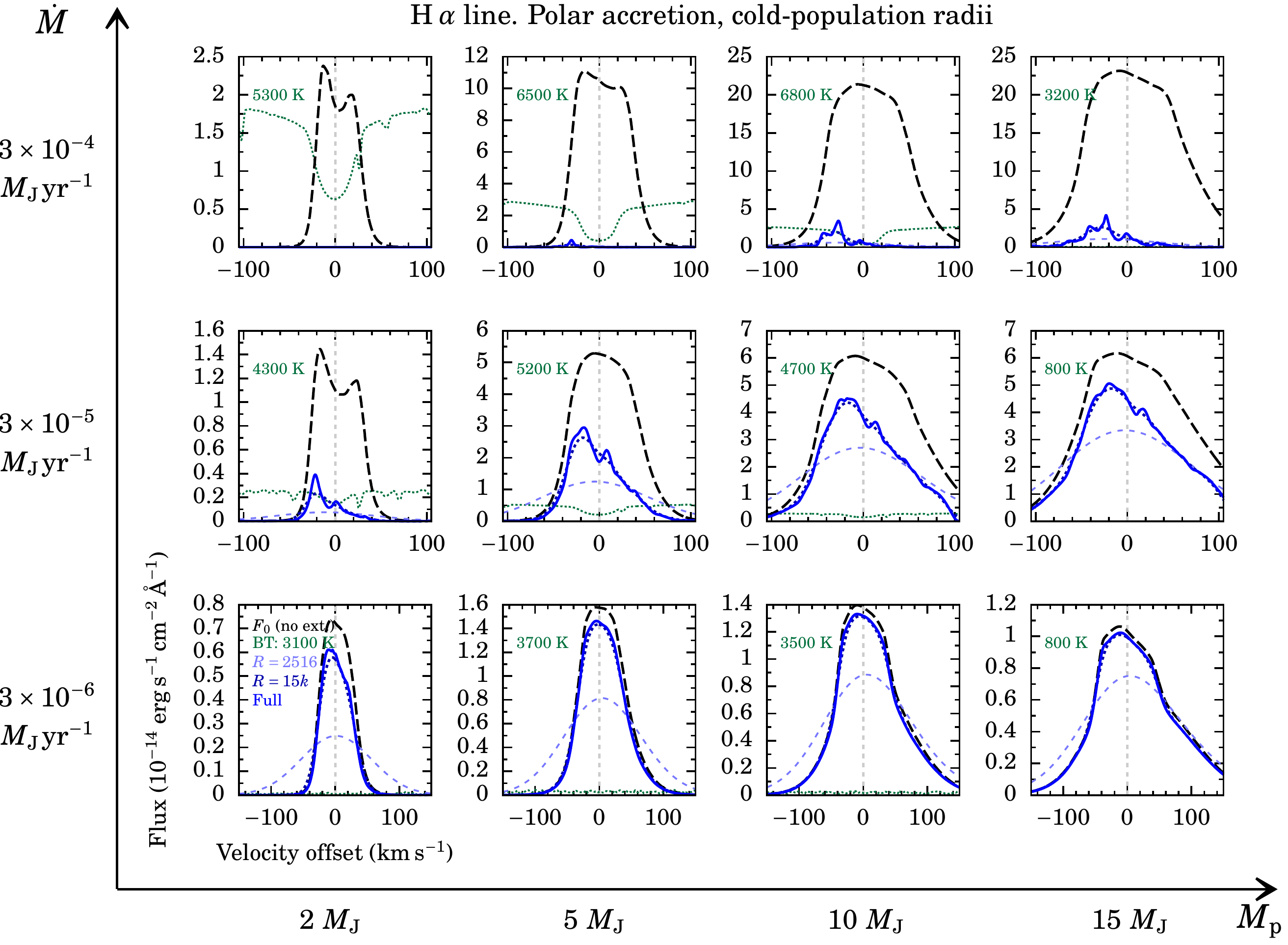}
\caption{
As in Figure~\ref{fig:profilesWARM100}, but for \texttt{Polar-Cold}.
}
\label{fig:profilesCOLD30}
\end{figure*}

Figures~\ref{fig:profilesCOLD100} and~\ref{fig:profilesCOLD30} show a grid of line profiles for the \texttt{SpherAcc-Cold} and the \texttt{Polar-Cold} scenarios, as in Figures~\ref{fig:profilesWARM100} and~\ref{fig:profilesWARM30} respectively. See the description in Section~\ref{sec:line profiles}. The \texttt{MagAcc-Cold} line profiles are shown in Figure~\ref{fig:profilesMagAkk}.

\section{Using measurements of other lines}
 \label{sec:mit anderen Linien}

In this section, we briefly recall the upper limits available on the fluxes of lines other than \Ha, and derive from the \Hb upper limit of \citet{Hashimoto+2020} constraints on the total extinction.
We follow the approach of \citet{Hashimoto+2020} but provide the expression for the extinction explicitly.

\citet{Stolker+20b} recovered for the first time \PDSb\ in the narrow filter NB4.05 of VLT/NACO centred at \Bra
($\lambda_{\mathrm{c}}=4.05~\upmu$m, effective width\footnote{From the SVO at \url{http://svo2.cab.inta-csic.es/theory/fps3/index.php?id=Paranal/NACO.NB405}. The width of 0.02~$\upmu$m mentioned by several (e.g.\ \citealp{janson08,quanz10,meshkat14fomal,kervella14,Stolker+20a,Stolker+20b})
is not correct. However, this does not change their results. The incorrect value was from \url{http://www.eso.org/sci/facilities/paranal/decommissioned/naco/inst/filters.html} (up to 2021 January), which nevertheless provides the right filter transmission profile.} $\Delta\lambda=0.0616~\upmu$m)  %
as part of the MIRACLES survey \citep{Stolker+20a}.
However, there is no evidence for a shock excess at NB4.05, where the thermal emission from the atmosphere is significant. %
Also, using VLT/SINFONI, \citet{Christiaens+2019b} and \citet{wang21vlti} did not detect a shock excess in \Brg at 2.166~$\upmu$m, nor did \citet{uyama21b} in \Pab with Keck/OSIRIS.
We note that this is
consistent with the models of \citet{Aoyama+2020}, which predict that these lines should be much lower than the photospheric emission. The moderate resolution of the instruments dilutes any shock excess into the continuum.

The measured upper limit on the flux ratio between \Hb and \Ha, $\fluxratobs$, can be converted to a lower limit on $\AHa$
through
\begin{equation}
\label{eq:AvonRundp}
 \AHa > 2.5~\mathrm{mag}\times\frac{\log_{10}\left(\fluxratth/\fluxratobs\right)}{\left(\lmbdHa/\lmbdHb\right)^{-\pUV}-1},
\end{equation}
where $\pUV=\Delta\log\kappa/\Delta\log\lambda$ is the logarithmic average opacity slope (the extinction law) between \Hb ($\lmbdHb=486$~nm) and \Ha, %
a UV opacity index.
The ratio of the \Ha and \Hb surface fluxes is $\fluxratth\equiv\FHb/\FHa$,
with typically $\fluxratth\approx1$--1.1.
This may seem surprisingly high given that \Hb is a higher-energy transition than \Ha and thus in principle weaker, but $\FHb\approx\FHa$ is due to the saturation of \Ha in the postshock region at high densities.
Equation~(\ref{eq:AvonRundp}) follows exactly and straightforwardly from %
the pure-absorption solution $F=F_0\exp(-\Delta\tau)$ (Equation~(\ref{eq:L pure abs}), with the radial dependence cancelling out) of the radiative transfer equation. Equation~(\ref{eq:AvonRundp})
is independent of the source (gas or dust) of the opacity
and %
of $\fpg$ or the absolute opacity; only the slope of the extinction law matters.
The shallower the absorption slope (the smaller $|\pUV|$ is), the larger the extinction has to be for the differential extinction to hide the signal at \Hb, and for $\pUV=0$ or positive, there is no solution.

Thus, combining the upper limit on $\FHb/\FHa$ %
with our extinction law
yields
$\AHa\gtrsim4$--8~mag for \PDSb  %
and %
$\AHa\gtrsim2$--4~mag for \PDSc.  %
This is marginally consistent with the constraints from Figure~\ref{fig:MPktAHa},
which is satisfactory given the different assumptions.
For comparison,
\citet{Hashimoto+2020} used 
$\pUV=-1.75$ from \citet{draine89},
leading to
$\AHa>2.0$~mag %
for \PDSb,
and $\AHa>1.0$~mag for \PDSc.
However, as they noted, the \citet{draine89} fit holds only down to 700~nm and the slope between \Hb and \Ha\xspace is
gentler (i.e.\ $|\pUV|$ smaller; see the end of Section~\ref{sec:dustopac}).
Thus our higher estimate should be more realistic.

\end{appendix}

\end{document}